\documentclass[fleqn,usenatbib]{mnras}

\usepackage{newtxtext,newtxmath}
\usepackage[T1]{fontenc}

\DeclareRobustCommand{\VAN}[3]{#2}
\let\VANthebibliography\thebibliography
\def\thebibliography{\DeclareRobustCommand{\VAN}[3]{##3}\VANthebibliography}

\usepackage{graphicx}	% Including figure files
\usepackage{amsmath}	% Advanced maths commands
\usepackage{subcaption}
\usepackage{caption}
\usepackage[T1]{fontenc}
\usepackage{float}

\newcommand{\figref}[1]{Fig.~\ref{#1}}
\newcommand{\tabref}[1]{Table~\ref{#1}}

\title[Characterising the centre of the galaxy
]{The inner galaxy as a window on its early assembly: lessons from a cosmological Milky Way analogue}

\author[T. Tomlinson et al.]{Thomas Tomlinson$^{1}$\thanks{E-mail: thomas.j.tomlinson@durham.ac.uk},
Francesca Fragkoudi$^{1, 2}$,
Andreia Carrillo$^{1, 3, 4}$,
Alis Deason$^{1,3}$, 
Azadeh Fattahi$^{1,5}$,
\newauthor
Rüdiger Pakmor$^{6}$, 
Robert J. J. Grand$^{7}$, 
Facundo A. Gómez$^{8}$, 
Freeke van de Voort$^{9}$, 
Rebekka Bieri$^{10}$\\
$^{1}$Institute for Computational Cosmology, Department of Physics, University of Durham, South Road, Durham DH1 3LE, UK\\
$^{2}$School of Sciences, European University Cyprus, Diogenes Street, Engomi, 1516 Nicosia, Cyprus \\
$^{3}$Centre for Extragalactic Astronomy, Department of Physics, University of Durham, South Road, Durham DH1 3LE, UK\\
$^{4}$Department of Physics and Astronomy, Carleton College, 1 North College St., Northfield, MN 55057, USA\\
$^{5}$The Oskar Klein Centre, Department of Physics, Stockholm University, Albanova University Center, 106 91 Stockholm, Sweden\\
$^{6}$Max-Planck-Institut für Astrophysik, Karl-Schwarzschild-Str. 1, D-85748, Garching, Germany\\
$^{7}$Astrophysics Research Institute, Liverpool John Moores University, 146 Brownlow Hill, Liverpool, L3 5RF, UK\\
$^{8}$Departamento de Astronomía, Universidad de La Serena, Av. Juan Cisternas 1200 Norte, La Serena, Chile\\
$^{9}$Cardiff Hub for Astrophysics Research and Technology, Department of Physics and Astronomy, School of Physical, Chemical and Environmental Sciences, \\Cardiff University, Queen’s Buildings, Cardiff, CF24 3AA, UK\\
$^{10}$Institut für Astrophysik, Universität Zürich, Winterthurerstrasse 190, 8057 Zürich, Switzerland
}

\date{Accepted XXX. Received YYY; in original form ZZZ}

\pubyear{2026}

\begin{document}
\label{firstpage}
\pagerange{\pageref{firstpage}--\pageref{lastpage}}
\maketitle


\begin{abstract}
Characterising the assembly history of the inner Milky Way is a key goal of upcoming spectroscopic surveys, yet disentangling its overlapping stellar components remains challenging. We investigate the chemodynamics of the centre of a Milky Way analogue in the Auriga Superstars cosmological hydrodynamical simulation suite, focusing on a high-resolution halo ($\approx 800\,\rm{M}_\odot$ star particles) with a realistic galactic bar and accretion history. We classify stars within the inner 4\,kpc into proto-galaxy, subsequent in-situ, and accreted components to assess their distinguishability using kinematics, dynamics, ages, and metallicities. Stellar net rotation declines sharply between $\rm [Fe/H]=0$ and $\rm [Fe/H]=-1$, marking the transition from disc-like in-situ stars to the proto-galaxy. While significant chemical and dynamical overlaps complicate the search for structures in energy–angular momentum ($E-L_z$) space, metal-poor stars ($\rm [Fe/H]<-1$) at low energy form a remarkably pure ($>97\%$) sample of proto-galaxy stars. Conversely, the high-energy, metal-poor tail offers the best chance of detecting accreted debris. However, at high energies the inner accreted halo is dominated by low-mass systems ($M_{\text{star}} \approx 10^7 - 10^9\,\rm{M}_\odot$), not the earliest massive mergers. Dynamically, few clues remain from early mergers, and early enrichment is rapid and similar across progenitors. In-situ contamination is an issue for $\rm [Fe/H]>-1$ throughout $E-L_z$ space. Overall, while accurate stellar ages and high-precision chemical abundances may help further constrain the Milky Way’s history, we should be cautious when associating stars with individual merger events in the inner galaxy.
\end{abstract}

\begin{keywords}
Galaxy: centre -- Galaxy: kinematics and dynamics -- Galaxy: formation
\end{keywords}

%%%%%%%%%%%%%%%%%%%%%%%%%%%%%%%%%%%%%%%%%%%%%%%%%%

%%%%%%%%%%%%%%%%% BODY OF PAPER %%%%%%%%%%%%%%%%%%

\section{Introduction}

One of the key questions in astronomy is how galaxies such as the Milky Way came to be. Observations of external galaxies provide glimpses into galaxies at different evolutionary stages. This can help inform us on the hierarchical assembly of galaxies and the frequency of mergers at a range of redshifts \citep{conselice_2003, bluck_2012, duncan_2019, dalmasso_2024}, as well as the timescales of secular processes (e.g. the formation of galactic bars, see \citealt{abraham_1999, guo_2025, le_conte_2026}). However, the galaxy that provides us with the most detailed view remains our own Milky Way, which allows for observations on a star-by-star basis. While we can only observe it in its current state, we can still infer from the present how it has evolved in the past.

% needs to be a smoother transition
The formation of galaxies, according to the theory of hierarchical structure formation, is shaped by a series of successive mergers \citep{white_1978, white_1991}. The search for traces of past merger events in the Milky Way constitutes the field of Galactic archaeology. The search for merger remnants is usually done using a combination of stellar chemistry and dynamics. Its premise is the assumption that with knowledge of a star's current position and velocity, as well as the composition of its atmosphere, we can infer the environment in which the star formed. We can make use of differences in the chemical evolution of stars, which have been found to differ for stars born in the Milky Way and accreted from dwarf satellites \citep{venn_2004, nissen_2010, nissen_2011, hawkins_2014, hawkins_2015}. This can be combined with dynamical information allowing us to distinguish accreted stars by their current orbits (see reviews by \citealt{binney_2013, hunt_2025} and references therein).

Since the launch of the \textit{Gaia} mission \citep{gaia_2016, gaia_2018, gaia_2023}, the search for past mergers has been revolutionised. In combination with dedicated large spectroscopic surveys such as APOGEE \citep{majewski_2017}, DESI \citep{cooper_2023}, and upcoming surveys on instruments such as 4MOST \citep{de_jong_2019}, it is now possible to study the dynamics and chemistry of the stellar halo of the Milky Way on a star-by-star basis for a significant fraction of stars. The number of known stellar streams discovered surged from around 60 before Gaia, to over 100 since \textit{Gaia} DR2 \citep{gaia_2018, bonaca_2025}. The most significant single discovery of \textit{Gaia} is arguably the Gaia-Sausage/Enceladus merger (GSE) \citep{belokurov_2018, helmi_2018, haywood_2018}. Debris from this progenitor dominates the halo of the Milky Way within 15-20 kpc from the centre. This merger event was a highly radial accretion of a massive dwarf galaxy ($M_{\mathrm{star}} \; \approx \; 10^{8} - 10^{9} \;M_{\odot}$) around 8-11 Gyrs ago.

For the discovery of merger remnants, a frequently used space is energy and angular momentum in the z-direction ($E-L_z$) \citep{helmi_2000}. These are conserved quantities in a static axisymmetric potential, and approximately conserved for a non-axisymmetric system \citep{gomez_2010}. Caveats with using energy and angular momentum are the effects of the bar (e.g. overdensities caused by bar resonances, see \citealt{dillamore_2023, tomlinson_2026}), as well as the creation of several overdensities by individual mergers \citep{jean_baptiste_2017}. 

Studies looking for traces of mergers in $E-L_z$ space have identified several possible accreted remnants. Among these is GSE, which is found in a region around $L_z = 0$. Other mergers thought to have been identified are Thamnos \citep{koppelman_2019}, Kraken \citep{kruijssen_2020}, as well as Heracles \citep{horta_2021, horta_2024}. These three proposed merger remnants are postulated to have occurred before GSE and are found at lower energies closer to the centre of the Galaxy \citep{horta_2024}. 

Since the initial successes of finding signatures of past accretion using \textit{Gaia} data, there have been efforts to supplement \textit{Gaia} observations with spectroscopic data, in order to get better chemical abundances as well as more accurate radial velocities for distant stars. This is motivated by simulations, which show that the oldest stars in the galaxy are preferentially found either in the outer halo or close to the centre of the galaxy \citep{starkenburg_2017a}. This motivated the Pristine survey \citep{starkenburg_2017b} to find very metal-poor stars in the halo, as well as its sub-survey PIGS \citep{arentsen_2020}, which searches for metal-poor stars in the bulge of the Milky Way. Additionally, the COMBS survey \citep{lucey_2019, lucey_2021, lucey_2022}, has been using high-resolution spectroscopy to get detailed chemical abundances of metal-poor bulge stars, in order to understand the origin of this population. Other surveys such as SDSS-V \citep{kollmeier_2026} will further improve our understanding of the stars in the bulge.

% I think this paragraph is mixing observations of building blocks and those in simulations, and not clearly distinguishing between the two. e.g. Horta et al. does not study the same BBs seen in the MW in the FIRE simulations, which are not the same thing.
% phrasing a little awkward
With these future observations we hope to reconstruct the formation of the Milky Way up to its earliest building blocks, which make up the proto-galaxy. This is thought to be a metal-poor population of stars with little net rotation and is associated with the so-called ``Aurora'' population in the Milky Way, as postulated by \citet{belokurov_2022}. This may very well be the same population as the metal-poor concentrated spheroid around the centre of the Milky Way identified by \citet{rix_2022}. In 13 haloes of the FIRE-2 simulations, \citet{horta_2024} studied the properties of the proto-galaxy and its building blocks. While this study finds that it may be possible to separate out the dominant component of the proto-galaxy, the centre of the galaxy where remnants of the proto-galaxy are predominantly found is a challenging environment for observations.

% smoother transition
The density of stars in the inner Galaxy is vastly higher than in the stellar halo, where most efforts to find accreted debris have been focused. The inner galaxy is dominated by in-situ star formation and the fractions of old metal-poor and accreted stars are very low at the centre of the Galaxy \citep{rix_2022, orkney_2023}. For simulations this is an issue, since the stellar mass resolution has to be very high in order to be able to resolve these populations and estimate their properties. Dynamical structures at the centre further need to be well resolved such that their effects are modelled accurately. We therefore use the Auriga Superstars suite of simulations, which has a very high stellar mass resolution of 800\,$M_\odot$, to gain a better understanding of the properties of the proto-galaxy, early accretion and how to best disentangle observations of the inner Galaxy. The paper is organised as follows: In Section \ref{sec:simulation} we introduce the simulated halo we analyse; in Section \ref{sec: observations} we present the observational data we use; in Section \ref{sec: results} we present our results, which we discuss further in Section \ref{sec: discussion}; and we finish with our summary and conclusions in Section \ref{sec: conclusions}.

%%%%%%%%%%%%%%%%%%%%%%%%%%%%%%%%%%%%%%%%%%%%%%%%%%%%%

\begin{figure}
     \includegraphics[width=\columnwidth]{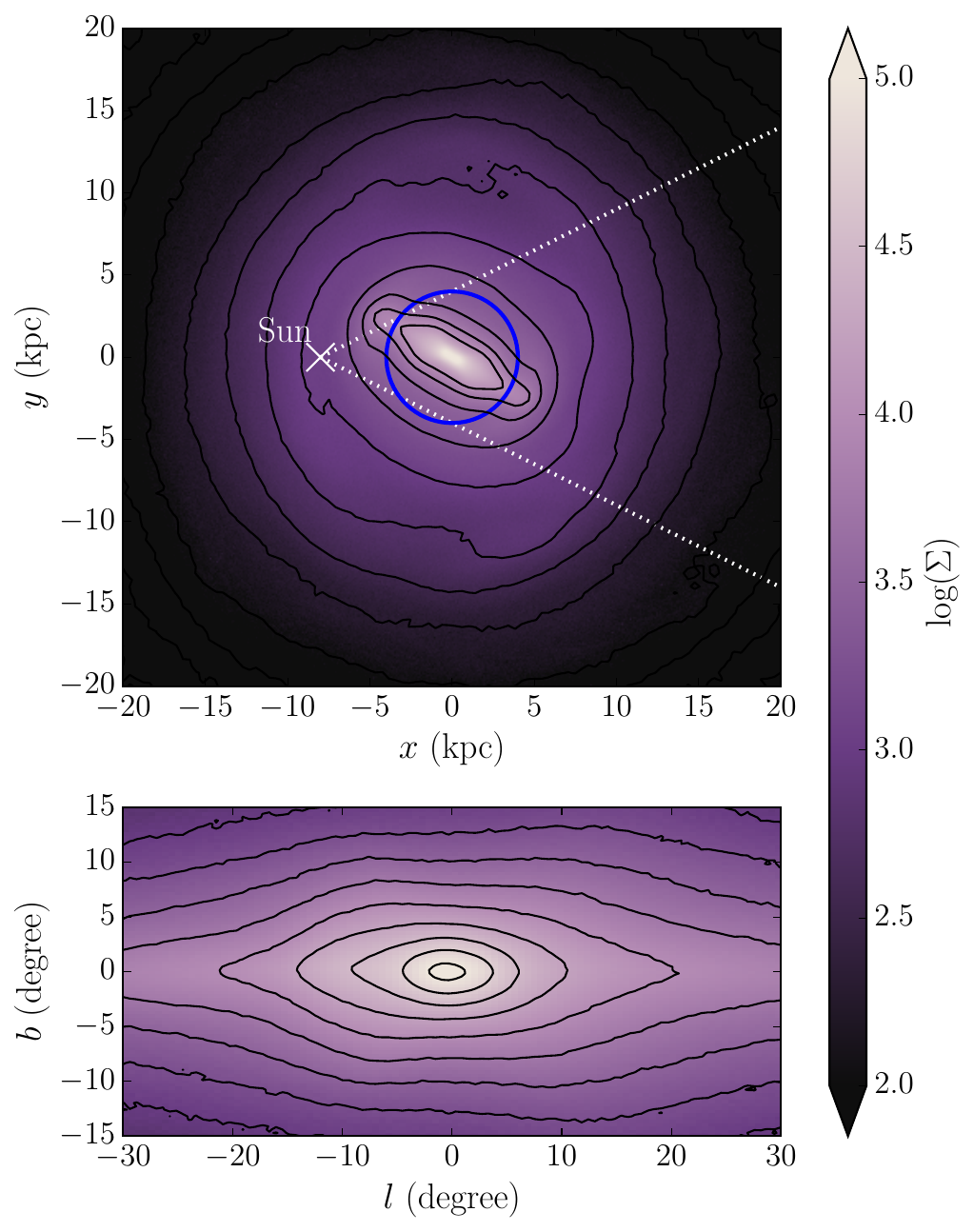}
     \caption{Face-on and edge-on stellar surface density projections of halo 18. The central region which is the main focus of this paper is outlined in blue in the top face-on view, what we take as the location of the Sun is marked by the white cross. The edge-on view is given in galactic longitude, $l$,  and galactic latitude, $b$, from the position of the Sun, showing how it would appear in observations. $\Sigma$ corresponds to the number of particles per bin, here in $x$ and $y$ (top) and $l$ and $b$ (bottom). }
     \label{fig:surface density of galaxy}
\end{figure}

\section{Simulation and Methods}
\label{sec:simulation}

% I found the description of Auriga, Auriga Superstars, and the particular Milky Way analogue somewhat repetitive. I think this whole part could perhaps be sharpened a little. My feeling would be to first introduce Auriga and the relevant properties of the simulation/halo, and then introduce Superstars and explain specifically what it adds. At the moment, some information on the resolution and methodology appears more than once.

% I found the wording in the opening description of the original Auriga simulations a little awkward. Some of the information in the following Superstars paragraphs seems to have already been stated, in particular the resolution and parts of the methodological description. I think some of this could be shortened to avoid repetition.

\subsection{Auriga Superstars Milky Way analogue}
The simulations used in this work are part of the Auriga Superstars suite of simulations (\citealt{grand_2023}; \citealt{pakmor_2025}; Fragkoudi et al., in prep.). These are hydrodynamical cosmological zoom-in simulations based on the original, hereafter called fiducial, Auriga suite of simulations \citep{grand_2017, grand_2024}. 

The fiducial Auriga suite contains simulations of 40 Milky Way-mass haloes, which are sampled from a cubic volume with side lengths of 100\;Mpc taken from the EAGLE dark matter only simulation \citep{schaye_2015}. The simulations are run with the \textsc{arepo} code \citep{springel_2010, pakmor_2016}, in a standard $\mathrm{\Lambda CDM}$ cosmology, with parameters $\Omega_{\rm m} = 0.307$, $\Omega_{\rm{b}} = 0.048$, $\Omega_{\Lambda} = 0.693$ and $H_{0} \mathrm{= 67.77 \; km \; s^{-1} \; Mpc^{-1}}$ \citep{planck_2013}. They make use of the zoom-in technique to achieve a baryonic mass resolution of $\sim 5 \times 10^{4} \; M_{\odot}$, with a typical dark matter resolution of $\sim 4 \times 10^{5} \; M_{\odot}$ . 

The physics model includes: primordial and metal-line cooling and heating from a spatially uniform, redshift-dependent UV background; a sub-grid model for the multiphase interstellar medium; stochastic star formation and stellar feedback driving galactic-scale outflows; chemical enrichment from core-collapse and Type Ia supernovae and AGB stars; the seeding and growth of supermassive black holes with thermal AGN feedback; and magnetic-field effects. The physics model is described in detail in \citet{grand_2017}. \par

The Auriga Superstars suite re-simulates the Auriga simulations with the Superstars method, keeping the same physics model and initial conditions. The Superstars method creates a larger number of lower-mass ($\sim 800 \; M_{\odot}$) star particles from a single gas cell. These stars have the same velocity and position as their parent gas cell with an added random isotropic component based on a Gaussian distribution with a width set by the minimum of the local sound speed and the velocity dispersion of the neighbouring gas cells\footnote{We note that the simulation studied here uses a variation of the Superstars method, with a factor of $\sim 2$ larger kick to the birth velocity of stars. However, we highlight that this does not have a significant impact on the properties of the galaxy -- see the discussion and Appendix A in \citealt{pakmor_2025} for more details.} \citep{pakmor_2025}. The mass-weighted average kick velocity vector is subtracted from all star particles born together, so that the total added momentum from the kicks is zero. Re-simulating a halo using the Superstars method largely preserves galaxy properties, with variations falling within the range of intrinsic variations found among realisations run on a different machine or with a different random number seed \citep{pakmor_2025_a}. The changes that do occur are smaller than the systematic shifts caused by changing the mass resolution of the gas. For more details on the method, we refer the reader to \citet{pakmor_2025}. Additionally, the mass resolution of dark matter particles in the Auriga Superstars suite is increased by a factor of 8 to $\sim5\times10^4 \;M_{\odot}$ compared to the fiducial Auriga simulations. The Auriga Superstars suite will be presented in Fragkoudi et al. (in prep.).\par 

\subsubsection{Halo 18}
\label{sec: halo 18}

%For halo 18, I think it would help to bring together the information that establishes it as a particularly close Milky Way analogue. At the moment, this information is somewhat spread through the section. Related to this, perhaps the discussion of the bar properties could come earlier, as part of the initial description of why halo 18 is a good Milky Way analogue.

In this study, we focus on halo 18 from the fiducial Auriga suite of simulations, re-simulated using the Superstars method. The simulation produces full outputs every $\sim50$\,Myrs. Metal abundances are taken from the simulation, and we use [Fe/H] values as a proxy for metallicity; we reduce [Fe/H] by 0.5 dex to bring the metallicities approximately in line with metallicities in the Milky Way bulge, using the bulge sample of \citet{ness_2016}, to facilitate comparisons with data from the Milky Way. This elevated metallicity in Auriga is a well-known phenomenon and agrees with the results of \citet{grand_2021}. The total halo mass (defined as the mass within the virial radius) is $\sim 1.1\times10^{12}M_{\odot}$ whereas the stellar component (the stellar mass within a sphere of 0.1 times the virial radius) has a mass of $\sim 7\times10^{10}M_{\odot}$. The virial radius is the radius enclosing a sphere whose mean matter density equals 200 times the critical density, given by $\rho_{\mathrm{crit}} = 3H^2(z)/(8\pi G)$. 

For the study of the inner galaxy, halo 18 is among the closest Milky Way analogues of the Superstars haloes. Like the Milky Way, it has a prominent bar and boxy/peanut bulge, with chemodynamical properties similar to the Milky~Way \citep{fragkoudi_2020}. The bar forms at a lookback time of  $\sim$8\,Gyrs (see also \citealt{merrow_2024}), and has a length of $\sim$5.6 kpc, based on the radius at which the $m=2$ Fourier mode of stars in the disc reaches 60\% of its maximum (see e.g. \citealt{fragkoudi_2021} for the method). This is close to recent estimates of the length of the Milky Way's bar of $\approx5$\,kpc \citep{portail_2017, hilmi_2020, lucey_2023}. The corotation radius is $\sim 10$\,kpc, which is slightly larger than estimates for the Milky Way that place its corotation radius between $4.5-9.6\,$kpc, albeit with significant uncertainty \citep{bland_hawthorn_2016,sanders_2019,chiba_2021,horta_2025}. Halo 18 further has an analogue for the GSE merger, with a mass and accretion time that fall within the estimates for GSE, and shared kinematic properties \citep{fattahi_2019}. The GSE analogue dominates the stellar halo of the galaxy; however, as found by \citet{orkney_2023} for the fiducial Auriga suite, the contribution of stars from the GSE analogues decreases close to the galactic centre while accreted fractions remain high at low metallicities. They argue that this makes the central galaxy a promising region for the discovery of early merger events. However, their definition of accreted stars includes very early mergers which we classify as part of the proto-galaxy. We build on this research, among others, in our analysis of the central galaxy of halo 18 of the Auriga Superstars suite.

We focus on the inner regions of the halo, a sphere with a radius of 4\,kpc around the centre of the galaxy, as shown in \figref{fig:surface density of galaxy}. In this figure we place the Sun at 8\,kpc to model what an observer at the position of the Sun would measure when looking at the centre. The bar is rotated to have an angle of 30 degrees ahead of the `Sun'-galactic centre line, similar to the Milky Way \citep{bland_hawthorn_2016}. However, we note that the scale length of the disc of our simulated galaxy is 3.7 kpc, which is on the high end of estimates for the Milky Way scale length (for which studies find values between $2.6-3.9\,$kpc; see e.g. \citealt{bland_hawthorn_2016}).

% Orkeney et al, 2023 is not discussed. Considering that it also searches for signatures of early accretion event in the GC using Au18, we should include it.

\begin{table}
\centering
\begin{tabular}{ |p{.7cm}|p{1.cm}|p{1.2cm}|p{.7cm}|p{.7cm}|p{.7cm}|p{.7cm}|}
\hline
Merger & Peak stellar mass ($M_{\odot}$) & Peak halo mass ($M_{\odot}$) & Accreted \nobreak{fraction} & Mean [Fe/H] & Infall time (Gyrs) & Infall redshift\\
\hline
M1 & $1.5\times10^{8}$ & $3.0\times10^{9}$ & 0.28 & -1.16 & 12.52 & 4.6\\
M2 & $1.9\times10^{8}$ & $9.6\times10^{9}$ & 0.10 & -1.18 & 12.11 & 3.7\\
M3 & $5.6\times10^{8}$ & $1.6\times10^{10}$ & 0.07 & -0.89 & 9.47 & 1.5\\
M4 & $1.5\times10^{9}$ & $3.6\times10^{10}$ & 0.53 & -0.63 & 9.29 & 1.4\\
\hline
\end{tabular}
\caption{\label{tab:1} Properties of the four main mergers in halo 18. The peak stellar and halo mass are based on all the bound particles, as identified by \textsc{subfind}. The accreted fraction describes what fraction of accreted stars within 4\,kpc of the centre of the main halo come from each merger. The mean metallicity given is for stars found within 20\,kpc of the centre of the main halo at $z=0$, calculated as the mean of values of [Fe/H] after taking the logarithm. The infall time (and redshift) refers to the first time the satellite crossed the virial radius of the main halo. The time is given as the lookback time. The start of the simulation occurs at a lookback time of 13.83\,Gyrs.}
\end{table}

\subsection{Separating into components}
\label{sec: separating components}

\begin{figure}
	\includegraphics[width=0.999 \columnwidth]{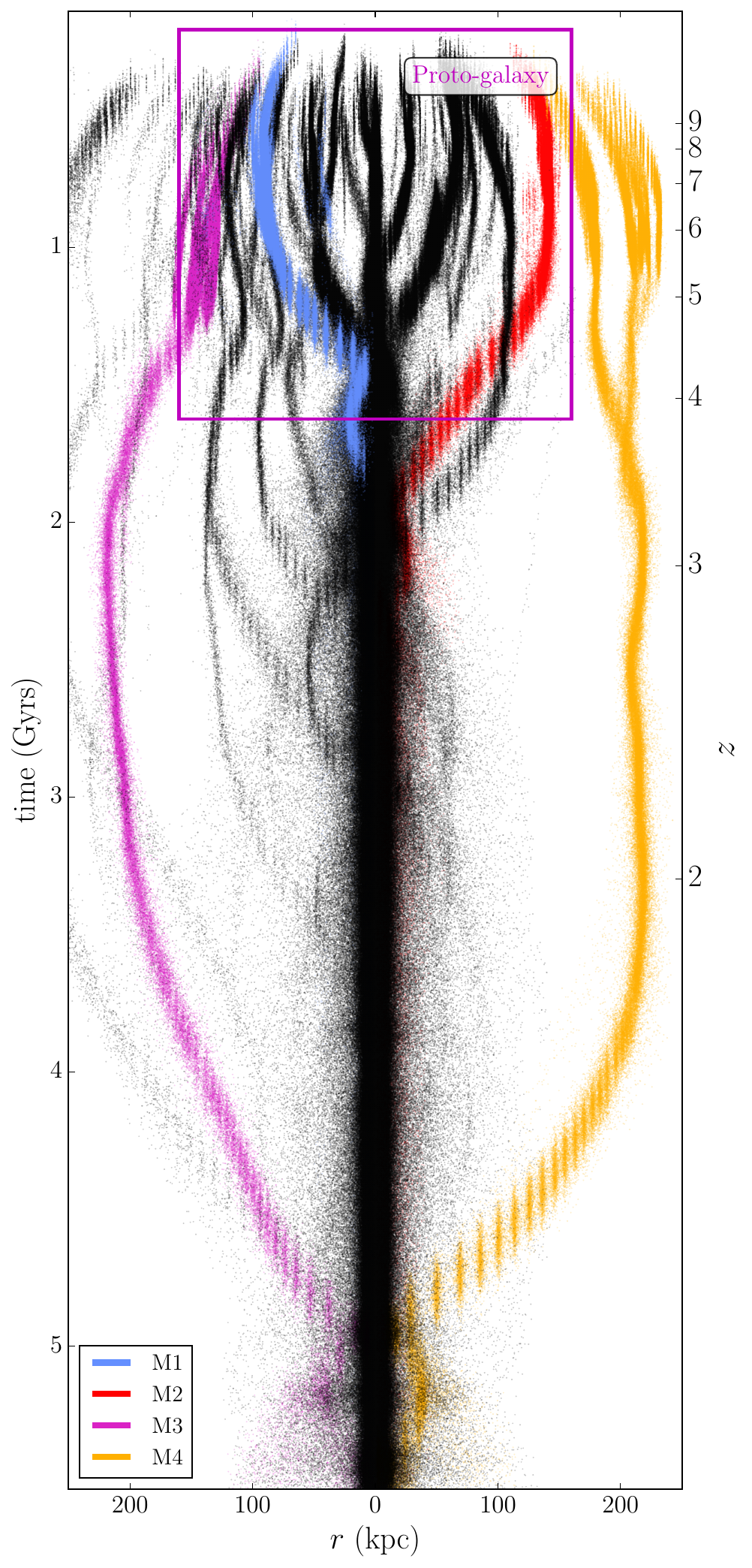}
    \caption{The assembly of the proto-galaxy of halo 18. We show the distance of particles from the centre of the main halo at different snapshots, with a Gaussian spread in lookback time to smooth the distribution of particles in between snapshots. We plot a subsample of all stars found at every snapshot, with a radial distribution that is required to reproduce the distribution of all stars. The proto-galaxy can be found inside the purple box. The four main mergers in the simulation are shown in different colours. We can see that before the cut-off time of the proto-galaxy, several objects of similar stellar mass are merging, with none of the merging subhaloes being the largest halo in all snapshots. The frequency of mergers falls drastically after the end of the proto-galaxy phase.}
    \label{fig:proto_galaxy_plot}
\end{figure}

\begin{figure*}
    \begin{subfigure}{\textwidth}
    \centering
    \includegraphics[width=0.9\linewidth]{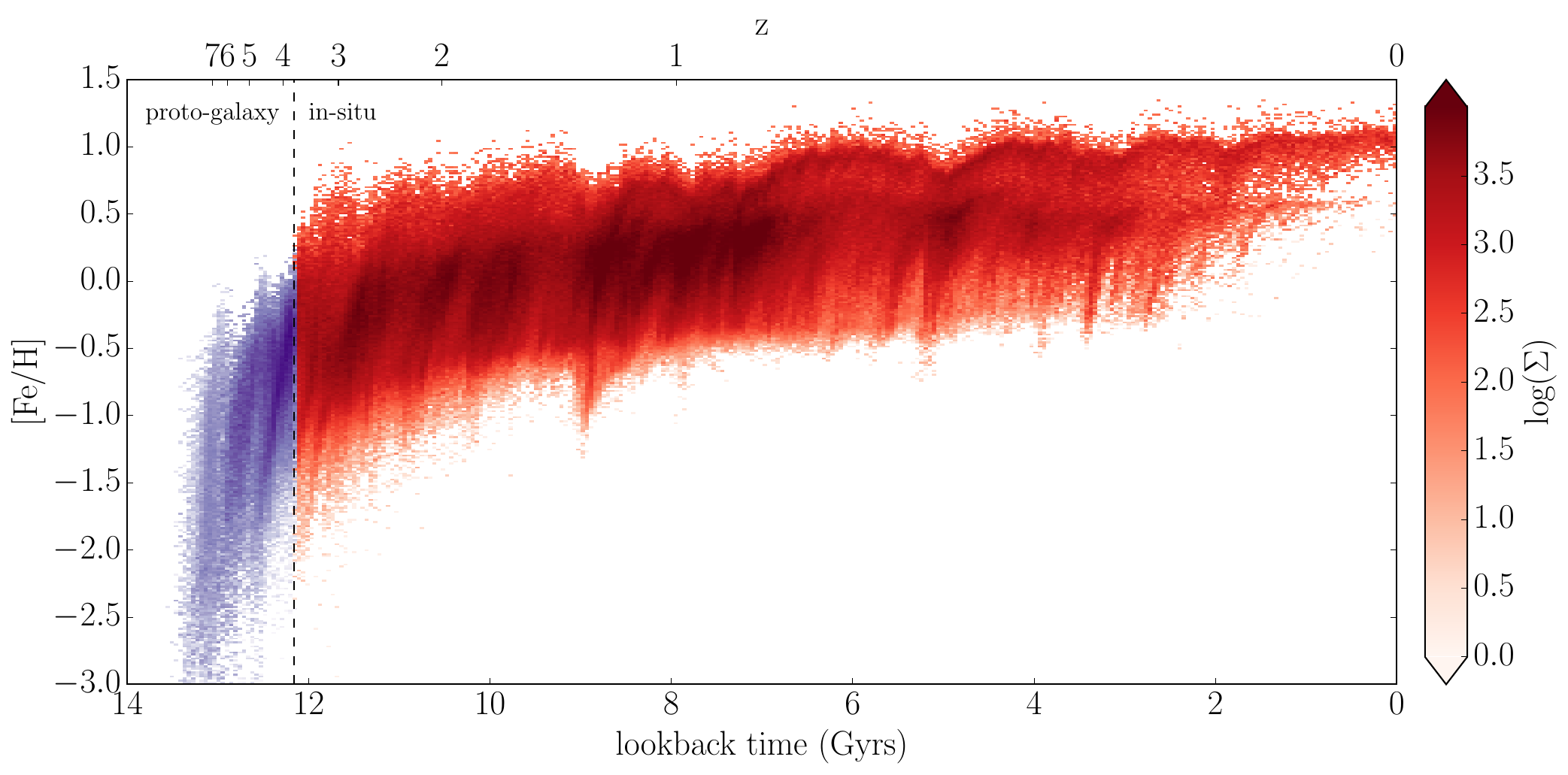}
    \end{subfigure}
    
    \begin{subfigure}{\textwidth}
    \centering
    \includegraphics[width=0.9\linewidth]{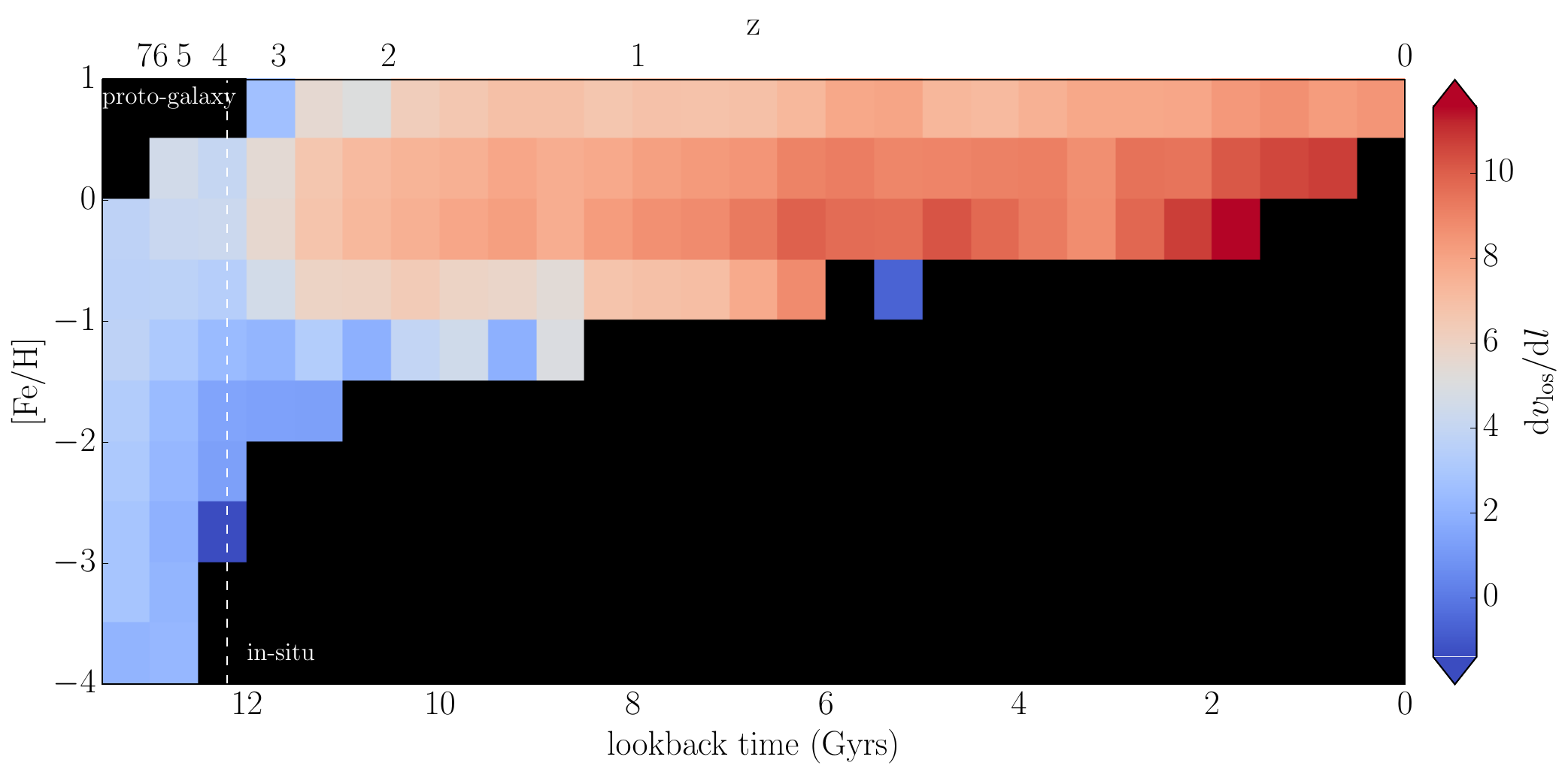}
    \end{subfigure}
    \caption{The upper panel shows the logarithmic stellar density of in-situ and proto-galaxy stars in age-metallicity space for halo 18. $\Sigma$ corresponds to the number of particles per bin, here in lookback time and $\rm [Fe/H]$. The bins coloured purple show stars that are part of the proto-galaxy, while the bins coloured red contain stars born in-situ. The cut-off time for this halo is at $z=3.8$. [Fe/H] enrichment occurs quickly in the proto-galaxy stage and plateaus after the cut-off time. The lower panel shows all stars in age-metallicity space. However, here the bins are coloured by the mean slope of the line-of-sight velocity curve of stars as a function of galactic longitude, which is discussed in more detail in Section \ref{sec: kinematics}. This essentially shows how the amount of rotation changes with stellar age. The proto-galaxy is characterised by slower rotation, which increases as the main halo forms at $z=3.8$.}
    \label{fig:age_metallicity_ins_proto}
\end{figure*}

For our analysis, we split our simulated halo into three components: proto-galaxy, in-situ, and accreted. In the FIRE simulations this remnant of the earliest hierarchical mass assembly was studied by \citet{horta_2024}. In the Milky Way, this could correspond to a population of stars identified by \citet{belokurov_2022} -- believed to be a remnant of the Galaxy before the formation of the disc, named Aurora. After this cut-off time, stars are counted either as in-situ if they formed in the main halo, or as accreted if they are born in a different halo that merges with the main halo. This follows the methodology used in \citet{fattahi_2019}, which makes use of the friends-of-friends and \textsc{subfind} algorithms \citep{davis_1985, springel_2001a, springel_2001b}. Star particles are assigned to a progenitor based on which subhalo they are bound to in the snapshot after their formation. Thus, stars formed from gas stripped from a progenitor will be flagged as in-situ if the star that forms is bound to the main Milky Way-like progenitor.

All stars that are found in the main halo at a redshift of 3.8 are counted as part of the proto-galaxy. This cut-off time we use to define the proto-galaxy is based on the definition in \citet{horta_2024}, but not identical. \citet{horta_2024} choose the time based on the snapshot at which the main halo reaches a stellar mass ratio of 3:1 with the next-most luminous halo in the simulation volume. In this study, we choose the time based on the snapshot after the last merger with a stellar mass ratio larger than 1:3 is completed. These definitions are similar, as by construction the main halo after the cut-off time is at least three times larger than any subhalo subsequently merging with the main halo of the simulation.

%Describe figure 2: Plot showing the assembly of the proto-galaxy

In \figref{fig:proto_galaxy_plot}, we show the distance of stars from the centre of mass of the main progenitor of the Milky Way analogue at each snapshot from the start of the simulation to a time of around 5.5\,Gyrs after the big bang. The proto-galaxy is shown inside a box at the top of the figure. The difficulty of labelling just one of these merging haloes as the main halo at any given time before the cut-off is apparent. The most massive halo in one snapshot is not necessarily the most massive one in the next one, due to the stochasticity of mergers. After the formation of the proto-galaxy the frequency of mergers drops off quickly and the following mergers generally have small merger-mass ratios. Star formation induced by the very early mergers characterising the proto-galaxy phase can leave imprints in the star formation history of the main halo at $z=0$. However, even in the simulation it is very challenging to associate these episodes with any given merger, as there is usually more than one ongoing merger at any given time.

%Describe figure 3: Age-metallicity relation split into proto-galaxy and in-situ

In the upper panel of \figref{fig:age_metallicity_ins_proto} we show the age-metallicity relation of stars tagged as either in-situ (red) or proto-galaxy (purple). We can see that the [Fe/H] enrichment is very fast in the proto-galaxy stage. Within the first few hundred Myrs stars reach [Fe/H] values of $-0.5$ dex. In the following Gyrs enrichment slows down to reach solar and super-solar levels. The stage where the enrichment plateaus falls approximately at the boundary between proto-galaxy and in-situ stars. The lower panel shows stars in bins of age and [Fe/H], where we colour each bin by the mean slope of the line-of-sight velocity curve of stars as a function of galactic longitude. We discuss this in more detail in Section \ref{sec: kinematics}. The trend we see in this panel essentially tells us that after the proto-galaxy stage, the galaxy begins to rotate faster. This further reassures us that our definition of the proto-galaxy stage is sensible and has a physical significance.

The accreted population of halo 18 is dominated by four main mergers that combined constitute over $98\%$ of accreted stars within 4\,kpc of the centre of the halo. Some properties of these mergers are summarised in \tabref{tab:1}. Two of these mergers, M1 and M2, merged very early on, shortly after the end of proto-galaxy phase. Although M1 has an infall time before the proto-galaxy cut-off time at $z=3.8$, it is still recognised as a separate subhalo by \textsc{subfind} when the proto-galaxy phase ends. We therefore count its stars as accreted. The other two mergers occurred about 2.5 Gyrs after M2. M3 and M4 merged within just a few hundred Myrs of each other and M4 was identified to be one of the GSE analogues in \citet{fattahi_2019}. M4 dominates the accreted population at the centre of the galaxy; however, at low metallicities M1 becomes dominant. In \figref{fig:proto_galaxy_plot} we show the distance of each of these main mergers from the main galaxy as a function of time. 

\begin{figure*}
    \includegraphics[width=\textwidth]{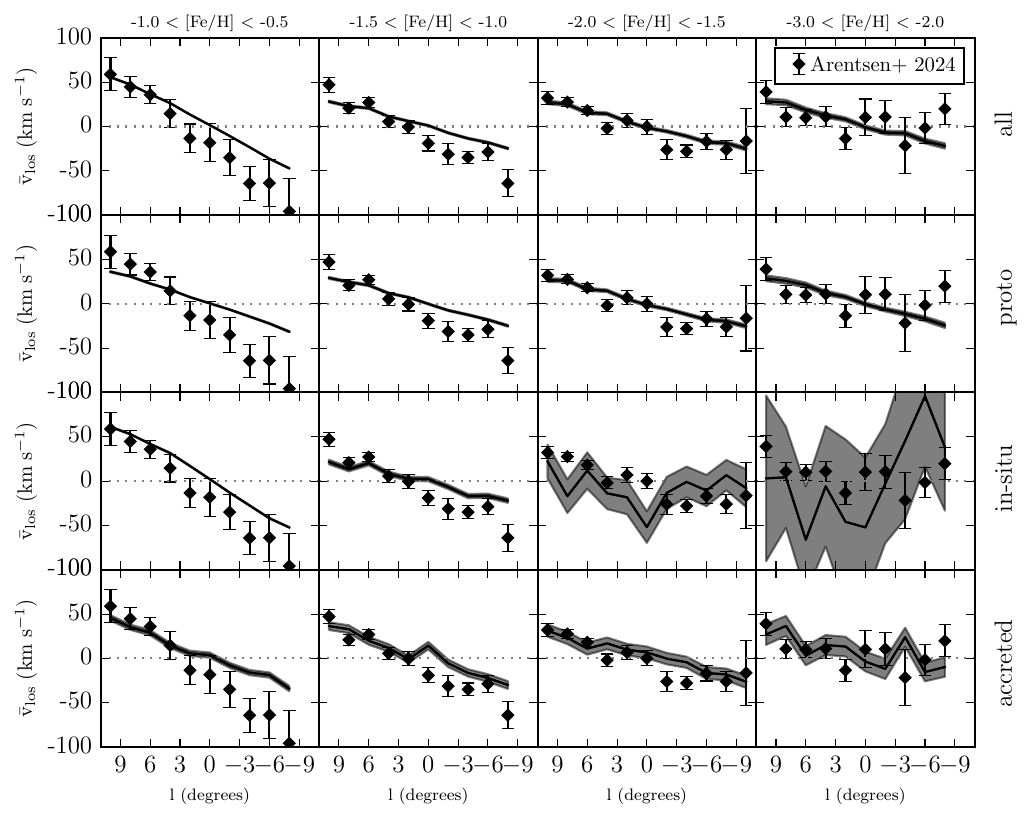}
    \caption{The line-of-sight velocity from the location of the Sun as a function of the galactic longitude $l$ in different bins of [Fe/H]. We compare data from halo 18 to observational PIGS data, as described in Section \ref{sec: pigs}. The observational data are shown as points in the plot, whereas the data from halo 18 is plotted as a line. In both cases the error is calculated as $\sigma/\sqrt{N}$, where $\sigma$ is the standard deviation and $N$ the number of particles for each data point. The simulation data are further split into all (first row), proto-galaxy (second row), in-situ stars (third row), and accreted stars (fourth row). Stars are further separated by their metallicity into different columns. The points remain the same in all rows. Halo 18 generally matches the observations well. We see a decrease in $v_{\rm los}$ from the first to the second metallicity bin, which stays approximately constant as we go to lower metallicities, pointing to a decrease in the rotation of stars. This is the case even for the proto-galaxy and accreted stars.}
    \label{fig:kinematics}
\end{figure*}

\subsection{Integrals of motion space}

In the following, we analyse the distribution of stars in the simulation in $E-L_z$ space. For every simulation particle, we compute the specific energy as $E = v^2/2 + \Phi(x, y, z)$. Here, $v$ is the absolute value of the velocity measured in the galaxy's centre-of-mass frame -- with the bulk motion of the halo subtracted -- while the gravitational potential $\Phi$ is derived via the tree code algorithm described by \citet{springel_2010}. We further shift the potential such that all particles situated at the virial radius have a potential value of zero. The specific angular momentum is calculated using the cross product $\hat{\textbf{k}} \, \mathrm{L_z} = \textbf{r}_{xy} \times \textbf{v}_{xy}$.

\section{Observational Data}
\label{sec: observations}

%PIGS

\subsection{The Pristine Inner Galaxy Survey}
\label{sec: pigs}

% I think a short introduction at the beginning would be helpful. When I reached this section, the observational comparison came somewhat out of the blue, as I had not realised from the preceding sections that you were going to compare the simulation to observations. Perhaps just a couple of sentences explaining what observational data you are introducing, why you are introducing them, and how you will use the comparison would help orient the reader.

To place our results in context with the Milky Way and help interpret the implications of our results for observations, we include a comparison with data from the Pristine Inner Galaxy Survey (PIGS), which is a sub-survey of the Pristine survey \citep{starkenburg_2017b}. This survey provides kinematic data for metal-poor stars close to the Galactic centre to test the simulation against. This survey was first introduced in \citet{arentsen_2020}. The Pristine survey uses a narrow-band CaHK filter for MegaCam on the 3.6-m Canada–France–Hawaii Telescope (CFHT) to photometrically search for the most metal-poor stars. Combined with broad-band photometry from PanSTARRS-1 (PS1; \citealt{chambers_2016}) or \textit{Gaia} DR2 \citep{gaia_2018}, candidates were selected for a more detailed spectroscopic follow-up for PIGS using the AAOmega + 2dF multifibre spectrograph on the 3.9-m Anglo-Australian Telescope (AAT) at low/intermediate resolution, which has 400 fibres in a two-degree field of view. 
A cut on the \textit{Gaia} DR2 parallax and its uncertainty was applied to remove foreground stars, which are mostly metal-rich. 

We took the PIGS data from the VizieR catalogue associated with \citet{arentsen_2024} \footnote{This catalogue can be accessed here: \citet{vizier_data}.}. Distances were derived using the Bayesian inference code \textsc{starhorse} \citep{santiago_2016, queiroz_2018}, while the values of [Fe/H] we use were derived using \textsc{ferre} \citep{allende_2006}. In our analysis we only consider stars that are within a 4\,kpc radius of the Galactic centre. We further make cuts based on the errors in radial velocity and \textsc{starhorse} distances. In both cases, we only use stars with errors less than $20\%$. This leaves us with a total sample of 8207 stars.

The measured line-of-sight velocities are converted to the Galactic standard of rest using \textsc{astropy} \citep{astropy_2022}. We assume that the Sun is at a distance of 8.2\,kpc from the centre of the Milky Way \citep{bland_hawthorn_2016}, has peculiar velocities $(U_\odot, W_\odot)=(11.1, 7.25) \; \mathrm{km \; s^{-1}}$ \citep{schonrich_2010}, and a total tangential velocity of $248 \; \mathrm{km \; s^{-1}}$ \citep{bland_hawthorn_2016}. In the following, whenever we refer to line-of-sight velocities of stars in the Milky Way, we mean the velocity in the Galactic standard of rest.

\section{Results}
\label{sec: results}

We initially discuss the kinematics of the inner galaxy in halo 18 in Section \ref{sec: kinematics}, where we also compare results from the simulation to PIGS data. We then proceed to look at the distribution of proto-galaxy, in-situ, and accreted stars in $E-L_z$ space, which we combine with their metallicity distribution in Section \ref{sec: chemistry_dynamics}. We additionally look at what we can learn with precise stellar ages in Section \ref{sec: age_dynamics}. Since a promising region of $E-L_z$ space for the discovery of accreted stars is the region of high energies, we look more closely into the nature of the accreted population at these high energies in Section \ref{sec: high_energy}. We finally take a look at the properties of the four main mergers in halo 18 in Section \ref{sec: mergers} to understand how well these can be separated from each other and the proto-galaxy and in-situ components.

\subsection{Kinematics}
\label{sec: kinematics}
%Describe figure 4: LoS velocities compared to observations (Pigs)

\begin{figure}
	\includegraphics[width=\columnwidth]{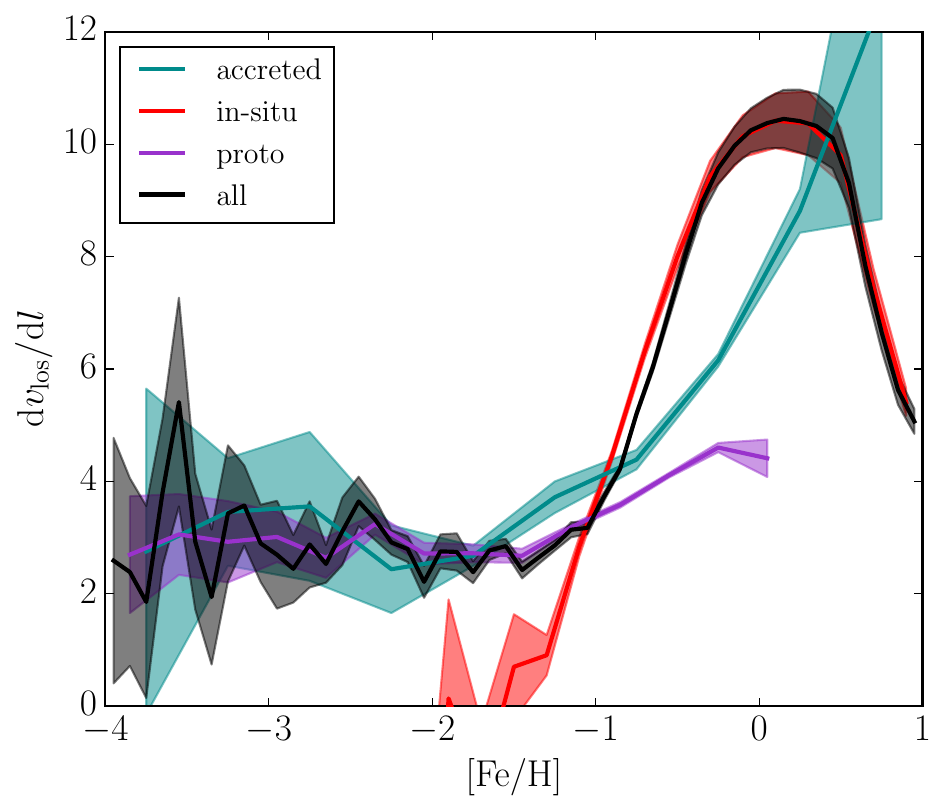}
    \caption{The slope of the line-of-sight velocity ($v_\mathrm{los}$) as a function of galactic longitude, $l$, in bins of [Fe/H] for all, proto-galaxy, in-situ, and accreted stars. The shaded regions show the $1\sigma$ standard error derived from the least-squares covariance matrix. There is a steep rise in this slope as [Fe/H] increases from $[\rm Fe/H]=-1$ to  $[\rm Fe/H]=0$. At lower metallicities the slope remains approximately constant. This is largely driven by the transition from in-situ to proto-galaxy stars.}
    \label{fig:slope_v_feh}
\end{figure}

\begin{figure*}
    \includegraphics[width=\textwidth]{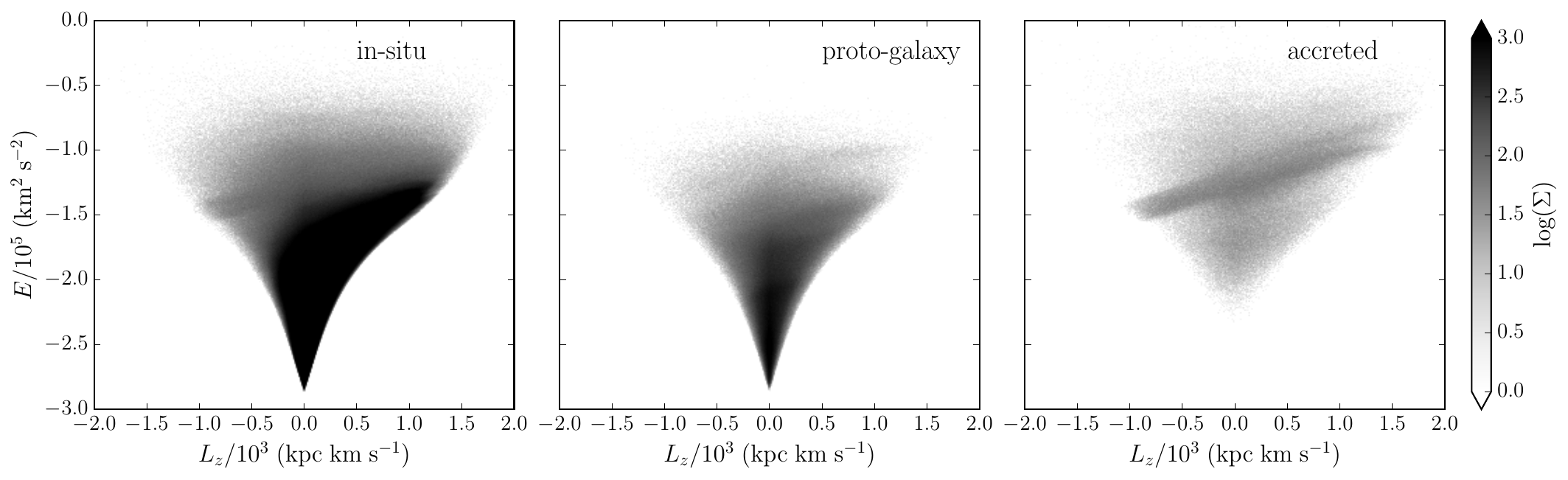}
    \caption{The three panels show logarithmic density plots of stars in $E-L_z$ space. $\Sigma$ corresponds to the number of particles per bin, here in $L_z$ and $E$. The first panel shows in-situ stars, the second panel proto-galaxy stars, and the third panel accreted stars. In all cases, stars were selected to be within 4\,kpc of the centre of the galaxy.
    In-situ stars are most densely concentrated at low energies and at positive $L_z$, where the disc resides; however, a significant fraction is also found at negative $L_z$ and at high energies.
     Proto-galaxy stars are mainly confined to energies below $E= -1 \times 10^5 \; \rm km^2 \, s^{-2}$. Accreted stars do not reach the bottom of the potential well and are mainly found at higher energies. The very apparent diagonal ridge is caused by stars trapped in resonance with the bar as discussed in \citet{tomlinson_2026}.}
    \label{fig:e_lz_comp}
\end{figure*}

Having defined the three different components we are interested in  -- proto-galaxy, in-situ, and accreted --  we can now study the properties of stars in the central region of this simulated halo. We initially turn our attention to the kinematics of stars. In \figref{fig:kinematics}, we show the line-of-sight velocities, $v_\mathrm{los}$, as seen from an observer 8\,kpc away from the centre of the galaxy, the location we place the `Sun' in \figref{fig:surface density of galaxy}. The four different rows show $v_{\rm los}$ for all stars as well as split into proto-galaxy, in-situ, and accreted stars. We further split stars into four [Fe/H] bins between $-0.5$ and $-3$ dex. We also compare to observational data from PIGS \citep{arentsen_2024}, which we describe in Section \ref{sec: pigs}. The PIGS data are shown as points in the figure, and are the same in each row. We use $v_{\rm los}$ as a proxy of the rotation of the galaxy. Higher values of $v_{\rm los}$, and a steeper slope of $v_{\rm los}$ as a function of galactic longitude, correspond to faster rotation. Values that are consistently close to zero point to a system with no ordered rotation. \citet{sipma_2022} performed a similar comparison between PIGS data and the fiducial Auriga simulations. They find that selection effects do not meaningfully change the result of the kinematic analysis.

% At the end of the first column, around “These trends are very comparable…”, I would specify more clearly that you mean the trends in the PIGS data.
In the PIGS data, the trend appears to be a steady decrease of $v_{\rm los}$ with decreasing metallicity. At the lowest metallicity, the $v_{\rm los}$ values are consistent with net rotation of the order of very few tens of $\rm km\,s^{-1}$ \citep{arentsen_2024}. The trends in the PIGS data are very comparable to what we see in all stars of halo 18 (top row), with line-of-sight velocities in the simulation largely falling within the error bars of the observations. Similarly to the data from the Milky Way, in halo 18 we see a drop-off in the net rotation from the highest to the second-highest metallicity bin. However, beyond this point in the simulation, rotation appears to stay constant with a small amount of net rotation. We want to note, though, that this is a trend not inconsistent with the PIGS observations, within errors.

% When introducing Ardern-Arentsen et al. in the second column, perhaps explicitly mention that these are observations. I had to stop for a moment to work out what was being compared here.
In their analysis of the PIGS data, \citet{arentsen_2024} wanted to determine whether there is evidence of past accretion in the kinematics of very metal-poor stars at the centre of the Galaxy. They find some evidence for two components in the distribution of azimuthal velocities of metal-poor stars. One that is slowly rotating, which they associate with the proto-Milky Way, and one which has almost no net rotation, which they see as a possible sign of an accreted origin. Since we are studying a simulation, we can explicitly classify stars as proto-galaxy, accreted, or in-situ.

We show the kinematics of these three components in the three bottom rows of \figref{fig:kinematics}. We find that there is only a small difference between these different components in terms of how fast they are rotating, which is most pronounced in the highest metallicity bin between -0.5 and -1 dex. At lower metallicities, the contribution from the in-situ population decreases\footnote{At these low metallicities the in-situ population is largely replaced by the proto-galaxy component.}, leading to higher Poisson noise; however, aside from the increased noise, in-situ stars are not distinguishable from proto-galaxy stars in this regime. This is unsurprising, as the boundary between in-situ and proto-galaxy is not a clear-cut distinction. Intriguingly, the accreted population has prograde rotation at all metallicities, with values that are, for the most part, within Poisson noise of the proto-galaxy component. However, this is a result that depends on the particular accretion history of the simulation; the accreted population of halo 27 of the Auriga Superstars suite for example shows no net rotation regardless of the metallicity.

We further find that the kinematics of the metal-poor inner galaxy are dominated by the proto-galaxy. This is a robust result, which we also see in two other Superstars haloes, which are introduced in Appendix \ref{sec:appendix a}. The decreasing rotation as we go to lower metallicities is linked to the transition between proto-galaxy and in-situ. In halo 18, rotation decreases until it stays approximately constant for stars with $\rm [Fe/H] < -1$. This can be seen in \figref{fig:slope_v_feh}, where we plot the changing slope of $v_{\rm los}$ as a function of $l$, at different metallicities. We can see very clearly that for $[\mathrm{Fe/H}]>-1$, the total stellar population follows the curve of the in-situ stars. However, for $[\mathrm{Fe/H}]<-1$, the curve of all stars follows the proto-galaxy. This happens at precisely the transition between a changing slope $\mathrm{d}v_{\mathrm{los}}/\mathrm{d}l$ and a roughly constant slope. This is a promising indication that we will be able to see a similar trend in data from the centre of our Galaxy, as the number of observations increases. This would allow us to make a well-motivated cut to select stars which are part of the proto-Milky Way. In combination with stellar ages this could help us estimate when the transition from a merger-dominated assembly of the Milky Way to a more quiescent merger history occurs.

% Finally, I wasn't sure I completely understood the last part of this paragraph, particularly the statement that this would allow us to determine the timing of the transition away from the proto-Milky Way. I understand the change in the kinematic behaviour, but I don't yet see clearly how that translates into a timing. Perhaps this could be explained a little more.

\subsection{Chemistry and dynamics}
\label{sec: chemistry_dynamics}

\begin{figure*}
	\includegraphics[width=\textwidth]{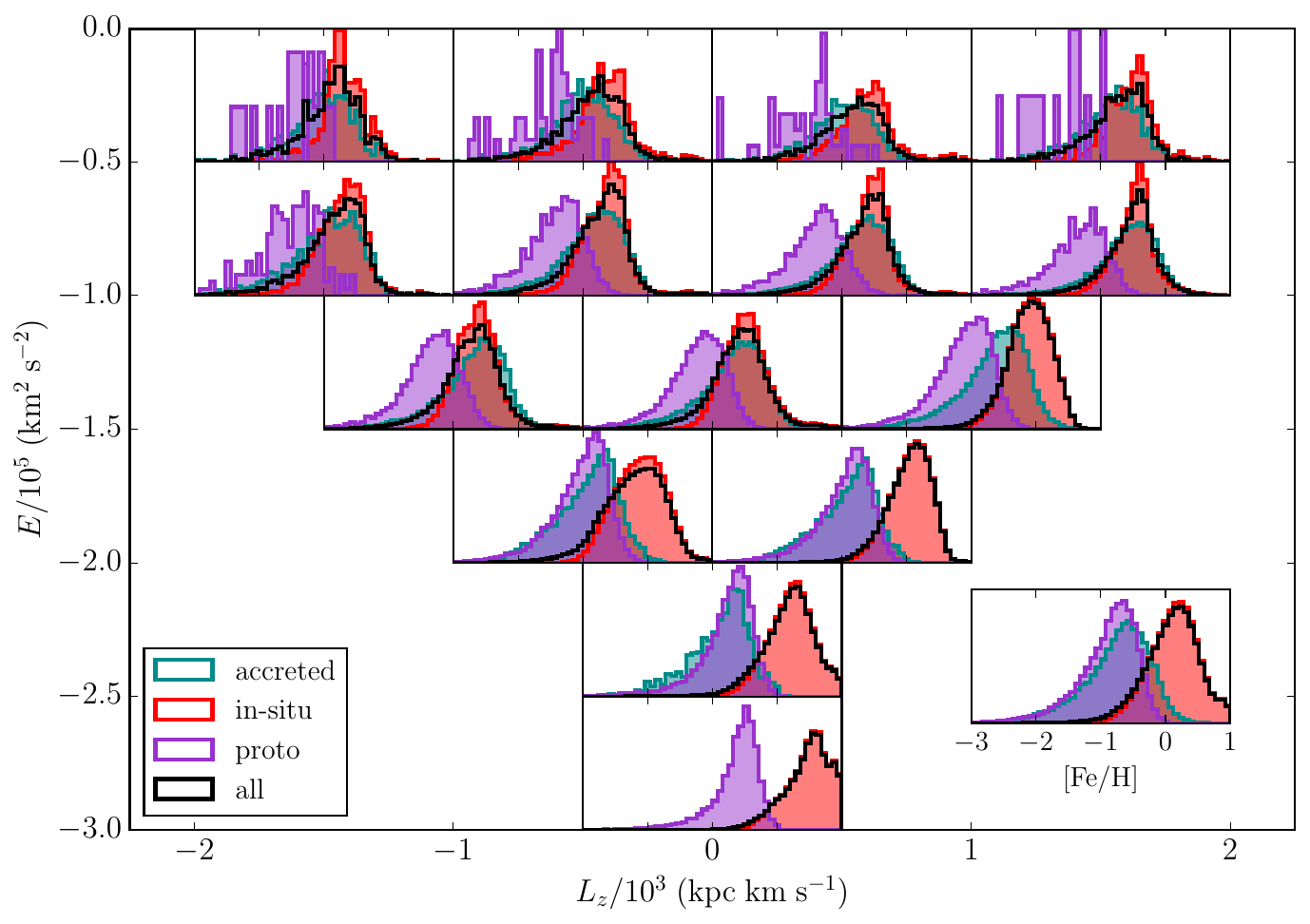}
    \caption{The normalised metallicity distribution functions (MDF) of stars, within 4\,kpc of the centre, in different regions of $E-L_z$ space of halo 18. We separate stars into different bins in $E-L_z$ of equal size with an extent of $1 \times 10^3 \rm \;kpc \,km \, s^{-1}$ in $L_z$ and $0.5 \times 10^5 \; \rm km^2 \, s^{-2}$ in $E$. The borders of the subplots in $E-L_z$ correspond to the boundaries of the bin in $E-L_z$. In each subplot we separate stars into accreted (turquoise), in-situ (red), proto-galaxy (purple), and all stars (black). Stars of each component are individually normalised and thus this plot gives us an idea of the distribution in $E-L_z$ space of each component but does not serve as a direct comparison of their total contributions. The inset in the lower right of the plot shows the MDFs of stars in the inner 4\,kpc, with no cuts in $E$ and $L_z$. While in-situ and proto-galaxy stars seem to have distinct peaks, the proto-galaxy appears as a low-metallicity tail in the MDF of all stars, with no peak visible here. Accreted stars have large overlaps with both of these components.}
    \label{fig:normalised e_lz bin plot}
\end{figure*}

\begin{figure*}
	\includegraphics[width=\textwidth]{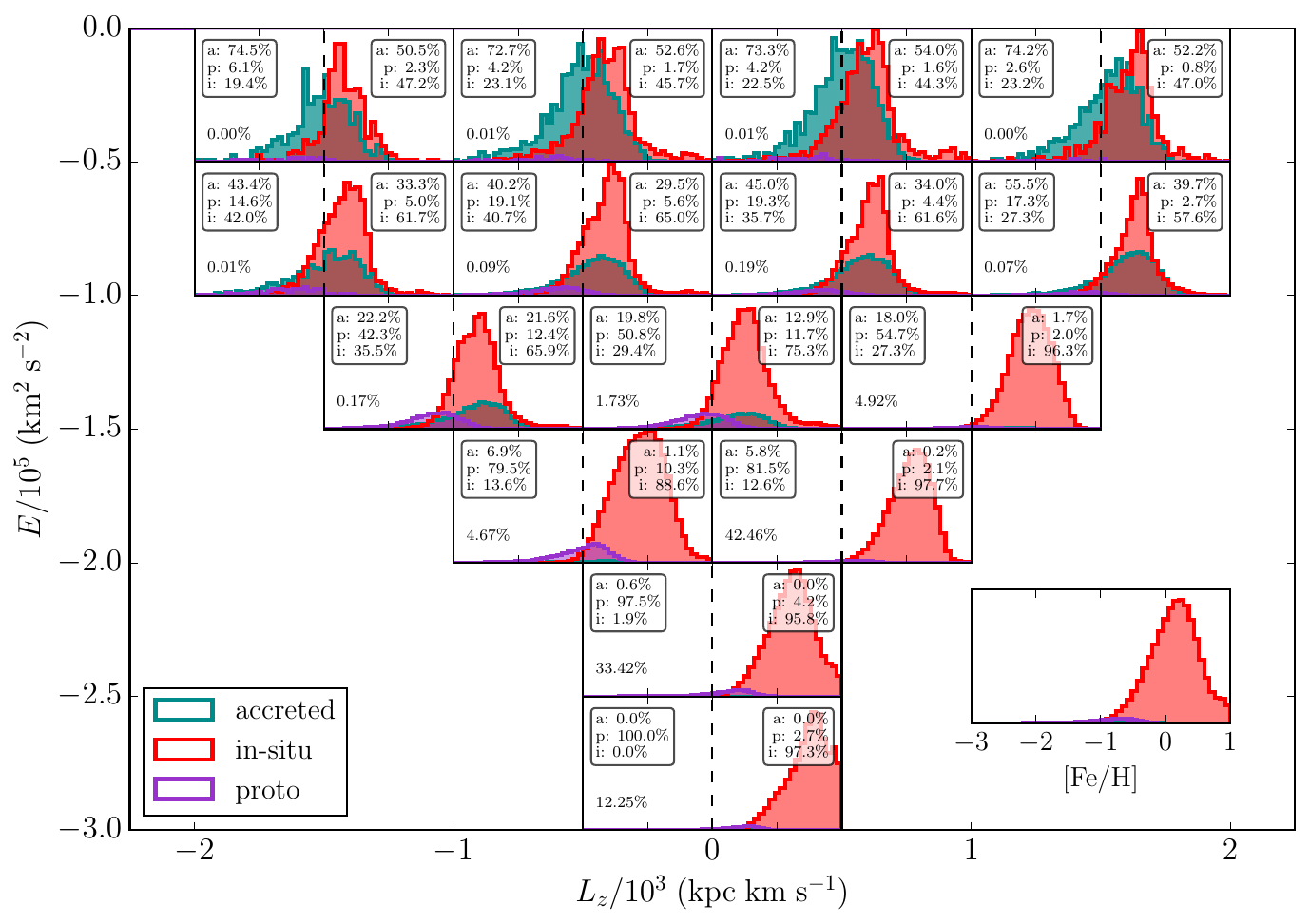}
    \caption{The metallicity distribution functions (MDF) of stars, within 4\,kpc of the centre, in different bins of $E-L_z$ space of halo 18, normalised with respect to the number of stars in each bin. We separate stars into different bins in $E-L_z$ of equal size with an extent of $1 \times 10^3 \rm \;kpc \,km \, s^{-1}$ in $L_z$ and $0.5 \times 10^5 \; \rm km^2 \, s^{-2}$ in $E$. The borders of the subplots in $E-L_z$ correspond to the boundaries of the bin in $E-L_z$. In each subplot we separate stars into accreted (turquoise), in-situ (red), and proto-galaxy (purple). The range of the y-axes of the different subplots changes between subplots, and thus the size of the peaks in the subplots should not be used to compare the number of stars present in different regions of $E-L_z$ space. To do that we include the percentage of stars that are found in each bin in $E-L_z$ in the bottom left of each subplot. The percentages in the top right show the fractional contributions of accreted (a), proto-galaxy (p), and in-situ (i) to the total number of stars in this $E-L_z$ bin. The percentages in the top left show the same fractions for all stars in this bin in $E-L_z$ with $[\rm Fe/H] < -1$. The inset in the lower right of the plot shows the MDFs of stars in the inner 4\,kpc, with no cuts in $E$ and $L_z$. Even with a metallicity cut, accreted stars make up a majority of stars only in bins with $E>-1 \times \rm 10^5 \; km^2 \, s^{-2}$. However, even combined these bins hold less than $0.4 \%$ of stars in the central 4\,kpc of this halo.}
    \label{fig:e_lz_bin_plot}
\end{figure*}

%Describe figure 5: E-L_z plot comparisons for in-situ, proto-galaxy, and accreted

In addition to the kinematics, we can use $E-L_z$ space to look for ways to separate the galaxy into its different components. We show the distribution of in-situ, proto-galaxy, and accreted stars in $E-L_z$ space in \figref{fig:e_lz_comp}. Here, we plot all stars from each component found within 4\,kpc of the centre of the galaxy. While in-situ stars are preferentially found at lower energies and within the disc of the halo, in-situ stars can also be excited to higher energies and/or retrograde orbits \citep{bugueno_2026}. In fact, the density of in-situ stars at the retrograde and high-energy regions of $E-L_z$ space is comparable to that of the accreted stars. 

The accreted component is absent at very low energies and is more uniformly distributed in $L_z$. A diagonal ridge is very apparent in the accreted population; however, it is also visible in the in-situ component. We have previously shown in \citet{tomlinson_2026} that this overdensity is caused by stars trapped in resonance with the bar in the simulation. Apart from this overdensity caused by secular processes within the halo, we see no obvious features in the accreted stars. At very high energies there are hints of horizontal features, caused by stars being lost during mergers. However, even here stars from several different merger events overlap. No overdensity stands out that would point to any single significant merger. 

The proto-galaxy is largely confined to energies below $-1 \times 10^5 \, \rm km^2 \, s^{-2}$. While it may look like proto-galaxy stars reach surprisingly high energies, 90\% of proto-galaxy stars stay confined within 6\,kpc from the centre of the galaxy. At low energies, the proto-galaxy is more prevalent than the accreted population. The distribution of proto-galaxy stars in $L_z$ is not clearly skewed to positive or negative $L_z$. There are no prominent features visible in this component. 

%Describe figure 6: E-L_z plot in bins with normalised MDF

Since there is no easy pathway to split stars into in-situ, accreted, and proto-galaxy through dynamics or metallicity alone, we combine $E-L_z$ information with the metallicity of stars to try to disentangle these components. In \figref{fig:normalised e_lz bin plot}, we split $E-L_z$ space into different bins of equal size, which follow the distribution of stars in this space. The location of each subplot matches the boundaries of each bin in $E-L_z$. In the subplots in the figure we show normalised metallicity distribution functions (MDF) of [Fe/H], where we separate stars into proto-galaxy (purple), in-situ (red), and accreted (turquoise), which are individually normalised. Thus, this plot gives us an idea of the distribution in $E-L_z$ space of each component but does not serve as a direct comparison of their total contributions. 
% How does it look if you keep the same normalisation? Also, add this exact line to the figure caption. From it it's not obvious to me that the normalisation is per component and I would have interpreted it differently.

The in-situ and proto-galaxy components generally have peaks separated by around 1 dex. However, this is partly by design. The separation between proto-galaxy and in-situ is defined by a cut in ages, which means that very old and metal-poor stars are by design excluded from the in-situ population. We further note that the proto-galaxy component has a significant low-metallicity tail. The accreted population falls between the in-situ and proto-galaxy components in most regions of $E-L_z$. At lower energies ($E < -1.5 \times 10^5 \, \rm km^2 \, s^{-2}$), the accreted MDF closely follows the MDF of the proto-galaxy component, whereas at higher energies the accreted MDF more closely resembles the in-situ population, but with a more prominent tail towards low metallicity. 

%Describe figure 7: E-L_z bin plot with unnormalised MDF

In \figref{fig:e_lz_bin_plot}, we show the MDFs, normalised with respect to the local number of stellar particles (in each bin), of the accreted, proto-galaxy, and in-situ stars in the same bins in $E-L_z$. The numbers in the top right corner of each sub-panel give the fractional contribution from the different components to the total number of stars present in that region of $E-L_z$ space. The numbers in the top left give the fractional contribution from each component to stars in this region of $E-L_z$ space with $[\rm Fe/H] < -1$. Finally, the number in the bottom left shows the fraction of the total number of stars found in this region of $E-L_z$ space compared to all stars within 4\,kpc of the centre.

At first glance, the most notable feature of this plot is the abundance of in-situ stars in every region of the $E-L_z$ plane. While most in-situ stars are found to be prograde, they also contribute significantly at higher energies and retrograde angular momentum. For $E < -1.5 \times 10^5 \, \rm km^2 \, s^{-2}$, in-situ stars consistently make up 88\% or more of the total number of stars in these $E-L_z$ bins. If we only consider prograde stars this fraction increases to over 95\% of stars. Although this appears initially daunting, we can make a cut in [Fe/H] to remove most of these in-situ stars. Selecting only stars with $[\rm Fe/H] < -1$, in the same regions of $E-L_z$ space, the fraction of in-situ stars decreases to 13.6\% at most, leaving mostly proto-galaxy stars.

Proto-galaxy stars are a significant presence in all but the highest energy bins at low metallicities. In the two lowest energy bins, over 97\% of stars with $\mathrm{[Fe/H]}<-1$ belong to the proto-galaxy component.  Thus, in order to study stars that were part of the proto-galaxy as it was forming, the low metallicity tails of stars found at low energy are the best region to probe, with little difference for stars that are prograde or retrograde. To study the accreted population,  regions with $E > -1 \times 10^5 \, \rm km^2 \, s^{-2}$ and low metallicity provide the highest likelihood of detecting an accreted star. In the region with $ -1.5 \times 10^5 \, \rm km^2 \, s^{-2} > E > -1 \times 10^5 \, \rm km^2 \, s^{-2}$, it is much more difficult to separate accreted, in-situ, and proto-galaxy stars.

\subsection{Dynamics and age}
\label{sec: age_dynamics}

\begin{figure*}
	\includegraphics[width=\textwidth]{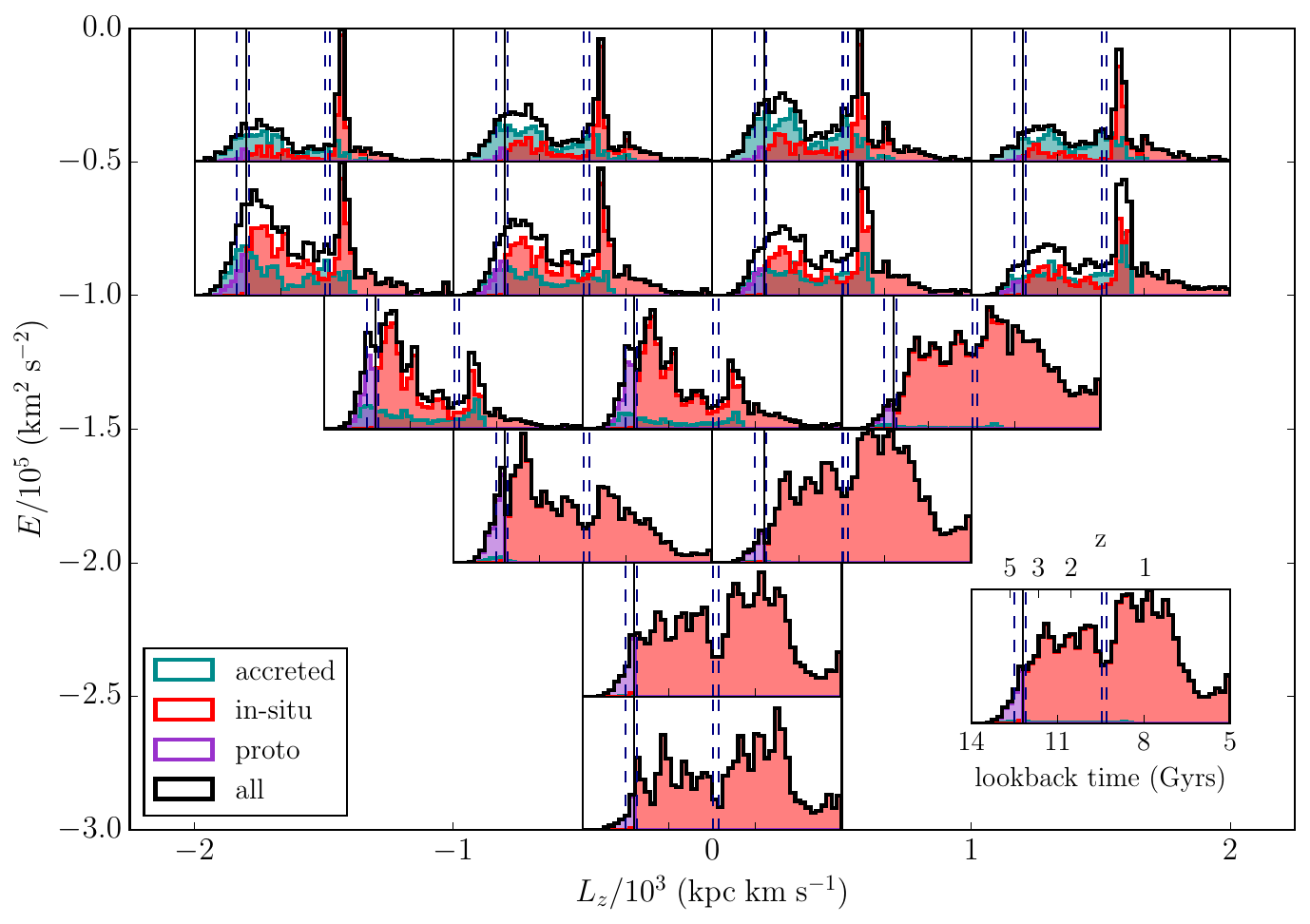}
    \caption{The distribution in the ages of stars, within 4\,kpc of the centre, in different regions of $E-L_z$ space. We separate stars into different bins in $E-L_z$ of equal size with an extent of $1 \times 10^3 \rm \;kpc \,km \, s^{-1}$ in $L_z$ and $0.5 \times 10^5 \; \rm km^2 \, s^{-2}$ in $E$. The borders of the subplots in $E-L_z$ correspond to the boundaries of the bin in $E-L_z$. In each subplot we separate stars into accreted (turquoise), in-situ (red), proto-galaxy (purple), and all stars (black). The dashed vertical lines show the infall times of the most significant mergers M1 to M4. The inset in the lower right of the plot shows the age distribution of stars in the inner 4\,kpc, with no cuts in $E$ and $L_z$.}
    \label{fig:e_lz_age_plot}
\end{figure*}

Determining the ages of stars is extremely useful for Galactic archaeology, but is observationally very difficult. Studies have made stellar age estimates using asteroseismology \citep{sharma_2019}, isochrone modelling \citep{nataf_2024}, as well as using spectrophotometric data in combination with distances to infer ages from stellar evolutionary models \citep{queiroz_2018}, among others. Errors can be large (up to $20\%$), and sample sizes are small compared to spectroscopic surveys. Estimates also tend to be less accurate at older ages. Nonetheless, there has been great progress in recent years and stellar ages will likely become an even more important tool in Galactic archaeology in the future. 

We therefore include stellar ages in our analysis to determine if these can be used to identify past mergers at the centre of the galaxy. In \figref{fig:e_lz_age_plot}, we show unnormalised histograms in age of stars in different bins in $E-L_z$ space, where stars are split into proto-galaxy, in-situ and accreted as before. We only explore stars older than 5\,Gyrs. After this time, star formation in most regions of $E-L_z$ space is smoothly decreasing with no notable star formation peaks. We further show the infall times of the most significant mergers, M1 to M4, as vertical dashed lines.

Although the proto-galaxy stage is characterised by a continuous increase in star formation in all $E-L_z$ bins, the age distribution of the in-situ and accreted components varies significantly in different bins of $E-L_z$ space. By itself, this is not entirely surprising, as we would expect most star formation to occur in the disc. We do indeed see that star formation is higher for longer time periods in the most prograde bins of $E-L_z$ space up to $E = -1 \times 10^5 \; \rm km^2 \; s^{-2}$. The stellar halo, i.e., stars with $L_z < 0$ or at high energies, is more dominated by individual peaks of star formation and sees a larger drop at recent times.

We previously mentioned in Section \ref{sec: separating components} that four mergers dominate the accreted population in the inner galaxy. In the age distribution of stars in the halo we find signatures of these mergers. Gas-rich mergers have been found in simulations to trigger starbursts \citep{kim_2009}; however, in observations the picture is a bit more complicated. There is evidence for merger-induced starbursts, but mergers do not necessarily lead to higher star formation rates \citep{knapen_2015}. 

In halo 18, there is a noticeable drop in star formation -- mainly in the in-situ population -- at a lookback time of 9.5 Gyrs. This is close to the infall times of mergers M3 and M4, the GSE analogue in halo 18. 
Following this dip, there is a pronounced peak in star formation -- also seen by \citet{grand_2020} -- which is especially noticeable at $E > -1 \times 10^5 \; \rm km^2 \; s^{-2}$. This enhanced star formation is also seen in other bins in $E-L_z$ space; however, it is far less pronounced in bins at lower energies. This peak is found predominantly in the in-situ population. By tracing back the location where these stars are born, we find that these are consistent with the orbit of M4 as it is falling into the main halo. This leads us to conclude that this star formation is occurring in gas that is brought in by the merger M4. \citet{orkney_2022} found similar evidence for star formation induced by gas-rich mergers in their analysis of halo 24 of the fiducial Auriga suite.

% I think some of the recent papers by Orkney et al. on mergers in auriga should be cited somewhere around here.
Since we see a starburst triggered by merger M4, one might wonder if a similar starburst occurred during mergers M1 and M2. In the retrograde $E-L_z$ bins with $-2 \times 10^5 \mathrm {\; km^2 \; s^{-2}} < E < -1 \times 10^5 \rm \; km^2 \; s^{-2}$, there is a peak in star formation after the infall of M1 and M2. But it is difficult to associate this peak with these mergers, since it is consistent with smoothly increasing star formation following on from the proto-galaxy phase. 

Nevertheless, studying the star formation rate in different regions of $E-L_z$ space could be a powerful tool for detecting past mergers, as stars formed inside a merging galaxy after infall preserve some of the angular momentum and energy of the merger. While a merger-induced star formation peak in the host galaxy may be washed out by secularly induced star formation, a peak in the gas of the infalling satellite can be preserved in the stellar halo. This is especially the case for mergers that deposit debris at high energy and in the retrograde part of $E-L_z$ space.

However, even with perfect age measurements, disentangling the effects of early mergers will be difficult. Other processes lead to peaks in star formation, and the effects of any single merger may be washed out due to other merger processes, as well as the secular evolution of galaxies. Changes in the potential over time also dilute the information we can gain from the present-day location of stars in $E-L_z$. This holds both for stars directly accreted in a merger and for star formation induced in infalling gas.

\subsection{The high energy accreted component}
\label{sec: high_energy}

\begin{figure*}
    \includegraphics[width=\textwidth]{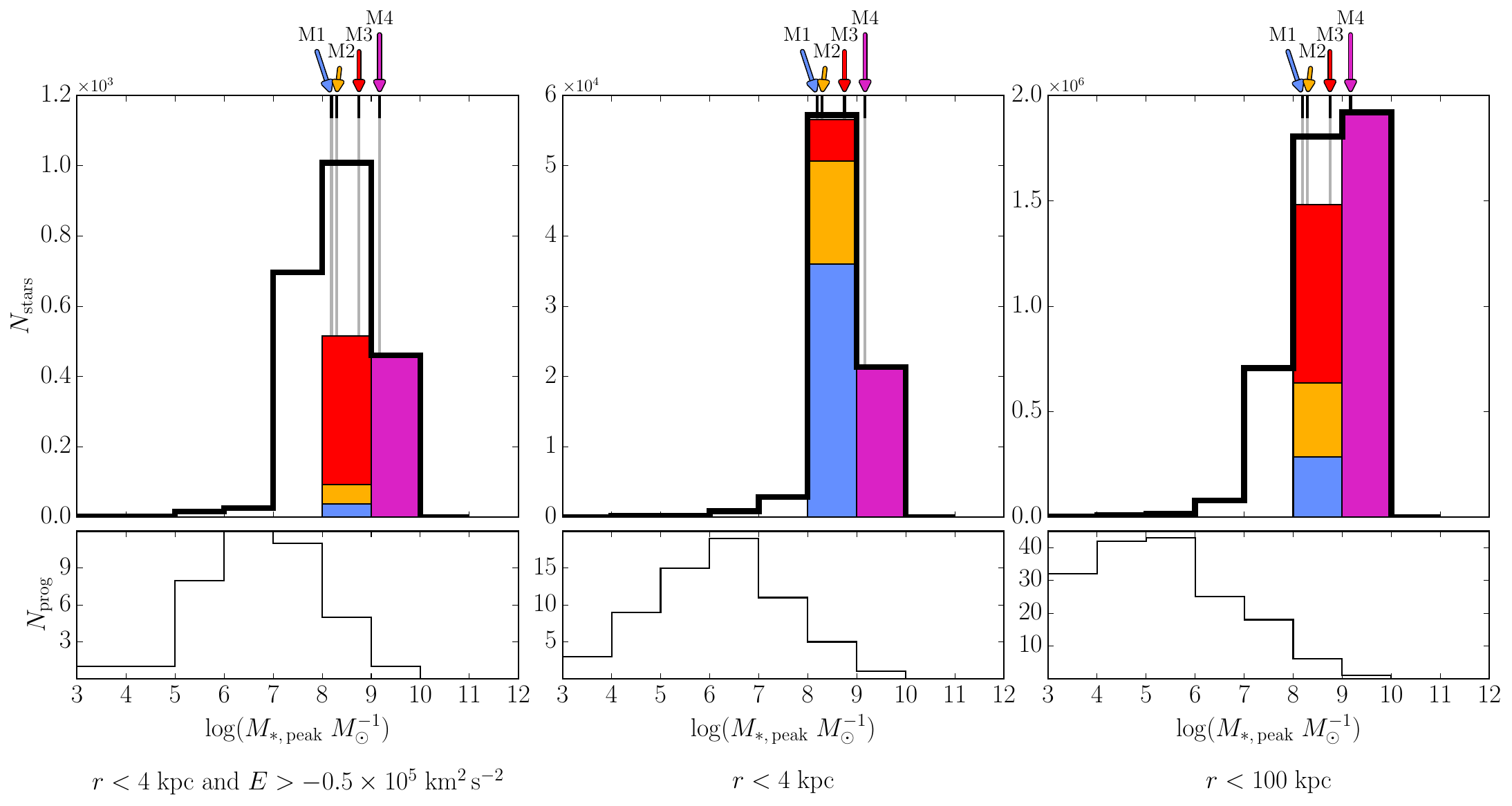}
    \caption{The three panels shown here represent three different selections of accreted stars in halo 18, all with $[\rm Fe/H] < -1$. The first panel shows accreted stars in the central 4\,kpc that have $E>-0.5 \times 10^5 \rm \; km^2 \, s^{-2}$, which is the selection that leaves us with the highest fraction of accreted stars. The second panel shows all accreted stars within 4\,kpc of the centre, but with no selection in energy. The third panel shows all accreted stars within 100\,kpc. The top histograms show how many accreted stars were brought in by progenitors in different peak stellar mass bins, for each selection of accreted stars. We highlight stars from the four main mergers in halo 18 (M1 to M4), whose peak stellar masses are indicated by the arrows above the top axis and vertical lines in the histogram. In the histogram stars from these mergers are stacked on top of each other and split into different colours, which correspond to M1, M2, M3, M4 as shown by the colours of the arrows at the top of the panel. The bottom histograms show how many progenitors contribute stars for each mass bin. In the inner regions, identifying accreted stars is easiest at high energies; however the contribution from the early mergers M1 \& M2 is much lower here than in the inner regions overall.}
    %We can see in the left panel, that at high energy stars from the four main mergers no longer dominate the accreted population, with a larger proportion of stars from progenitors with $10^7 < M_{\rm *, peak} < 10^8$.}
    \label{fig:merger_contribution_highE}
\end{figure*}

%Describe figure 10: merger contribution of high energy bin

Since it is largely the high-energy, high-velocity tails that have large accreted fractions, we want to take a closer look at these to determine which mergers make up the accreted population in this space. We do this in \figref{fig:merger_contribution_highE}, where we show the stellar contribution from merged satellites of different masses in halo 18. In the following analysis, we make a metallicity cut of $\rm [Fe/H] < -1 $. We look at accreted stars in the central 4\,kpc that have  $E>-0.5 \times 10^5 \rm \; km^2 \, s^{-2}$ (left panel), as well as accreted stars in the central 4\,kpc without a cut in energy (middle panel). We further look at accreted stars within 100\,kpc (right panel) to better understand how the contribution of stars from mergers of different stellar masses varies between the very inner accreted halo and the extended stellar halo of the galaxy.

The total number of mergers that have deposited metal-poor stars in the spatial and dynamical regions of interest increases as we go from the central 4\,kpc with an energy cut, to the central 4\,kpc without an energy cut, and ultimately to the halo within 100\,kpc. There are 33 progenitors with a stellar mass larger than $10^3 \, \rm M_{\odot}$ that deposit stars within 4\,kpc of the centre at $z=0$ with $E>-0.5 \times 10^5 \rm \; km^2 \, s^{-2}$. There are 65 progenitors in the central 4\,kpc with no energy cut, and 176 when just making a radius cut of 100\,kpc (without a metallicity cut these numbers are 36, 69, and 178 respectively). 

% Without a metallicity cut the numbers are 36, 69, 178

The contribution of stars from different mergers changes based on the cut we make (however for all of these we still consider only stars with $\rm [Fe/H] < -1 $). Looking at the central 4\,kpc without further cuts, we see that the vast majority of accreted stars ($>90\%$) come from the four most significant mergers M1 to M4. Comparing this to the selection within 100\,kpc, we see that here we have a larger contribution from less massive mergers. We further see that M4 dominates within 100\,kpc, whereas within 4\,kpc M1 is dominant. This picture changes drastically as we introduce the energy cut in the selection within 4\,kpc. Mergers M1 to M4 now contribute less than $50\%$. We see both a higher number of stars from mergers whose progenitors had masses between $10^8 - 10^9 \; M_{\odot}$ that are not M1, M2, or M3, and also an increased contribution from lower-mass progenitors with masses of $10^7 - 10^8 \; M_{\odot}$. 

\begin{figure*}
	\includegraphics[width=.94\textwidth]{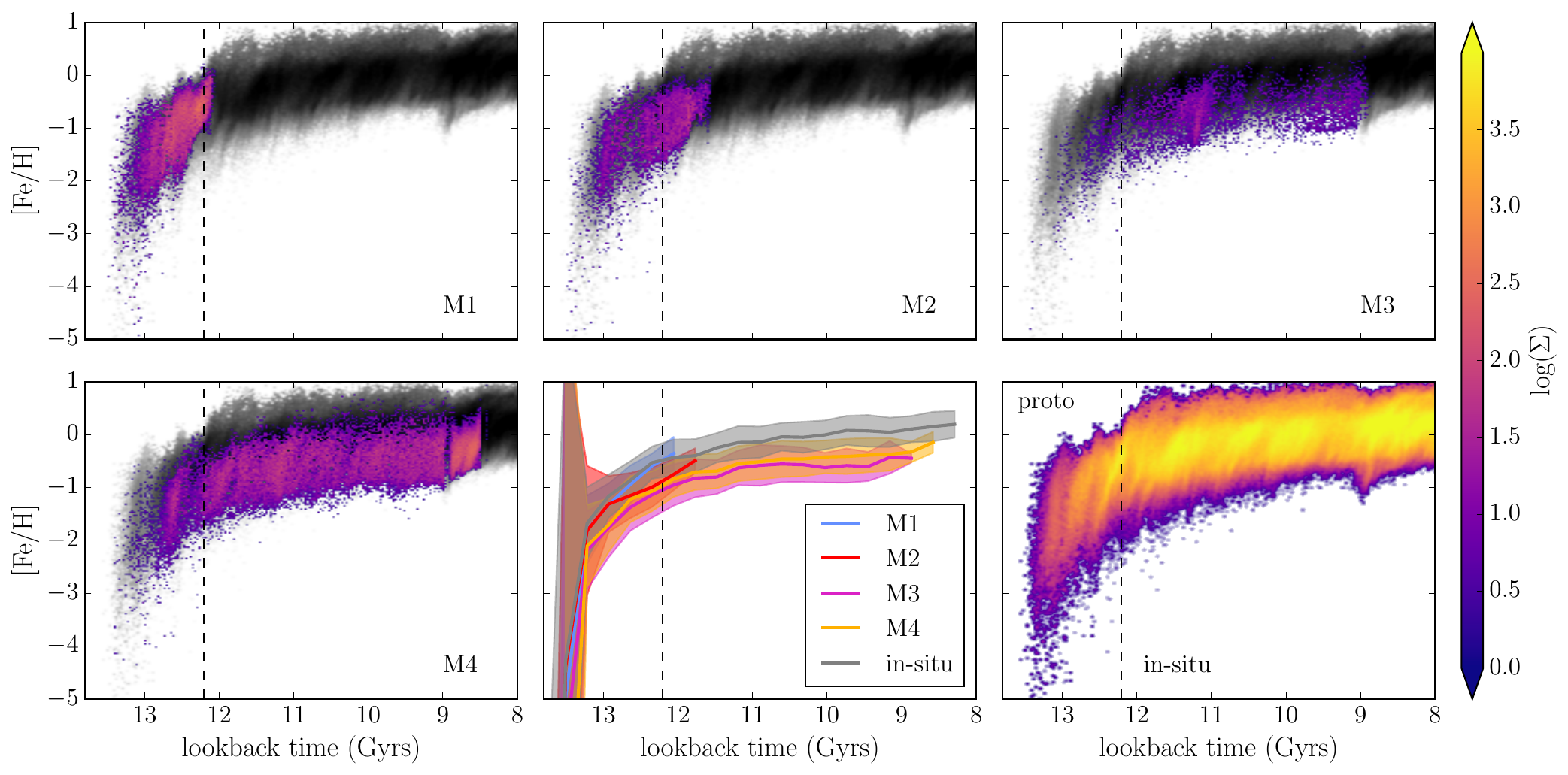}
    \caption{Age-metallicity relations of the 4 mergers and the in-situ population, for stars within 4\,kpc of the centre. The bins are coloured by logarithmic density, $\Sigma$ corresponds to the number of particles per bin, here in lookback time and $\rm [Fe/H]$. In panels 1 to 4 (left to right, top to bottom), the coloured bins show the distribution of stars from the four main mergers, arranged by infall time. The grey background shows the total population of stars. Panel 5 shows the mean of the distribution, with a shaded region of $1\sigma$ standard deviation. Panel 6 shows the proto-galaxy and in-situ populations.}
    \label{fig:age_metallicity_mergers}
\end{figure*}

The take-away from this Subsection is that accreted fractions, even at low metallicities, are much higher at high energies. While at high energies, with a metallicity cut, accreted stars contribute more than 50\% of stars in the inner regions (see \figref{fig:e_lz_bin_plot}), it is not the large early mergers that dominate, but later or smaller accretion events. This is important to keep in mind when searching for traces of mergers in our own Galaxy. Even at the very centre there are likely to be remnants of later mergers, especially at high energies. Further, as we will see in the next Subsection, while the inner regions of the galaxy do contain a higher fraction of stars from very early accretion events (M1 \& M2), they are likely to be found in lower-energy regions of phase-space that are dominated by in-situ and proto-galaxy stars.

\begin{figure*}
	\includegraphics[width=.8\textwidth]{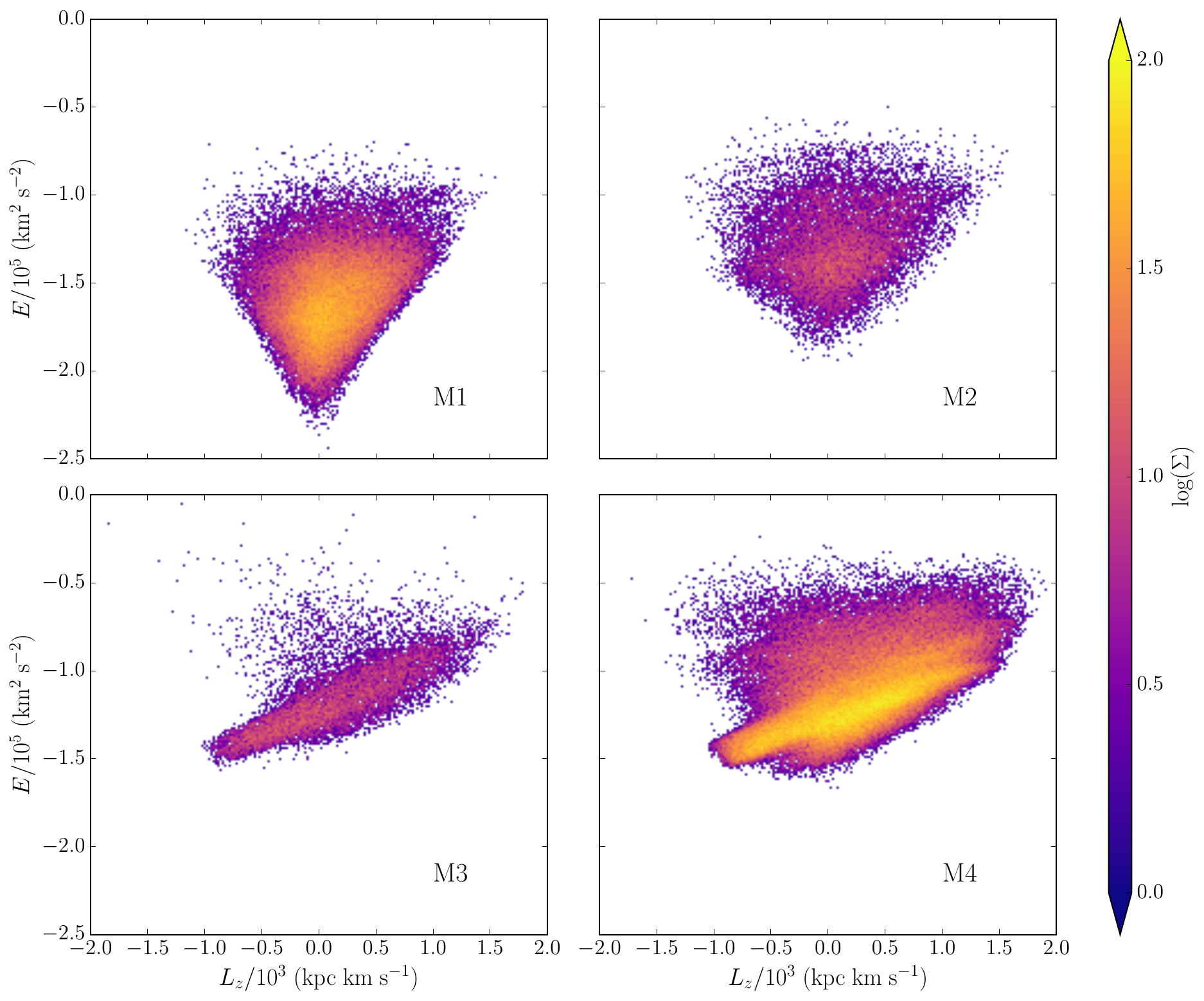}
    \caption{Distribution of stars from the four main mergers in $E-L_z$ space, for stars within 4\,kpc of the centre. $\Sigma$ corresponds to the number of particles per bin, here in $L_z$ and $E$. The earlier mergers M1 and M2 are found at lower energies than M3 and M4, but there is no visible substructure that is a remnant of the merger. The prominent ridge seen in M3 and M4 is the result of stars being trapped in resonance with the bar.}
    \label{fig:e_lz_different mergers}
\end{figure*}

\subsection{The properties of the most significant mergers}
\label{sec: mergers}

As mentioned in Section \ref{sec: separating components}, in the central 4\,kpc there are four mergers that contribute more than $98\%$ to the accreted population in this region. Two of these, M1 and M2, merged shortly after the end of the proto-galaxy stage, while M3 and M4 merged at a time of around 4\,Gyrs since the start of the simulation. A summary of some of the properties of these mergers is provided in \tabref{tab:1}.

%Describe figure 11: Age-metallicity relation of the 4 main mergers

We want to understand where these main mergers lie in terms of their properties and how they compare to each other as well as the proto-galaxy and in-situ components. For this purpose, we show the age-metallicity relation of stars from mergers M1 to M4 in \figref{fig:age_metallicity_mergers}. We also compare with the age-metallicity relation of the proto-galaxy and in-situ stars. 

M1 and M2 show rapid early enrichment, similar to the proto-galaxy. M1 has essentially the same mean [Fe/H] value as the proto-galaxy at all times. M2 begins deviating from the proto-galaxy at a lookback time of around 12.7\,Gyrs and is at most 0.5 dex lower in metallicity than the proto-galaxy. M3 and M4 are lower still in metallicity, with a difference in the mean of largely $0.5-0.7$ dex to the in-situ population for each age. Thus, it is only after the first billion years of enrichment that noticeable differences appear between the galaxy and its building blocks and subsequent significant mergers. However, after the cut-off time of the proto-galaxy (z=3.8), M3 and M4 have mean [Fe/H] consistently $\sim$0.5 dex lower than the in-situ population. Thus, once the main halo has formed from its building blocks, it may be possible to distinguish stars accreted in subsequent mergers from those formed inside the main galaxy, with the addition of precise ages.

% is it? I thought you said it was really hard above?  Maybe if you knew their exact ages, but otherwise the accreted star metallicities just overlap with the protogalaxy metallicities.... so how do you distinguish them?

%Describe figure 12: E-L_z distributions of the 4 main mergers

In \figref{fig:e_lz_different mergers}, we show the distribution of stars from M1 - M4 in $E-L_z$ space. The general trend for these mergers is that the earlier the merger, the lower its energy at the present day. It does not follow that these mergers are identifiable as separate accretion events from their distribution in $E-L_z$. Indeed, large overlaps in $E-L_z$ space complicate disentangling these mergers. In addition to these overlaps, not much information is preserved beyond different spreads in energy. The early mergers M1 and M2 are redistributed in $E-L_z$ space in such a way that erases information on the properties of the merger. M3 and M4 show features in $E-L_z$ space; however the most prominent overdensities are caused by the accreted stars being trapped in bar resonances \citep{tomlinson_2026}. 
% I assume that looking at the mean Lz would tell you if something about its orbit?
 
%\captionsetup[subfigure]{justification=raggedright, singlelinecheck=false}

\section{Discussion}
\label{sec: discussion}

%What does it mean to define this proto-galaxy component? What are the implications for observations?

\subsection{The proto-galaxy}
\label{sec:proto-galaxy_discussion}

%-> At these early times it is difficult to distinguish where stars where formed. 

In recent years, there has been an increased interest in the proto-galaxy, driven both by the discovery of old and metal-poor stellar populations in the Milky Way and through the study of cosmological simulations. In simulations, one possible definition of the proto-galaxy is the point at which it is no longer sensible to describe any halo as the main halo. At early times, Milky Way-like galaxies do not have just one progenitor that is clearly the most massive at any time. They could have three or more haloes of similar masses, and the most massive halo in one snapshot may not be the most massive halo in the next snapshot due to the frequent mergers that characterise this early phase of the universe. 

Due to the constant mergers and similar properties of the earliest progenitors, differentiating where stars formed is extremely challenging. Early enrichment in halo 18 is nearly indistinguishable for a large range of different progenitors, which we can see in \figref{fig:age_metallicity_mergers}. M1 and M2 have much smaller masses than the main halo at the time of their merger; however, their age-metallicity relation overlaps entirely with that of the proto-galaxy. This agrees with the results of previous studies of the fiducial Auriga suite, which identify the inner galaxy as a promising region for the discovery of early mergers, but also point out that chemical abundances of early, Kraken-like mergers are almost indistinguishable from the metal-poor in-situ and proto-galaxy population \citep{orkney_2023}. It may be possible to associate bursts in star formation that occurred in one of the proto-galaxy's building blocks but not in the other. However, the close proximity of progenitors means that merger-induced starbursts can occur at the same time in different haloes. Identifying individual starbursts also requires ages with higher precision than what is currently possible for the oldest stars.

However, simply identifying stars associated with the proto-galaxy appears fairly straightforward. The recipe is to look for metal-poor stars found at very low energies. Finding the correct metallicity cut to isolate the proto-galaxy could be based on a kinematic selection that identifies the metallicity at which rotation increases significantly. In this simulation, this places a boundary of $\rm [Fe/H]<-1$. However, massive radial mergers can disrupt the kinematics of the galaxy and make it harder to recover the timing of the formation of the disc \citep{orkney_2026}. Nonetheless, here we find that metal-poor stars at low energy are consistently contributed by the proto-galaxy at a very high fraction. Even for another Superstars halo (halo 27, see Appendix \ref{sec:appendix a}) with a much higher accreted fraction, the proto-galaxy makes up over $60\%$ of stars with $\rm [Fe/H]<-1$ in the three lowest energy bins of \figref{fig:mdf_grid_au27}. At slightly higher energies, where there is a larger contribution from early accretion events, separating these from the proto-galaxy is still very difficult due to similar early enrichment. At later times, when metal enrichment is more gradual, the prospect of differentiating subsequent mergers from the proto-galaxy is more achievable.

\subsection{What can we learn about the central galaxy?}

What can we learn in the future about the innermost parts of the Milky Way? The centre of the Galaxy is interesting in part due to the presence of some of the oldest stars. We see this in halo 18 as a high proto-galaxy fraction at low metallicity. We further see that early significant accretion events are generally more confined to the inner galaxy, on lower energy orbits. But where merger remnants are preferentially found and where we can actually identify them as distinct accretion events are two very different things. 

% Did you test this with a larger sample than M1-4?
We find that at low energies, where early significant mergers are found, the stellar distribution is dominated by proto-galaxy stars at low metallicities, with very similar metallicity distributions of the accreted and proto-galaxy populations. There are no overdensities preserving information on the orbits of these mergers. Where they end up in $E-L_z$ space is largely predetermined by the time of the merger and the mass of the infalling satellites. The features we do see in the accreted population are not caused by properties of the merger, but by secular processes, in particular the trapping of stars by the bar. Thus, features we see in our Galaxy, even in their chemistry (see \citealt{tomlinson_2026}), may not inform us about particular mergers, but the evolution of the bar in the Galaxy.

At high energies, the likelihood of a star being accreted is much higher. However, the stars found at these high energies ($E > -0.5 \times 10^5 \; \rm km^2 \; s^{-2}$ in this simulation) constitute only around $0.02\%$ of stars in the inner 4\;kpc of this simulated galaxy. Additionally, stars at these high energies are much more likely to have been brought in by mergers of less massive progenitors. In particular, mergers M1 and M2, which together dominate the overall accreted population in the inner 4\;kpc, are barely represented at high energies. Nonetheless, high-energy stars in the inner galaxy are a promising object for the search for past mergers.

Further, while current clustering algorithms may struggle to accurately identify stars from single progenitors, especially in crowded regions with a lot of in-situ contamination \citep{thomas_2025}, in the future this may be overcome with models informed by simulations and provided with detailed chemical and dynamical information. A new algorithm, based on simulation-based inference in the fiducial Auriga simulations, developed by \citet{sante_2025}, appears to be more accurate at recovering debris from single progenitors. An analysis of the local halo using this algorithm further raised the exciting prospect that the number of early accretion events is significantly higher than the number of previously identified progenitors \citep{sante_2026}.

Thus, not all is lost. New methods may allow us to identify merger remnants in regions with higher in-situ contamination. Moreover, accurate ages for a larger number of stars in the central Galaxy will allow us to better constrain its evolution. It would be possible to identify early starbursts, which may be used in the search for past mergers, as also proposed by \citet{orkney_2022}. More detailed chemical abundances, which we have not discussed in this paper, can also provide further insight into the different environments that stars formed in. In observations, abundances such as aluminium, manganese, and europium have been used in attempts to separate in-situ and accreted stars in the Milky Way \citep{horta_2021, monty_2024}. Thus, while there may be a point beyond which identifying the progenitor of a star with confidence is no longer possible, we can nonetheless keep pushing this boundary to earlier and earlier times and delve deeper into the Galaxy's past.

% is it worth to discuss a bit about the low-Z, high E region is a good place to search for accreted stars?

\section{Summary and Conclusions}
\label{sec: conclusions}

How much information on the Galaxy's past can be found as we look deeper and deeper into its heart? Using the Auriga Superstars suite of high-resolution simulations, we have analysed the centre of a Milky Way analogue to understand whether we can uncover the formation of the Milky Way up to its earliest building blocks. To do this, we split stars in the simulation into three components, proto-galaxy, in-situ, and accreted. We use information encoded in the kinematics, dynamical spaces, metallicity, as well as ages, to attempt to disentangle the origin of stars in the central 4\,kpc of the simulated galaxy. Our main conclusions are summarized as follows:
% Could use a different word maybe? You just used disentangle for formation of the MW.

\begin{description}
    \item[(i)] As shown in \figref{fig:kinematics}, the line-of-sight velocities of halo 18 are in good agreement with observational data from PIGS \citep{arentsen_2020, arentsen_2024}. We find a steep decline in stellar net rotation between $\text{[Fe/H]} = 0$ and $\text{[Fe/H]} = -1$. Below $\text{[Fe/H]} = -1$, the rotation plateaus and remains approximately constant. This transition marks a shift between the slowly rotating proto-galaxy and the strongly rotating in-situ stars. Using kinematics of the metal-poor stars at the centre of the Milky Way, it may be possible to find a similar cut in metallicity that allows for the selection of proto-galaxy stars. It is not just the proto-galaxy that is rotating, however. We find that the accreted stars have an almost indistinguishable amount of net rotation from the proto-galaxy.
    \item[(ii)] In this simulation, the proto-galaxy component can be robustly identified by combining a metallicity cut of $\text{[Fe/H]} < -1$ with a cut in energy, targeting the most bound stars at the lowest energies. In the two lowest energy bins of halo 18, over $97\%$ of the metal-poor stars belong to the proto-galaxy. There is halo-to-halo variation in the exact numbers; however at very low energies these are consistently above $96\%$.
% Did you not just say in point (i) that one may be able to find proto-galaxy stars with just a Z cut?
    \item[(iii)] While it is possible to accurately identify the proto-galaxy at low energies, elsewhere in $E-L_z$ space disentangling the components of the inner galaxy is very challenging. In-situ stars are found in large numbers across $E-L_z$ space, contributing significantly even on retrograde and high-energy orbits. In halo 18, applying a metallicity cut of $\text{[Fe/H]} < -1$ mitigates this in-situ contamination, but also removes a significant fraction of accreted stars. Furthermore, the prominent diagonal overdensities seen in the accreted population within $E-L_z$ space are not intrinsic signatures of the individual mergers themselves, but are instead a consequence of secular processes -- specifically, accreted stars being trapped in resonance with the galactic bar \citep{tomlinson_2026}.
% Is this not similar to some of point (i)?
    
% Maybe then the "may be possible to find a similar cut..." in point (i) is a bit weak compared to this?
    \item[(iv)] During mergers, star formation in the infalling gas can lead to pronounced peaks in the star formation rate. These are especially noticeable when looking at stars at high energies. Variation in the star formation history in different regions of $E-L_z$ space can provide information on the $E-L_z$ space distribution of a merger and help identify individual merger events. However, this applies only to stars formed during the merger and does not provide us with information on other stars that were part of the progenitor.
% I don't follow how the peaks are at high energies?
    \item[(v)] In $E-L_z$ space, the highest accreted fractions in the inner galaxy are found within high-energy regions ($E > -0.5 \times 10^5\,\mathrm{km}^2\,\mathrm{s}^{-2}$) combined with a metallicity cut of $\text{[Fe/H]} < -1$. At these high energies, observers are more likely to encounter stars from smaller, less massive progenitors ($M_{\text{star}} \approx 10^7 - 10^9\,\rm{M}_\odot$) rather than the earliest massive mergers. To find the most ancient, massive accretion events, we need to probe lower energies. Here the contribution from in-situ and proto-galaxy stars increases, and additionally these progenitors are heavily phase-mixed. This prevents the earliest massive mergers from being cleanly separated into discrete events by dynamics and metallicity alone.
%  thought your results were that this was really hard. I just went back to the section and you write something that seems like the opposite of this: "It does not follow that these mergers are identifiable as separate accretion events from their distribution in E-L_z."
    \item[(vi)] Combining metallicity with accurate ages will likely provide the best avenue to discover and correctly identify individual mergers. At later times, after the proto-galaxy phase, accreted systems have metallicities noticeably lower than the in-situ population (around 0.5 dex in halo 18). However, the rapid metal enrichment during the proto-galaxy phase leads to a large scatter that obscures differences between different mergers and the proto-galaxy.
%also because they are probably all quite similar in mass at these earlier times?
    
\end{description}

% Need to think about final paragraph

The centre of our Galaxy holds the knowledge of its early assembly. We do, however, need to be careful in how we interpret the chemistry and dynamics of these oldest, most metal-poor stars. Here, high-resolution spectra, as well as stellar ages, will be crucial to provide us with information on the earliest assembly of our Galaxy.

\section*{Acknowledgements}

TT acknowledges funding from the Science and Technology Facilities Council (grant no. 2876865). 
FF is supported by a UKRI Future Leaders Fellowship (grant no. MR/X033740/1). 
AF acknowledges support by a UKRI Future Leaders Fellowship (grant no. MR/T042362/1) and a Swedish Wallenberg Academy Fellowship. 
AC and AD acknowledge support from the Science and Technology Facilities Council (STFC) [grant numbers ST/T000244/1 and ST/X001075/1] and the Leverhulme Trust. AD is supported by a Royal Society University Research Fellowship. 
RJJG acknowledges support from an STFC Ernest Rutherford Fellowship (ST/W003643/1).
FAG acknowledges support from the ANID BASAL project FB210003, from the ANID FONDECYT Regular grant 1251493 and from the HORIZON-MSCA-2021-SE-01 Research and Innovation Programme under the Marie Sklodowska-Curie grant agreement number 101086388.
FvdV is supported by a Royal Society University Research Fellowship (URF/R1/191703 and URF/R/241005).
RB is supported by the SNSF through the Ambizione Grant PZ00P2\textunderscore223532.
This work used the DiRAC@Durham facility managed by the Institute for Computational Cosmology on behalf of the STFC DiRAC HPC Facility (www.dirac.ac.uk). The equipment was funded by BEIS capital funding via STFC capital grants ST/P002293/1, ST/R002371/1 and ST/S002502/1, Durham University and STFC operations grant ST/R000832/1. DiRAC is part of the National e-Infrastructure.

%%%%%%%%%%%%%%%%%%%%%%%%%%%%%%%%%%%%%%%%%%%%%%%%%%
\section*{Data Availability}

The data underlying this article will be shared on reasonable request to the corresponding author.



\bibliographystyle{mnras}
\bibliography{main} % if your bibtex file is called example.bib

@ARTICLE{abraham_1999,
       author = {{Abraham}, R.~G. and {Merrifield}, M.~R. and {Ellis}, R.~S. and {Tanvir}, N.~R. and {Brinchmann}, J.},
        title = "{The evolution of barred spiral galaxies in the Hubble Deep Fields North and South}",
      journal = {\mnras},
         year = 1999,
        month = sep,
       volume = {308},
       number = {2},
        pages = {569-576},
          doi = {10.1046/j.1365-8711.1999.02766.x},
archivePrefix = {arXiv},
       eprint = {astro-ph/9811476},
 primaryClass = {astro-ph},
       adsurl = {https://ui.adsabs.harvard.edu/abs/1999MNRAS.308..569A}
}

@ARTICLE{allende_2006,
       author = {{Allende Prieto}, Carlos and {Beers}, Timothy C. and {Wilhelm}, Ronald and {Newberg}, Heidi Jo and {Rockosi}, Constance M. and {Yanny}, Brian and {Lee}, Young Sun},
        title = "{A Spectroscopic Study of the Ancient Milky Way: F- and G-Type Stars in the Third Data Release of the Sloan Digital Sky Survey}",
      journal = {\apj},
         year = 2006,
        month = jan,
       volume = {636},
       number = {2},
        pages = {804-820},
          doi = {10.1086/498131},
archivePrefix = {arXiv},
       eprint = {astro-ph/0509812},
 primaryClass = {astro-ph},
       adsurl = {https://ui.adsabs.harvard.edu/abs/2006ApJ...636..804A}
}

@ARTICLE{arentsen_2020,
       author = {{Arentsen}, A. and {Starkenburg}, E. and {Martin}, N.~F. and {Hill}, V. and {Ibata}, R. and {Kunder}, A. and {Schultheis}, M. and {Venn}, K.~A. and {Zucker}, D.~B. and {Aguado}, D. and {Carlberg}, R. and {Gonz{\'a}lez Hern{\'a}ndez}, J.~I. and {Lardo}, C. and {Longeard}, N. and {Malhan}, K. and {Navarro}, J.~F. and {S{\'a}nchez-Janssen}, R. and {Sestito}, F. and {Thomas}, G. and {Youakim}, K. and {Lewis}, G.~F. and {Simpson}, J.~D. and {Wan}, Z.},
        title = "{The Pristine Inner Galaxy Survey (PIGS) I: tracing the kinematics of metal-poor stars in the Galactic bulge}",
      journal = {\mnras},
         year = 2020,
        month = jan,
       volume = {491},
       number = {1},
        pages = {L11-L16},
          doi = {10.1093/mnrasl/slz156},
archivePrefix = {arXiv},
       eprint = {1910.06337},
 primaryClass = {astro-ph.GA},
       adsurl = {https://ui.adsabs.harvard.edu/abs/2020MNRAS.491L..11A}
}

@ARTICLE{arentsen_2024,
       author = {{Ardern-Arentsen}, Anke and {Monari}, Giacomo and {Queiroz}, Anna B.~A. and {Starkenburg}, Else and {Martin}, Nicolas F. and {Chiappini}, Cristina and {Aguado}, David S. and {Belokurov}, Vasily and {Carlberg}, Ray and {Monty}, Stephanie and {Myeong}, GyuChul and {Schultheis}, Mathias and {Sestito}, Federico and {Venn}, Kim A. and {Vitali}, Sara and {Yuan}, Zhen and {Zhang}, Hanyuan and {Buder}, Sven and {Lewis}, Geraint F. and {Oliver}, William H. and {Wan}, Zhen and {Zucker}, Daniel B.},
        title = "{The Pristine Inner Galaxy Survey - VIII. Characterizing the orbital properties of the ancient, very metal-poor inner Milky Way}",
      journal = {\mnras},
         year = 2024,
        month = may,
       volume = {530},
       number = {3},
        pages = {3391-3411},
          doi = {10.1093/mnras/stae1049},
archivePrefix = {arXiv},
       eprint = {2312.03847},
 primaryClass = {astro-ph.GA},
       adsurl = {https://ui.adsabs.harvard.edu/abs/2024MNRAS.530.3391A}
}

@ARTICLE{belokurov_2018,
       author = {{Belokurov}, V. and {Erkal}, D. and {Evans}, N.~W. and {Koposov}, S.~E. and {Deason}, A.~J.},
        title = "{Co-formation of the disc and the stellar halo}",
      journal = {MNRAS},
         year = 2018,
        month = jul,
       volume = {478},
       number = {1},
        pages = {611-619},
          doi = {10.1093/mnras/sty982},
archivePrefix = {arXiv},
       eprint = {1802.03414},
 primaryClass = {astro-ph.GA},
       adsurl = {https://ui.adsabs.harvard.edu/abs/2018MNRAS.478..611B}
}

@ARTICLE{belokurov_2022,
       author = {{Belokurov}, Vasily and {Kravtsov}, Andrey},
        title = "{From dawn till disc: Milky Way's turbulent youth revealed by the APOGEE+Gaia data}",
      journal = {\mnras},
         year = 2022,
        month = jul,
       volume = {514},
       number = {1},
        pages = {689-714},
          doi = {10.1093/mnras/stac1267},
archivePrefix = {arXiv},
       eprint = {2203.04980},
 primaryClass = {astro-ph.GA},
       adsurl = {https://ui.adsabs.harvard.edu/abs/2022MNRAS.514..689B}
}

@article{binney_2013,
	title = {Dynamics for galactic archaeology},
	volume = {57},
	doi = {10.1016/j.newar.2013.08.001},
	number = {3-4},
	journal = {New Astron. Rev.},
	author = {Binney, James},
	month = sep,
	year = {2013},
	note = {\_eprint: 1309.2794},
	pages = {29--51},
}

@ARTICLE{bland_hawthorn_2016,
       author = {{Bland-Hawthorn}, Joss and {Gerhard}, Ortwin},
        title = "{The Galaxy in Context: Structural, Kinematic, and Integrated Properties}",
      journal = {ARA&A},
         year = 2016,
        month = sep,
       volume = {54},
        pages = {529-596},
          doi = {10.1146/annurev-astro-081915-023441},
archivePrefix = {arXiv},
       eprint = {1602.07702},
 primaryClass = {astro-ph.GA},
       adsurl = {https://ui.adsabs.harvard.edu/abs/2016ARA&A..54..529B}
}

@ARTICLE{bonaca_2025,
       author = {{Bonaca}, Ana and {Price-Whelan}, Adrian M.},
        title = "{Stellar streams in the Gaia era}",
      journal = {\nar},
         year = 2025,
        month = jun,
       volume = {100},
          eid = {101713},
        pages = {101713},
          doi = {10.1016/j.newar.2024.101713},
archivePrefix = {arXiv},
       eprint = {2405.19410},
 primaryClass = {astro-ph.GA},
       adsurl = {https://ui.adsabs.harvard.edu/abs/2025NewAR.10001713B}
}

@ARTICLE{bluck_2012,
       author = {{Bluck}, Asa F.~L. and {Conselice}, Christopher J. and {Buitrago}, Fernando and {Gr{\"u}tzbauch}, Ruth and {Hoyos}, Carlos and {Mortlock}, Alice and {Bauer}, Amanda E.},
        title = "{The Structures and Total (Minor + Major) Merger Histories of Massive Galaxies up to z \raisebox{-0.5ex}\textasciitilde 3 in the HST GOODS NICMOS Survey: A Possible Solution to the Size Evolution Problem}",
      journal = {\apj},
         year = 2012,
        month = mar,
       volume = {747},
       number = {1},
          eid = {34},
        pages = {34},
          doi = {10.1088/0004-637X/747/1/34},
archivePrefix = {arXiv},
       eprint = {1111.5662},
 primaryClass = {astro-ph.CO},
       adsurl = {https://ui.adsabs.harvard.edu/abs/2012ApJ...747...34B}
}

@ARTICLE{bugueno_2026,
       author = {{Bugue{\~n}o}, M.~C. and {G{\'o}mez}, Facundo A. and {Dolfi}, Arianna and {Tissera}, Patricia B.},
        title = "{Origins of coplanar counterrotating stellar disk components in late-type galaxies: I. Identification, characterization, and formation mechanisms of coplanar counterrotating stellar disk components}",
      journal = {\aap},
         year = 2026,
        month = feb,
       volume = {707},
          eid = {A51},
        pages = {A51},
          doi = {10.1051/0004-6361/202557544},
archivePrefix = {arXiv},
       eprint = {2510.03106},
 primaryClass = {astro-ph.GA},
       adsurl = {https://ui.adsabs.harvard.edu/abs/2026A&A...707A..51C}
}

@ARTICLE{chambers_2016,
       author = {{Chambers}, K.~C. and {Magnier}, E.~A. and {Metcalfe}, N. and {Flewelling}, H.~A. and {Huber}, M.~E. and {Waters}, C.~Z. and {Denneau}, L. and {Draper}, P.~W. and {Farrow}, D. and {Finkbeiner}, D.~P. and {Holmberg}, C. and {Koppenhoefer}, J. and {Price}, P.~A. and {Rest}, A. and {Saglia}, R.~P. and {Schlafly}, E.~F. and {Smartt}, S.~J. and {Sweeney}, W. and {Wainscoat}, R.~J. and {Burgett}, W.~S. and {Chastel}, S. and {Grav}, T. and {Heasley}, J.~N. and {Hodapp}, K.~W. and {Jedicke}, R. and {Kaiser}, N. and {Kudritzki}, R.-P. and {Luppino}, G.~A. and {Lupton}, R.~H. and {Monet}, D.~G. and {Morgan}, J.~S. and {Onaka}, P.~M. and {Shiao}, B. and {Stubbs}, C.~W. and {Tonry}, J.~L. and {White}, R. and {Ba{\~n}ados}, E. and {Bell}, E.~F. and {Bender}, R. and {Bernard}, E.~J. and {Boegner}, M. and {Boffi}, F. and {Botticella}, M.~T. and {Calamida}, A. and {Casertano}, S. and {Chen}, W.-P. and {Chen}, X. and {Cole}, S. and {Deacon}, N. and {Frenk}, C. and {Fitzsimmons}, A. and {Gezari}, S. and {Gibbs}, V. and {Goessl}, C. and {Goggia}, T. and {Gourgue}, R. and {Goldman}, B. and {Grant}, P. and {Grebel}, E.~K. and {Hambly}, N.~C. and {Hasinger}, G. and {Heavens}, A.~F. and {Heckman}, T.~M. and {Henderson}, R. and {Henning}, T. and {Holman}, M. and {Hopp}, U. and {Ip}, W.-H. and {Isani}, S. and {Jackson}, M. and {Keyes}, C.~D. and {Koekemoer}, A.~M. and {Kotak}, R. and {Le}, D. and {Liska}, D. and {Long}, K.~S. and {Lucey}, J.~R. and {Liu}, M. and {Martin}, N.~F. and {Masci}, G. and {McLean}, B. and {Mindel}, E. and {Misra}, P. and {Morganson}, E. and {Murphy}, D.~N.~A. and {Obaika}, A. and {Narayan}, G. and {Nieto-Santisteban}, M.~A. and {Norberg}, P. and {Peacock}, J.~A. and {Pier}, E.~A. and {Postman}, M. and {Primak}, N. and {Rae}, C. and {Rai}, A. and {Riess}, A. and {Riffeser}, A. and {Rix}, H.~W. and {R{\"o}ser}, S. and {Russel}, R. and {Rutz}, L. and {Schilbach}, E. and {Schultz}, A.~S.~B. and {Scolnic}, D. and {Strolger}, L. and {Szalay}, A. and {Seitz}, S. and {Small}, E. and {Smith}, K.~W. and {Soderblom}, D.~R. and {Taylor}, P. and {Thomson}, R. and {Taylor}, A.~N. and {Thakar}, A.~R. and {Thiel}, J. and {Thilker}, D. and {Unger}, D. and {Urata}, Y. and {Valenti}, J. and {Wagner}, J. and {Walder}, T. and {Walter}, F. and {Watters}, S.~P. and {Werner}, S. and {Wood-Vasey}, W.~M. and {Wyse}, R.},
        title = "{The Pan-STARRS1 Surveys}",
      journal = {arXiv e-prints},
         year = 2016,
        month = dec,
          eid = {arXiv:1612.05560},
        pages = {arXiv:1612.05560},
          doi = {10.48550/arXiv.1612.05560},
archivePrefix = {arXiv},
       eprint = {1612.05560},
 primaryClass = {astro-ph.IM},
       adsurl = {https://ui.adsabs.harvard.edu/abs/2016arXiv161205560C}
}

@ARTICLE{chiba_2021,
       author = {{Chiba}, Rimpei and {Sch{\"o}nrich}, Ralph},
        title = "{Tree-ring structure of Galactic bar resonance}",
      journal = {MNRAS},
         year = 2021,
        month = aug,
       volume = {505},
       number = {2},
        pages = {2412-2426},
          doi = {10.1093/mnras/stab1094},
archivePrefix = {arXiv},
       eprint = {2102.08388},
 primaryClass = {astro-ph.GA},
       adsurl = {https://ui.adsabs.harvard.edu/abs/2021MNRAS.505.2412C}
}

@ARTICLE{conselice_2003,
       author = {{Conselice}, Christopher J. and {Bershady}, Matthew A. and {Dickinson}, Mark and {Papovich}, Casey},
        title = "{A Direct Measurement of Major Galaxy Mergers at z<\raisebox{-0.5ex}\textasciitilde3}",
      journal = {\aj},
         year = 2003,
        month = sep,
       volume = {126},
       number = {3},
        pages = {1183-1207},
          doi = {10.1086/377318},
archivePrefix = {arXiv},
       eprint = {astro-ph/0306106},
 primaryClass = {astro-ph},
       adsurl = {https://ui.adsabs.harvard.edu/abs/2003AJ....126.1183C}
}

@ARTICLE{le_conte_2026,
       author = {{Le Conte}, Zoe A. and {Gadotti}, Dimitri A. and {Ferreira}, Leonardo and {Conselice}, Christopher J. and {de S{\'a}-Freitas}, Camila and {Kim}, Taehyun and {Neumann}, Justus and {Fragkoudi}, Francesca and {Athanassoula}, E. and {Adams}, Nathan J.},
        title = "{The evolution of the bar fraction and bar lengths in the last 12 billion years}",
      journal = {\mnras},
         year = 2026,
        month = jan,
       volume = {545},
       number = {1},
          eid = {staf2010},
        pages = {staf2010},
          doi = {10.1093/mnras/staf2010},
archivePrefix = {arXiv},
       eprint = {2510.07407},
 primaryClass = {astro-ph.GA},
       adsurl = {https://ui.adsabs.harvard.edu/abs/2026MNRAS.545f2010L}
}

@ARTICLE{cooper_2023,
       author = {{Cooper}, Andrew P. and {Koposov}, Sergey E. and {Allende Prieto}, Carlos and {Manser}, Christopher J. and {Kizhuprakkat}, Namitha and {Myers}, Adam D. and {Dey}, Arjun and {G{\"a}nsicke}, Boris T. and {Li}, Ting S. and {Rockosi}, Constance and {Valluri}, Monica and {Najita}, Joan and {Deason}, Alis and {Raichoor}, Anand and {Wang}, M. -Y. and {Ting}, Y. -S. and {Kim}, Bokyoung and {Carrillo}, Andreia and {Wang}, Wenting and {Beraldo e Silva}, Leandro and {Han}, Jiwon Jesse and {Ding}, Jiani and {S{\'a}nchez-Conde}, Miguel and {Aguilar}, Jessica N. and {Ahlen}, Steven and {Bailey}, Stephen and {Belokurov}, Vasily and {Brooks}, David and {Cunha}, Katia and {Dawson}, Kyle and {de la Macorra}, Axel and {Doel}, Peter and {Eisenstein}, Daniel J. and {Fagrelius}, Parker and {Fanning}, Kevin and {Font-Ribera}, Andreu and {Forero-Romero}, Jaime E. and {Gazta{\~n}aga}, Enrique and {Gontcho a Gontcho}, Satya and {Guy}, Julien and {Honscheid}, Klaus and {Kehoe}, Robert and {Kisner}, Theodore and {Kremin}, Anthony and {Landriau}, Martin and {Levi}, Michael E. and {Martini}, Paul and {Meisner}, Aaron M. and {Miquel}, Ramon and {Moustakas}, John and {Nie}, Jundan J.~D. and {Palanque-Delabrouille}, Nathalie and {Percival}, Will J. and {Poppett}, Claire and {Prada}, Francisco and {Rehemtulla}, Nabeel and {Schlafly}, Edward and {Schlegel}, David and {Schubnell}, Michael and {Sharples}, Ray M. and {Tarl{\'e}}, Gregory and {Wechsler}, Risa H. and {Weinberg}, David H. and {Zhou}, Zhimin and {Zou}, Hu},
        title = "{Overview of the DESI Milky Way Survey}",
      journal = {ApJ},
         year = 2023,
        month = apr,
       volume = {947},
       number = {1},
          eid = {37},
        pages = {37},
          doi = {10.3847/1538-4357/acb3c0},
archivePrefix = {arXiv},
       eprint = {2208.08514},
 primaryClass = {astro-ph.GA},
       adsurl = {https://ui.adsabs.harvard.edu/abs/2023ApJ...947...37C}
}

@ARTICLE{dalmasso_2024,
       author = {{Dalmasso}, Nicol{\`o} and {Calabr{\`o}}, Antonello and {Leethochawalit}, Nicha and {Vulcani}, Benedetta and {Boyett}, Kristan and {Trenti}, Michele and {Treu}, Tommaso and {Castellano}, Marco and {Brada{\v{c}}}, Maru{\v{s}}a and {Metha}, Benjamin and {Santini}, Paola},
        title = "{The rate and contribution of mergers to mass assembly from NIRCam observations of galaxy candidates up to 13.3 billion years ago}",
      journal = {\mnras},
         year = 2024,
        month = oct,
       volume = {533},
       number = {4},
        pages = {4472-4484},
          doi = {10.1093/mnras/stae2064},
archivePrefix = {arXiv},
       eprint = {2403.11428},
 primaryClass = {astro-ph.GA},
       adsurl = {https://ui.adsabs.harvard.edu/abs/2024MNRAS.533.4472D}
}

@ARTICLE{davis_1985,
       author = {{Davis}, M. and {Efstathiou}, G. and {Frenk}, C.~S. and {White}, S.~D.~M.},
        title = "{The evolution of large-scale structure in a universe dominated by cold dark matter}",
      journal = {ApJ},
         year = 1985,
        month = may,
       volume = {292},
        pages = {371-394},
          doi = {10.1086/163168},
       adsurl = {https://ui.adsabs.harvard.edu/abs/1985ApJ...292..371D}
}

@article{dillamore_2023,
	title = {Stellar halo substructure generated by bar resonances},
	volume = {524},
	doi = {10.1093/mnras/stad2136},
	number = {3},
	journal = {MNRAS},
	author = {Dillamore, Adam M. and Belokurov, Vasily and Evans, N. Wyn and Davies, Elliot Y.},
	month = sep,
	year = {2023},
	note = {\_eprint: 2303.00008},
	pages = {3596--3608},
}

@ARTICLE{duncan_2019,
       author = {{Duncan}, Kenneth and {Conselice}, Christopher J. and {Mundy}, Carl and {Bell}, Eric and {Donley}, Jennifer and {Galametz}, Audrey and {Guo}, Yicheng and {Grogin}, Norman A. and {Hathi}, Nimish and {Kartaltepe}, Jeyhan and {Kocevski}, Dale and {Koekemoer}, Anton M. and {P{\'e}rez-Gonz{\'a}lez}, Pablo G. and {Mantha}, Kameswara B. and {Snyder}, Gregory F. and {Stefanon}, Mauro},
        title = "{Observational Constraints on the Merger History of Galaxies since z {\ensuremath{\approx}} 6: Probabilistic Galaxy Pair Counts in the CANDELS Fields}",
      journal = {\apj},
         year = 2019,
        month = may,
       volume = {876},
       number = {2},
          eid = {110},
        pages = {110},
          doi = {10.3847/1538-4357/ab148a},
archivePrefix = {arXiv},
       eprint = {1903.12188},
 primaryClass = {astro-ph.GA},
       adsurl = {https://ui.adsabs.harvard.edu/abs/2019ApJ...876..110D}
}

@ARTICLE{fattahi_2019,
       author = {{Fattahi}, Azadeh and {Belokurov}, Vasily and {Deason}, Alis J. and {Frenk}, Carlos S. and {G{\'o}mez}, Facundo A. and {Grand}, Robert J.~J. and {Marinacci}, Federico and {Pakmor}, R{\"u}diger and {Springel}, Volker},
        title = "{The origin of galactic metal-rich stellar halo components with highly eccentric orbits}",
      journal = {MNRAS},
         year = 2019,
        month = apr,
       volume = {484},
       number = {4},
        pages = {4471-4483},
          doi = {10.1093/mnras/stz159},
archivePrefix = {arXiv},
       eprint = {1810.07779},
 primaryClass = {astro-ph.GA},
       adsurl = {https://ui.adsabs.harvard.edu/abs/2019MNRAS.484.4471F}
}

@ARTICLE{fragkoudi_2020,
       author = {{Fragkoudi}, F. and {Grand}, R.~J.~J. and {Pakmor}, R. and {Bl{\'a}zquez-Calero}, G. and {Gargiulo}, I. and {Gomez}, F. and {Marinacci}, F. and {Monachesi}, A. and {Ness}, M.~K. and {Perez}, I. and {Tissera}, P. and {White}, S.~D.~M.},
        title = "{Chemodynamics of barred galaxies in cosmological simulations: On the Milky Way's quiescent merger history and in-situ bulge}",
      journal = {MNRAS},
         year = 2020,
        month = jun,
       volume = {494},
       number = {4},
        pages = {5936-5960},
          doi = {10.1093/mnras/staa1104},
archivePrefix = {arXiv},
       eprint = {1911.06826},
 primaryClass = {astro-ph.GA},
       adsurl = {https://ui.adsabs.harvard.edu/abs/2020MNRAS.494.5936F}
}

@ARTICLE{fragkoudi_2021,
       author = {{Fragkoudi}, F. and {Grand}, R.~J.~J. and {Pakmor}, R. and {Springel}, V. and {White}, S.~D.~M. and {Marinacci}, F. and {Gomez}, F.~A. and {Navarro}, J.~F.},
        title = "{Revisiting the tension between fast bars and the {\ensuremath{\Lambda}}CDM paradigm}",
      journal = {\aap},
         year = 2021,
        month = jun,
       volume = {650},
          eid = {L16},
        pages = {L16},
          doi = {10.1051/0004-6361/202140320},
archivePrefix = {arXiv},
       eprint = {2011.13942},
 primaryClass = {astro-ph.GA},
       adsurl = {https://ui.adsabs.harvard.edu/abs/2021A&A...650L..16F}
}

@ARTICLE{gaia_2016,
       author = {{Gaia Collaboration} and {Prusti}, T. and {de Bruijne}, J.~H.~J. and {Brown}, A.~G.~A. and {Vallenari}, A. and {Babusiaux}, C. and {Bailer-Jones}, C.~A.~L. and {Bastian}, U. and {Biermann}, M. and {Evans}, D.~W. and {Eyer}, L. and {Jansen}, F. and {Jordi}, C. and {Klioner}, S.~A. and {Lammers}, U. and {Lindegren}, L. and {Luri}, X. and {Mignard}, F. and {Milligan}, D.~J. and {Panem}, C. and {Poinsignon}, V. and {Pourbaix}, D. and {Randich}, S. and {Sarri}, G. and {Sartoretti}, P. and {Siddiqui}, H.~I. and {Soubiran}, C. and {Valette}, V. and {van Leeuwen}, F. and {Walton}, N.~A. and {Aerts}, C. and {Arenou}, F. and {Cropper}, M. and {Drimmel}, R. and {H{\o}g}, E. and {Katz}, D. and {Lattanzi}, M.~G. and {O'Mullane}, W. and {Grebel}, E.~K. and {Holland}, A.~D. and {Huc}, C. and {Passot}, X. and {Bramante}, L. and {Cacciari}, C. and {Casta{\~n}eda}, J. and {Chaoul}, L. and {Cheek}, N. and {De Angeli}, F. and {Fabricius}, C. and {Guerra}, R. and {Hern{\'a}ndez}, J. and {Jean-Antoine-Piccolo}, A. and {Masana}, E. and {Messineo}, R. and {Mowlavi}, N. and {Nienartowicz}, K. and {Ord{\'o}{\~n}ez-Blanco}, D. and {Panuzzo}, P. and {Portell}, J. and {Richards}, P.~J. and {Riello}, M. and {Seabroke}, G.~M. and {Tanga}, P. and {Th{\'e}venin}, F. and {Torra}, J. and {Els}, S.~G. and {Gracia-Abril}, G. and {Comoretto}, G. and {Garcia-Reinaldos}, M. and {Lock}, T. and {Mercier}, E. and {Altmann}, M. and {Andrae}, R. and {Astraatmadja}, T.~L. and {Bellas-Velidis}, I. and {Benson}, K. and {Berthier}, J. and {Blomme}, R. and {Busso}, G. and {Carry}, B. and {Cellino}, A. and {Clementini}, G. and {Cowell}, S. and {Creevey}, O. and {Cuypers}, J. and {Davidson}, M. and {De Ridder}, J. and {de Torres}, A. and {Delchambre}, L. and {Dell'Oro}, A. and {Ducourant}, C. and {Fr{\'e}mat}, Y. and {Garc{\'\i}a-Torres}, M. and {Gosset}, E. and {Halbwachs}, J.-L. and {Hambly}, N.~C. and {Harrison}, D.~L. and {Hauser}, M. and {Hestroffer}, D. and {Hodgkin}, S.~T. and {Huckle}, H.~E. and {Hutton}, A. and {Jasniewicz}, G. and {Jordan}, S. and {Kontizas}, M. and {Korn}, A.~J. and {Lanzafame}, A.~C. and {Manteiga}, M. and {Moitinho}, A. and {Muinonen}, K. and {Osinde}, J. and {Pancino}, E. and {Pauwels}, T. and {Petit}, J.-M. and {Recio-Blanco}, A. and {Robin}, A.~C. and {Sarro}, L.~M. and {Siopis}, C. and {Smith}, M. and {Smith}, K.~W. and {Sozzetti}, A. and {Thuillot}, W. and {van Reeven}, W. and {Viala}, Y. and {Abbas}, U. and {Abreu Aramburu}, A. and {Accart}, S. and {Aguado}, J.~J. and {Allan}, P.~M. and {Allasia}, W. and {Altavilla}, G. and {{\'A}lvarez}, M.~A. and {Alves}, J. and {Anderson}, R.~I. and {Andrei}, A.~H. and {Anglada Varela}, E. and {Antiche}, E. and {Antoja}, T. and {Ant{\'o}n}, S. and {Arcay}, B. and {Atzei}, A. and {Ayache}, L. and {Bach}, N. and {Baker}, S.~G. and {Balaguer-N{\'u}{\~n}ez}, L. and {Barache}, C. and {Barata}, C. and {Barbier}, A. and {Barblan}, F. and {Baroni}, M. and {Barrado y Navascu{\'e}s}, D. and {Barros}, M. and {Barstow}, M.~A. and {Becciani}, U. and {Bellazzini}, M. and {Bellei}, G. and {Bello Garc{\'\i}a}, A. and {Belokurov}, V. and {Bendjoya}, P. and {Berihuete}, A. and {Bianchi}, L. and {Bienaym{\'e}}, O. and {Billebaud}, F. and {Blagorodnova}, N. and {Blanco-Cuaresma}, S. and {Boch}, T. and {Bombrun}, A. and {Borrachero}, R. and {Bouquillon}, S. and {Bourda}, G. and {Bouy}, H. and {Bragaglia}, A. and {Breddels}, M.~A. and {Brouillet}, N. and {Br{\"u}semeister}, T. and {Bucciarelli}, B. and {Budnik}, F. and {Burgess}, P. and {Burgon}, R. and {Burlacu}, A. and {Busonero}, D. and {Buzzi}, R. and {Caffau}, E. and {Cambras}, J. and {Campbell}, H. and {Cancelliere}, R. and {Cantat-Gaudin}, T. and {Carlucci}, T. and {Carrasco}, J.~M. and {Castellani}, M. and {Charlot}, P. and {Charnas}, J. and {Charvet}, P. and {Chassat}, F. and {Chiavassa}, A. and {Clotet}, M. and {Cocozza}, G. and {Collins}, R.~S. and {Collins}, P. and {Costigan}, G.},
        title = "{The Gaia mission}",
      journal = {A&A},
         year = 2016,
        month = nov,
       volume = {595},
          eid = {A1},
        pages = {A1},
          doi = {10.1051/0004-6361/201629272},
archivePrefix = {arXiv},
       eprint = {1609.04153},
 primaryClass = {astro-ph.IM},
       adsurl = {https://ui.adsabs.harvard.edu/abs/2016A&A...595A...1G}
}

@ARTICLE{gaia_2018,
       author = {{Gaia Collaboration} and {Brown}, A.~G.~A. and {Vallenari}, A. and {Prusti}, T. and {de Bruijne}, J.~H.~J. and {Babusiaux}, C. and {Bailer-Jones}, C.~A.~L. and {Biermann}, M. and {Evans}, D.~W. and {Eyer}, L. and {Jansen}, F. and {Jordi}, C. and {Klioner}, S.~A. and {Lammers}, U. and {Lindegren}, L. and {Luri}, X. and {Mignard}, F. and {Panem}, C. and {Pourbaix}, D. and {Randich}, S. and {Sartoretti}, P. and {Siddiqui}, H.~I. and {Soubiran}, C. and {van Leeuwen}, F. and {Walton}, N.~A. and {Arenou}, F. and {Bastian}, U. and {Cropper}, M. and {Drimmel}, R. and {Katz}, D. and {Lattanzi}, M.~G. and {Bakker}, J. and {Cacciari}, C. and {Casta{\~n}eda}, J. and {Chaoul}, L. and {Cheek}, N. and {De Angeli}, F. and {Fabricius}, C. and {Guerra}, R. and {Holl}, B. and {Masana}, E. and {Messineo}, R. and {Mowlavi}, N. and {Nienartowicz}, K. and {Panuzzo}, P. and {Portell}, J. and {Riello}, M. and {Seabroke}, G.~M. and {Tanga}, P. and {Th{\'e}venin}, F. and {Gracia-Abril}, G. and {Comoretto}, G. and {Garcia-Reinaldos}, M. and {Teyssier}, D. and {Altmann}, M. and {Andrae}, R. and {Audard}, M. and {Bellas-Velidis}, I. and {Benson}, K. and {Berthier}, J. and {Blomme}, R. and {Burgess}, P. and {Busso}, G. and {Carry}, B. and {Cellino}, A. and {Clementini}, G. and {Clotet}, M. and {Creevey}, O. and {Davidson}, M. and {De Ridder}, J. and {Delchambre}, L. and {Dell'Oro}, A. and {Ducourant}, C. and {Fern{\'a}ndez-Hern{\'a}ndez}, J. and {Fouesneau}, M. and {Fr{\'e}mat}, Y. and {Galluccio}, L. and {Garc{\'\i}a-Torres}, M. and {Gonz{\'a}lez-N{\'u}{\~n}ez}, J. and {Gonz{\'a}lez-Vidal}, J.~J. and {Gosset}, E. and {Guy}, L.~P. and {Halbwachs}, J.-L. and {Hambly}, N.~C. and {Harrison}, D.~L. and {Hern{\'a}ndez}, J. and {Hestroffer}, D. and {Hodgkin}, S.~T. and {Hutton}, A. and {Jasniewicz}, G. and {Jean-Antoine-Piccolo}, A. and {Jordan}, S. and {Korn}, A.~J. and {Krone-Martins}, A. and {Lanzafame}, A.~C. and {Lebzelter}, T. and {L{\"o}ffler}, W. and {Manteiga}, M. and {Marrese}, P.~M. and {Mart{\'\i}n-Fleitas}, J.~M. and {Moitinho}, A. and {Mora}, A. and {Muinonen}, K. and {Osinde}, J. and {Pancino}, E. and {Pauwels}, T. and {Petit}, J.-M. and {Recio-Blanco}, A. and {Richards}, P.~J. and {Rimoldini}, L. and {Robin}, A.~C. and {Sarro}, L.~M. and {Siopis}, C. and {Smith}, M. and {Sozzetti}, A. and {S{\"u}veges}, M. and {Torra}, J. and {van Reeven}, W. and {Abbas}, U. and {Abreu Aramburu}, A. and {Accart}, S. and {Aerts}, C. and {Altavilla}, G. and {{\'A}lvarez}, M.~A. and {Alvarez}, R. and {Alves}, J. and {Anderson}, R.~I. and {Andrei}, A.~H. and {Anglada Varela}, E. and {Antiche}, E. and {Antoja}, T. and {Arcay}, B. and {Astraatmadja}, T.~L. and {Bach}, N. and {Baker}, S.~G. and {Balaguer-N{\'u}{\~n}ez}, L. and {Balm}, P. and {Barache}, C. and {Barata}, C. and {Barbato}, D. and {Barblan}, F. and {Barklem}, P.~S. and {Barrado}, D. and {Barros}, M. and {Barstow}, M.~A. and {Bartholom{\'e} Mu{\~n}oz}, S. and {Bassilana}, J.-L. and {Becciani}, U. and {Bellazzini}, M. and {Berihuete}, A. and {Bertone}, S. and {Bianchi}, L. and {Bienaym{\'e}}, O. and {Blanco-Cuaresma}, S. and {Boch}, T. and {Boeche}, C. and {Bombrun}, A. and {Borrachero}, R. and {Bossini}, D. and {Bouquillon}, S. and {Bourda}, G. and {Bragaglia}, A. and {Bramante}, L. and {Breddels}, M.~A. and {Bressan}, A. and {Brouillet}, N. and {Br{\"u}semeister}, T. and {Brugaletta}, E. and {Bucciarelli}, B. and {Burlacu}, A. and {Busonero}, D. and {Butkevich}, A.~G. and {Buzzi}, R. and {Caffau}, E. and {Cancelliere}, R. and {Cannizzaro}, G. and {Cantat-Gaudin}, T. and {Carballo}, R. and {Carlucci}, T. and {Carrasco}, J.~M. and {Casamiquela}, L. and {Castellani}, M. and {Castro-Ginard}, A. and {Charlot}, P. and {Chemin}, L. and {Chiavassa}, A. and {Cocozza}, G. and {Costigan}, G. and {Cowell}, S. and {Crifo}, F. and {Crosta}, M. and {Crowley}, C. and {Cuypers}, J. and {Dafonte}, C. and {Damerdji}, Y. and {Dapergolas}, A. and {David}, P. and {David}, M. and {de Laverny}, P. and {De Luise}, F.},
        title = "{Gaia Data Release 2. Summary of the contents and survey properties}",
      journal = {\aap},
         year = 2018,
        month = aug,
       volume = {616},
          eid = {A1},
        pages = {A1},
          doi = {10.1051/0004-6361/201833051},
archivePrefix = {arXiv},
       eprint = {1804.09365},
 primaryClass = {astro-ph.GA},
       adsurl = {https://ui.adsabs.harvard.edu/abs/2018A&A...616A...1G}
}

@ARTICLE{gaia_2023,
       author = {{Gaia Collaboration} and {Vallenari}, A. and {Brown}, A.~G.~A. and {Prusti}, T. and {de Bruijne}, J.~H.~J. and {Arenou}, F. and {Babusiaux}, C. and {Biermann}, M. and {Creevey}, O.~L. and {Ducourant}, C. and {Evans}, D.~W. and {Eyer}, L. and {Guerra}, R. and {Hutton}, A. and {Jordi}, C. and {Klioner}, S.~A. and {Lammers}, U.~L. and {Lindegren}, L. and {Luri}, X. and {Mignard}, F. and {Panem}, C. and {Pourbaix}, D. and {Randich}, S. and {Sartoretti}, P. and {Soubiran}, C. and {Tanga}, P. and {Walton}, N.~A. and {Bailer-Jones}, C.~A.~L. and {Bastian}, U. and {Drimmel}, R. and {Jansen}, F. and {Katz}, D. and {Lattanzi}, M.~G. and {van Leeuwen}, F. and {Bakker}, J. and {Cacciari}, C. and {Casta{\~n}eda}, J. and {De Angeli}, F. and {Fabricius}, C. and {Fouesneau}, M. and {Fr{\'e}mat}, Y. and {Galluccio}, L. and {Guerrier}, A. and {Heiter}, U. and {Masana}, E. and {Messineo}, R. and {Mowlavi}, N. and {Nicolas}, C. and {Nienartowicz}, K. and {Pailler}, F. and {Panuzzo}, P. and {Riclet}, F. and {Roux}, W. and {Seabroke}, G.~M. and {Sordo}, R. and {Th{\'e}venin}, F. and {Gracia-Abril}, G. and {Portell}, J. and {Teyssier}, D. and {Altmann}, M. and {Andrae}, R. and {Audard}, M. and {Bellas-Velidis}, I. and {Benson}, K. and {Berthier}, J. and {Blomme}, R. and {Burgess}, P.~W. and {Busonero}, D. and {Busso}, G. and {C{\'a}novas}, H. and {Carry}, B. and {Cellino}, A. and {Cheek}, N. and {Clementini}, G. and {Damerdji}, Y. and {Davidson}, M. and {de Teodoro}, P. and {Nu{\~n}ez Campos}, M. and {Delchambre}, L. and {Dell'Oro}, A. and {Esquej}, P. and {Fern{\'a}ndez-Hern{\'a}ndez}, J. and {Fraile}, E. and {Garabato}, D. and {Garc{\'\i}a-Lario}, P. and {Gosset}, E. and {Haigron}, R. and {Halbwachs}, J.-L. and {Hambly}, N.~C. and {Harrison}, D.~L. and {Hern{\'a}ndez}, J. and {Hestroffer}, D. and {Hodgkin}, S.~T. and {Holl}, B. and {Jan{\ss}en}, K. and {Jevardat de Fombelle}, G. and {Jordan}, S. and {Krone-Martins}, A. and {Lanzafame}, A.~C. and {L{\"o}ffler}, W. and {Marchal}, O. and {Marrese}, P.~M. and {Moitinho}, A. and {Muinonen}, K. and {Osborne}, P. and {Pancino}, E. and {Pauwels}, T. and {Recio-Blanco}, A. and {Reyl{\'e}}, C. and {Riello}, M. and {Rimoldini}, L. and {Roegiers}, T. and {Rybizki}, J. and {Sarro}, L.~M. and {Siopis}, C. and {Smith}, M. and {Sozzetti}, A. and {Utrilla}, E. and {van Leeuwen}, M. and {Abbas}, U. and {{\'A}brah{\'a}m}, P. and {Abreu Aramburu}, A. and {Avirial radiuserts}, C. and {Aguado}, J.~J. and {Ajaj}, M. and {Aldea-Montero}, F. and {Altavilla}, G. and {{\'A}lvarez}, M.~A. and {Alves}, J. and {Anders}, F. and {Anderson}, R.~I. and {Anglada Varela}, E. and {Antoja}, T. and {Baines}, D. and {Baker}, S.~G. and {Balaguer-N{\'u}{\~n}ez}, L. and {Balbinot}, E. and {Balog}, Z. and {Barache}, C. and {Barbato}, D. and {Barros}, M. and {Barstow}, M.~A. and {Bartolom{\'e}}, S. and {Bassilana}, J.-L. and {Bauchet}, N. and {Becciani}, U. and {Bellazzini}, M. and {Berihuete}, A. and {Bernet}, M. and {Bertone}, S. and {Bianchi}, L. and {Binnenfeld}, A. and {Blanco-Cuaresma}, S. and {Blazere}, A. and {Boch}, T. and {Bombrun}, A. and {Bossini}, D. and {Bouquillon}, S. and {Bragaglia}, A. and {Bramante}, L. and {Breedt}, E. and {Bressan}, A. and {Brouillet}, N. and {Brugaletta}, E. and {Bucciarelli}, B. and {Burlacu}, A. and {Butkevich}, A.~G. and {Buzzi}, R. and {Caffau}, E. and {Cancelliere}, R. and {Cantat-Gaudin}, T. and {Carballo}, R. and {Carlucci}, T. and {Carnerero}, M.~I. and {Carrasco}, J.~M. and {Casamiquela}, L. and {Castellani}, M. and {Castro-Ginard}, A. and {Chaoul}, L. and {Charlot}, P. and {Chemin}, L. and {Chiaramida}, V. and {Chiavassa}, A. and {Chornay}, N. and {Comoretto}, G. and {Contursi}, G. and {Cooper}, W.~J. and {Cornez}, T. and {Cowell}, S. and {Crifo}, F. and {Cropper}, M. and {Crosta}, M. and {Crowley}, C. and {Dafonte}, C. and {Dapergolas}, A. and {David}, M. and {David}, P. and {de Laverny}, P. and {De Luise}, F. and {De March}, R.},
        title = "{Gaia Data Release 3. Summary of the content and survey properties}",
      journal = {A&A},
         year = 2023,
        month = jun,
       volume = {674},
          eid = {A1},
        pages = {A1},
          doi = {10.1051/0004-6361/202243940},
archivePrefix = {arXiv},
       eprint = {2208.00211},
 primaryClass = {astro-ph.GA},
       adsurl = {https://ui.adsabs.harvard.edu/abs/2023A&A...674A...1G}
}

@ARTICLE{gomez_2010,
       author = {{G{\'o}mez}, Facundo A. and {Helmi}, Amina},
        title = "{On the identification of substructure in phase space using orbital frequencies}",
      journal = {\mnras},
         year = 2010,
        month = feb,
       volume = {401},
       number = {4},
        pages = {2285-2298},
          doi = {10.1111/j.1365-2966.2009.15841.x},
archivePrefix = {arXiv},
       eprint = {0904.1377},
 primaryClass = {astro-ph.GA},
       adsurl = {https://ui.adsabs.harvard.edu/abs/2010MNRAS.401.2285G}
}

@ARTICLE{grand_2017,
       author = {{Grand}, Robert J.~J. and {G{\'o}mez}, Facundo A. and {Marinacci}, Federico and {Pakmor}, R{\"u}diger and {Springel}, Volker and {Campbell}, David J.~R. and {Frenk}, Carlos S. and {Jenkins}, Adrian and {White}, Simon D.~M.},
        title = "{The Auriga Project: the properties and formation mechanisms of disc galaxies across cosmic time}",
      journal = {MNRAS},
         year = 2017,
        month = may,
       volume = {467},
       number = {1},
        pages = {179-207},
          doi = {10.1093/mnras/stx071},
archivePrefix = {arXiv},
       eprint = {1610.01159},
 primaryClass = {astro-ph.GA},
       adsurl = {https://ui.adsabs.harvard.edu/abs/2017MNRAS.467..179G}
}

@ARTICLE{grand_2020,
       author = {{Grand}, Robert J.~J. and {Kawata}, Daisuke and {Belokurov}, Vasily and {Deason}, Alis J. and {Fattahi}, Azadeh and {Fragkoudi}, Francesca and {G{\'o}mez}, Facundo A. and {Marinacci}, Federico and {Pakmor}, R{\"u}diger},
        title = "{The dual origin of the Galactic thick disc and halo from the gas-rich Gaia-Enceladus Sausage merger}",
      journal = {\mnras},
         year = 2020,
        month = sep,
       volume = {497},
       number = {2},
        pages = {1603-1618},
          doi = {10.1093/mnras/staa2057},
archivePrefix = {arXiv},
       eprint = {2001.06009},
 primaryClass = {astro-ph.GA},
       adsurl = {https://ui.adsabs.harvard.edu/abs/2020MNRAS.497.1603G}
}

@ARTICLE{grand_2021,
       author = {{Grand}, Robert J.~J. and {Marinacci}, Federico and {Pakmor}, R{\"u}diger and {Simpson}, Christine M. and {Kelly}, Ashley J. and {G{\'o}mez}, Facundo A. and {Jenkins}, Adrian and {Springel}, Volker and {Frenk}, Carlos S. and {White}, Simon D.~M.},
        title = "{Determining the full satellite population of a Milky Way-mass halo in a highly resolved cosmological hydrodynamic simulation}",
      journal = {\mnras},
         year = 2021,
        month = nov,
       volume = {507},
       number = {4},
        pages = {4953-4967},
          doi = {10.1093/mnras/stab2492},
archivePrefix = {arXiv},
       eprint = {2105.04560},
 primaryClass = {astro-ph.GA},
       adsurl = {https://ui.adsabs.harvard.edu/abs/2021MNRAS.507.4953G}
}

@ARTICLE{grand_2023,
       author = {{Grand}, Robert J.~J. and {Pakmor}, R{\"u}diger and {Fragkoudi}, Francesca and {G{\'o}mez}, Facundo A. and {Trick}, Wilma and {Simpson}, Christine M. and {van de Voort}, Freeke and {Bieri}, Rebekka},
        title = "{An ever-present Gaia snail shell triggered by a dark matter wake}",
      journal = {MNRAS},
         year = 2023,
        month = sep,
       volume = {524},
       number = {1},
        pages = {801-816},
          doi = {10.1093/mnras/stad1969},
archivePrefix = {arXiv},
       eprint = {2211.08437},
 primaryClass = {astro-ph.GA},
       adsurl = {https://ui.adsabs.harvard.edu/abs/2023MNRAS.524..801G}
}

@ARTICLE{grand_2024,
       author = {{Grand}, Robert J.~J. and {Fragkoudi}, Francesca and {G{\'o}mez}, Facundo A. and {Jenkins}, Adrian and {Marinacci}, Federico and {Pakmor}, R{\"u}diger and {Springel}, Volker},
        title = "{Overview and public data release of the augmented Auriga Project: cosmological simulations of dwarf and Milky Way-mass galaxies}",
      journal = {MNRAS},
         year = 2024,
        month = aug,
       volume = {532},
       number = {2},
        pages = {1814-1831},
          doi = {10.1093/mnras/stae1598},
archivePrefix = {arXiv},
       eprint = {2401.08750},
 primaryClass = {astro-ph.GA},
       adsurl = {https://ui.adsabs.harvard.edu/abs/2024MNRAS.532.1814G}
}

@ARTICLE{guo_2025,
       author = {{Guo}, Yuchen and {Jogee}, Shardha and {Wise}, Eden and {Pritchett}, Keith and {McGrath}, Elizabeth J. and {Finkelstein}, Steven L. and {Iyer}, Kartheik G. and {Arrabal Haro}, Pablo and {Bagley}, Micaela B. and {Dickinson}, Mark and {Kartaltepe}, Jeyhan S. and {Koekemoer}, Anton M. and {Papovich}, Casey and {Pirzkal}, Nor and {Yung}, L.~Y. Aaron and {Backhaus}, Bren E. and {Bell}, Eric F. and {Bhatawdekar}, Rachana and {Cheng}, Yingjie and {Costantin}, Luca and {de la Vega}, Alexander and {Giavalisco}, Mauro and {Hathi}, Nimish P. and {Holwerda}, Benne W. and {Kurczynski}, Peter and {Lucas}, Ray A. and {Mobasher}, Bahram and {P{\'e}rez-Gonz{\'a}lez}, Pablo G. and {Pacucci}, Fabio},
        title = "{The Abundance and Properties of Barred Galaxies out to z {\ensuremath{\sim}} 4 Using JWST CEERS Data}",
      journal = {\apj},
         year = 2025,
        month = jun,
       volume = {985},
       number = {2},
          eid = {181},
        pages = {181},
          doi = {10.3847/1538-4357/adc8a7},
archivePrefix = {arXiv},
       eprint = {2409.06100},
 primaryClass = {astro-ph.GA},
       adsurl = {https://ui.adsabs.harvard.edu/abs/2025ApJ...985..181G}
}

@ARTICLE{haywood_2018,
       author = {{Haywood}, M. and {Di Matteo}, P. and {Lehnert}, M.~D. and {Snaith}, O. and {Khoperskov}, S. and {G{\'o}mez}, A.},
        title = "{In Disguise or Out of Reach: First Clues about In Situ and Accreted Stars in the Stellar Halo of the Milky Way from Gaia DR2}",
      journal = {ApJ},
         year = 2018,
        month = aug,
       volume = {863},
       number = {2},
          eid = {113},
        pages = {113},
          doi = {10.3847/1538-4357/aad235},
archivePrefix = {arXiv},
       eprint = {1805.02617},
 primaryClass = {astro-ph.GA},
       adsurl = {https://ui.adsabs.harvard.edu/abs/2018ApJ...863..113H}
}

@article{hawkins_2014,
	title = {On the relative ages of the {\textbackslash}ensuremath{\textbackslash}alpha-rich and {\textbackslash}ensuremath{\textbackslash}alpha-poor stellar populations in the {Galactic} halo},
	volume = {445},
	doi = {10.1093/mnras/stu1910},
	number = {3},
	journal = {MNRAS},
	author = {Hawkins, K. and Jofré, P. and Gilmore, G. and Masseron, T.},
	month = dec,
	year = {2014},
	note = {\_eprint: 1409.3431},
	pages = {2575--2588},
}

@article{hawkins_2015,
	title = {Using chemical tagging to redefine the interface of the {Galactic} disc and halo},
	volume = {453},
	doi = {10.1093/mnras/stv1586},
	number = {1},
	journal = {MNRAS},
	author = {Hawkins, K. and Jofré, P. and Masseron, T. and Gilmore, G.},
	month = oct,
	year = {2015},
	pages = {758--774},
}

@ARTICLE{helmi_2000,
       author = {{Helmi}, Amina and {de Zeeuw}, P. Tim},
        title = "{Mapping the substructure in the Galactic halo with the next generation of astrometric satellites}",
      journal = {MNRAS},
         year = 2000,
        month = dec,
       volume = {319},
       number = {3},
        pages = {657-665},
          doi = {10.1046/j.1365-8711.2000.03895.x},
archivePrefix = {arXiv},
       eprint = {astro-ph/0007166},
 primaryClass = {astro-ph},
       adsurl = {https://ui.adsabs.harvard.edu/abs/2000MNRAS.319..657H}
}

@ARTICLE{helmi_2018,
       author = {{Helmi}, Amina and {Babusiaux}, Carine and {Koppelman}, Helmer H. and {Massari}, Davide and {Veljanoski}, Jovan and {Brown}, Anthony G.~A.},
        title = "{The merger that led to the formation of the Milky Way's inner stellar halo and thick disk}",
      journal = {Nature},
         year = 2018,
        month = oct,
       volume = {563},
       number = {7729},
        pages = {85-88},
          doi = {10.1038/s41586-018-0625-x},
archivePrefix = {arXiv},
       eprint = {1806.06038},
 primaryClass = {astro-ph.GA},
       adsurl = {https://ui.adsabs.harvard.edu/abs/2018Natur.563...85H}
}

@ARTICLE{hilmi_2020,
       author = {{Hilmi}, T. and {Minchev}, I. and {Buck}, T. and {Martig}, M. and {Quillen}, A.~C. and {Monari}, G. and {Famaey}, B. and {de Jong}, R.~S. and {Laporte}, C.~F.~P. and {Read}, J. and {Sanders}, J.~L. and {Steinmetz}, M. and {Wegg}, C.},
        title = "{Fluctuations in galactic bar parameters due to bar-spiral interaction}",
      journal = {MNRAS},
         year = 2020,
        month = sep,
       volume = {497},
       number = {1},
        pages = {933-955},
          doi = {10.1093/mnras/staa1934},
archivePrefix = {arXiv},
       eprint = {2003.05457},
 primaryClass = {astro-ph.GA},
       adsurl = {https://ui.adsabs.harvard.edu/abs/2020MNRAS.497..933H}
}

@ARTICLE{horta_2021,
       author = {{Horta}, Danny and {Schiavon}, Ricardo P. and {Mackereth}, J. Ted and {Pfeffer}, Joel and {Mason}, Andrew C. and {Kisku}, Shobhit and {Fragkoudi}, Francesca and {Allende Prieto}, Carlos and {Cunha}, Katia and {Hasselquist}, Sten and {Holtzman}, Jon and {Majewski}, Steven R. and {Nataf}, David and {O'Connell}, Robert W. and {Schultheis}, Mathias and {Smith}, Verne V.},
        title = "{Evidence from APOGEE for the presence of a major building block of the halo buried in the inner Galaxy}",
      journal = {\mnras},
         year = 2021,
        month = jan,
       volume = {500},
       number = {1},
        pages = {1385-1403},
          doi = {10.1093/mnras/staa2987},
archivePrefix = {arXiv},
       eprint = {2007.10374},
 primaryClass = {astro-ph.GA},
       adsurl = {https://ui.adsabs.harvard.edu/abs/2021MNRAS.500.1385H}
}

@ARTICLE{horta_2024,
       author = {{Horta}, Danny and {Cunningham}, Emily C. and {Sanderson}, Robyn and {Johnston}, Kathryn V. and {Deason}, Alis and {Wetzel}, Andrew and {McCluskey}, Fiona and {Garavito-Camargo}, Nicol{\'a}s and {Necib}, Lina and {Faucher-Gigu{\`e}re}, Claude-Andr{\'e} and {Arora}, Arpit and {Gandhi}, Pratik J.},
        title = "{The proto-galaxy of Milky Way-mass haloes in the FIRE simulations}",
      journal = {\mnras},
         year = 2024,
        month = feb,
       volume = {527},
       number = {4},
        pages = {9810-9825},
          doi = {10.1093/mnras/stad3834},
archivePrefix = {arXiv},
       eprint = {2307.15741},
 primaryClass = {astro-ph.GA},
       adsurl = {https://ui.adsabs.harvard.edu/abs/2024MNRAS.527.9810H}
}

@ARTICLE{horta_2025,
       author = {{Horta}, Danny and {Petersen}, Michael S. and {Pe{\~n}arrubia}, Jorge},
        title = "{Disentangling the Galaxy's Gordian knot: evidence from APOGEE{\textendash}Gaia for a knotted and slower bar in the Milky Way}",
      journal = {MNRAS},
         year = 2025,
        month = apr,
       volume = {538},
       number = {2},
        pages = {998-1018},
          doi = {10.1093/mnras/staf348},
archivePrefix = {arXiv},
       eprint = {2402.07986},
 primaryClass = {astro-ph.GA},
       adsurl = {https://ui.adsabs.harvard.edu/abs/2025MNRAS.538..998H}
}

@ARTICLE{hunt_2025,
       author = {{Hunt}, Jason A.~S. and {Vasiliev}, Eugene},
        title = "{Milky Way dynamics in light of Gaia}",
      journal = {\nar},
         year = 2025,
        month = jun,
       volume = {100},
          eid = {101721},
        pages = {101721},
          doi = {10.1016/j.newar.2024.101721},
archivePrefix = {arXiv},
       eprint = {2501.04075},
 primaryClass = {astro-ph.GA},
       adsurl = {https://ui.adsabs.harvard.edu/abs/2025NewAR.10001721H}
}

@article{jean_baptiste_2017,
	title = {On the kinematic detection of accreted streams in the {Gaia} era: a cautionary tale},
	volume = {604},
	url = {https://doi.org/10.1051%2F0004-6361%2F201629691},
	doi = {10.1051/0004-6361/201629691},
	journal = {A\&A},
	author = {Jean-Baptiste, I. and Di Matteo, P. and Haywood, M. and Gómez, A. and Montuori, M. and Combes, F. and Semelin, B.},
	month = aug,
	year = {2017},
	pages = {A106},
}

@ARTICLE{de_jong_2019,
       author = {{de Jong}, R.~S. and {Agertz}, O. and {Berbel}, A.~A. and {Aird}, J. and {Alexander}, D.~A. and {Amarsi}, A. and {Anders}, F. and {Andrae}, R. and {Ansarinejad}, B. and {Ansorge}, W. and {Antilogus}, P. and {Anwand-Heerwart}, H. and {Arentsen}, A. and {Arnadottir}, A. and {Asplund}, M. and {Auger}, M. and {Azais}, N. and {Baade}, D. and {Baker}, G. and {Baker}, S. and {Balbinot}, E. and {Baldry}, I.~K. and {Banerji}, M. and {Barden}, S. and {Barklem}, P. and {Barth{\'e}l{\'e}my-Mazot}, E. and {Battistini}, C. and {Bauer}, S. and {Bell}, C.~P.~M. and {Bellido-Tirado}, O. and {Bellstedt}, S. and {Belokurov}, V. and {Bensby}, T. and {Bergemann}, M. and {Bestenlehner}, J.~M. and {Bielby}, R. and {Bilicki}, M. and {Blake}, C. and {Bland-Hawthorn}, J. and {Boeche}, C. and {Boland}, W. and {Boller}, T. and {Bongard}, S. and {Bongiorno}, A. and {Bonifacio}, P. and {Boudon}, D. and {Brooks}, D. and {Brown}, M.~J.~I. and {Brown}, R. and {Br{\"u}ggen}, M. and {Brynnel}, J. and {Brzeski}, J. and {Buchert}, T. and {Buschkamp}, P. and {Caffau}, E. and {Caillier}, P. and {Carrick}, J. and {Casagrande}, L. and {Case}, S. and {Casey}, A. and {Cesarini}, I. and {Cescutti}, G. and {Chapuis}, D. and {Chiappini}, C. and {Childress}, M. and {Christlieb}, N. and {Church}, R. and {Cioni}, M. -R.~L. and {Cluver}, M. and {Colless}, M. and {Collett}, T. and {Comparat}, J. and {Cooper}, A. and {Couch}, W. and {Courbin}, F. and {Croom}, S. and {Croton}, D. and {Daguis{\'e}}, E. and {Dalton}, G. and {Davies}, L.~J.~M. and {Davis}, T. and {de Laverny}, P. and {Deason}, A. and {Dionies}, F. and {Disseau}, K. and {Doel}, P. and {D{\"o}scher}, D. and {Driver}, S.~P. and {Dwelly}, T. and {Eckert}, D. and {Edge}, A. and {Edvardsson}, B. and {Youssoufi}, D.~E. and {Elhaddad}, A. and {Enke}, H. and {Erfanianfar}, G. and {Farrell}, T. and {Fechner}, T. and {Feiz}, C. and {Feltzing}, S. and {Ferreras}, I. and {Feuerstein}, D. and {Feuillet}, D. and {Finoguenov}, A. and {Ford}, D. and {Fotopoulou}, S. and {Fouesneau}, M. and {Frenk}, C. and {Frey}, S. and {Gaessler}, W. and {Geier}, S. and {Gentile Fusillo}, N. and {Gerhard}, O. and {Giannantonio}, T. and {Giannone}, D. and {Gibson}, B. and {Gillingham}, P. and {Gonz{\'a}lez-Fern{\'a}ndez}, C. and {Gonzalez-Solares}, E. and {Gottloeber}, S. and {Gould}, A. and {Grebel}, E.~K. and {Gueguen}, A. and {Guiglion}, G. and {Haehnelt}, M. and {Hahn}, T. and {Hansen}, C.~J. and {Hartman}, H. and {Hauptner}, K. and {Hawkins}, K. and {Haynes}, D. and {Haynes}, R. and {Heiter}, U. and {Helmi}, A. and {Aguayo}, C.~H. and {Hewett}, P. and {Hinton}, S. and {Hobbs}, D. and {Hoenig}, S. and {Hofman}, D. and {Hook}, I. and {Hopgood}, J. and {Hopkins}, A. and {Hourihane}, A. and {Howes}, L. and {Howlett}, C. and {Huet}, T. and {Irwin}, M. and {Iwert}, O. and {Jablonka}, P. and {Jahn}, T. and {Jahnke}, K. and {Jarno}, A. and {Jin}, S. and {Jofre}, P. and {Johl}, D. and {Jones}, D. and {J{\"o}nsson}, H. and {Jordan}, C. and {Karovicova}, I. and {Khalatyan}, A. and {Kelz}, A. and {Kennicutt}, R. and {King}, D. and {Kitaura}, F. and {Klar}, J. and {Klauser}, U. and {Kneib}, J. -P. and {Koch}, A. and {Koposov}, S. and {Kordopatis}, G. and {Korn}, A. and {Kosmalski}, J. and {Kotak}, R. and {Kovalev}, M. and {Kreckel}, K. and {Kripak}, Y. and {Krumpe}, M. and {Kuijken}, K. and {Kunder}, A. and {Kushniruk}, I. and {Lam}, M.~I. and {Lamer}, G. and {Laurent}, F. and {Lawrence}, J. and {Lehmitz}, M. and {Lemasle}, B. and {Lewis}, J. and {Li}, B. and {Lidman}, C. and {Lind}, K. and {Liske}, J. and {Lizon}, J. -L. and {Loveday}, J. and {Ludwig}, H. -G. and {McDermid}, R.~M. and {Maguire}, K. and {Mainieri}, V. and {Mali}, S. and {Mandel}, H.},
        title = "{4MOST: Project overview and information for the First Call for Proposals}",
      journal = {The Messenger},
         year = 2019,
        month = mar,
       volume = {175},
        pages = {3-11},
          doi = {10.18727/0722-6691/5117},
archivePrefix = {arXiv},
       eprint = {1903.02464},
 primaryClass = {astro-ph.IM},
       adsurl = {https://ui.adsabs.harvard.edu/abs/2019Msngr.175....3D}
}

@ARTICLE{kim_2009,
       author = {{Kim}, Ji-hoon and {Wise}, John H. and {Abel}, Tom},
        title = "{Galaxy Mergers with Adaptive Mesh Refinement: Star Formation and Hot Gas Outflow}",
      journal = {\apjl},
         year = 2009,
        month = apr,
       volume = {694},
       number = {2},
        pages = {L123-L127},
          doi = {10.1088/0004-637X/694/2/L123},
archivePrefix = {arXiv},
       eprint = {0902.3001},
 primaryClass = {astro-ph.GA},
       adsurl = {https://ui.adsabs.harvard.edu/abs/2009ApJ...694L.123K}
}

@ARTICLE{knapen_2015,
       author = {{Knapen}, Johan H. and {Cisternas}, Mauricio and {Querejeta}, Miguel},
        title = "{Interacting galaxies in the nearby Universe: only moderate increase of star formation}",
      journal = {\mnras},
         year = 2015,
        month = dec,
       volume = {454},
       number = {2},
        pages = {1742-1750},
          doi = {10.1093/mnras/stv2135},
archivePrefix = {arXiv},
       eprint = {1509.05164},
 primaryClass = {astro-ph.GA},
       adsurl = {https://ui.adsabs.harvard.edu/abs/2015MNRAS.454.1742K}
}

@ARTICLE{kollmeier_2026,
       author = {{Kollmeier}, Juna A. and {Rix}, Hans-Walter and {Aerts}, Conny and {Aird}, James and {Vera Alfaro}, Pablo and {Almeida}, Andr{\'e}s and {Anderson}, Scott F. and {Arseneau}, Stefan M. and {Assef}, Roberto J. and {Aviram}, Shir and {Aydar}, Catarina and {Badenes}, Carles and {Bandyopadhyay}, Avrajit and {Barger}, Kat and {Barkhouser}, Robert H. and {Bauer}, Franz E. and {Behmard}, Aida and {Bender}, Chad and {Besser}, Felipe and {Bhattarai}, Binod and {Bilgi}, Pavaman and {Bird}, Jonathan and {Bizyaev}, Dmitry and {Blanc}, Guillermo A. and {Blanton}, Michael R. and {Bochanski}, John and {Bovy}, Jo and {Brandon}, Christopher and {Brandt}, William Nielsen and {Brownstein}, Joel R. and {Buchner}, Johannes and {Burchett}, Joseph N. and {Carlberg}, Joleen and {Casey}, Andrew R. and {Castaneda-Carlos}, Lesly and {Chakraborty}, Priyanka and {Chanam{\'e}}, Julio and {Chandra}, Vedant and {Cherinka}, Brian and {Chilingarian}, Igor and {Comparat}, Johan and {Cosens}, Maren and {Covey}, Kevin and {Crane}, Jeffrey D. and {Crumpler}, Nicole R. and {Cruz-Gonzalez}, Irene and {Cunha}, Katia and {Cunningham}, Tim and {Dai}, Xinyu and {Darling}, Jeremy and {Davidson}, Jr., James W. and {Davis}, Megan C. and {De Lee}, Nathan and {Deacon}, Niall and {M{\'e}ndez Delgado}, Jos{\'e} Eduardo and {Demasi}, Sebastian and {Demianenko}, Mariia and {Derwent}, Mark and {D'Onghia}, Elena and {Di Mille}, Francesco and {Dias}, Bruno and {Donor}, John and {Dow}, Peter N. and {Drory}, Niv and {Dwelly}, Tom and {Egorov}, Oleg and {Egorova}, Evgeniya and {El-Badry}, Kareem and {Engelman}, Mike and {Eracleous}, Mike and {Fan}, Xiaohui and {Farr}, Emily and {Fries}, Logan and {Frinchaboy}, Peter and {Froning}, Cynthia S. and {G{\"a}nsicke}, Boris T. and {Garc{\'\i}a}, Pablo and {Gelfand}, Joseph and {Gentile Fusillo}, Nicola Pietro and {Glover}, Simon and {Grabowski}, Katie and {Grebel}, Eva K. and {Green}, Paul J. and {Grier}, Catherine and {Gupta}, Pramod and {Gray}, Aidan C. and {H{\"a}berle}, Maximilian and {Hall}, Patrick B. and {Hammond}, Randolph P. and {Hawkins}, Keith and {Harding}, Albert C. and {Heged{\H{u}}s}, Viola and {Herbst}, Tom and {Hermes}, J.~J. and {Rodr{\'\i}guez Hidalgo}, Paola and {Hilder}, Thomas and {Hogg}, David W. and {Holtzman}, Jon A. and {Horta}, Danny and {Huang}, Yang and {Hwang}, Hsiang-Chih and {Ibarra-Medel}, Hector Javier and {Imig}, Julie and {Inight}, Keith and {Jana}, Arghajit and {Ji}, Alexander P. and {Jim{\'e}nez-Arranz}, {\'O}scar and {Jofre}, Paula and {Johns}, Matt and {Johnson}, Jennifer and {Johnson}, James W. and {Johnston}, Evelyn J. and {Jones}, Amy M. and {Katkov}, Ivan and {Knapp}, Gillian R. and {Koekemoer}, Anton M. and {Kounkel}, Marina and {Kreckel}, Kathryn and {Krishnarao}, Dhanesh and {Krumpe}, Mirko and {Kumari}, Nimisha and {Kupfer}, Thomas and {Lacerna}, Ivan and {Laporte}, Chervin and {Lepine}, Sebastien and {Li}, Jing and {Liu}, Xin and {Loebman}, Sarah and {Long}, Knox and {Roman-Lopes}, Alexandre and {Lu}, Yuxi and {Majewski}, Steven Raymond and {Maoz}, Dan and {McKinnon}, Kevin A. and {Medan}, Ilija and {Merloni}, Andrea and {Minniti}, Dante and {Morrison}, Sean and {Myers}, Natalie and {M{\'e}sz{\'a}ros}, Szabolcs and {Nandra}, Kirpal and {Nayak}, Prasanta K. and {Ness}, Melissa K. and {Nidever}, David L. and {O'Brien}, Thomas and {Oeur}, Micah and {Oravetz}, Audrey and {Oravetz}, Daniel and {Otto}, Jonah and {Pallathadka}, Gautham Adamane and {Palunas}, Povilas and {Pan}, Kaike and {Pappalardo}, Daniel and {Pandey}, Rakesh and {Negrete Pe{\~n}aloza}, Castalia Alenka and {Pinsonneault}, Marc H. and {Pogge}, Richard W. and {Taghizadeh Popp}, Manuchehr and {Price-Whelan}, Adrian M. and {Pulatova}, Nadiia and {Qiu}, Dan and {Ramirez}, Solange and {Rankine}, Amy and {Ricci}, Claudio and {Runnoe}, Jessie C. and {Sanchez}, Sebastian and {Salvato}, Mara and {Sarbadhicary}, Sumit K. and {Sattler}, Natascha and {Saydjari}, Andrew K. and {Sayres}, Conor and {Schinnerer}, Eva and {Schlaufman}, Kevin C. and {Schneider}, Donald P. and {Schreiber}, Matthias R. and {Schwope}, Axel and {Serna}, Javier and {Shen}, Yue and {Sif{\'o}n}, Crist{\'o}bal and {Singh}, Amrita and {Sinha}, Amaya and {Smee}, Stephen and {Song}, Ying-Yi and {Souto}, Diogo and {Stassun}, Keivan G. and {Steinmetz}, Matthias and {Stone-Martinez}, Alexander and {Stringfellow}, Guy and {Stutz}, Amelia and {S{\'a}nchez-Gallego}, Jos{\'e} and {Tan}, Jonathan C. and {Tayar}, Jamie and {Thai}, Riley and {Thakar}, Ani and {Ting}, Yuan-Sen and {Tkachenko}, Andrew and {Tovmassian}, Gagik and {Trakhtenbrot}, Benny and {Fern{\'a}ndez-Trincado}, Jos{\'e} G. and {Troup}, Nicholas},
        title = "{Sloan Digital Sky Survey. V. Pioneering Panoptic Spectroscopy}",
      journal = {\aj},
         year = 2026,
        month = jan,
       volume = {171},
       number = {1},
          eid = {52},
        pages = {52},
          doi = {10.3847/1538-3881/ae0576},
archivePrefix = {arXiv},
       eprint = {2507.06989},
 primaryClass = {astro-ph.IM},
       adsurl = {https://ui.adsabs.harvard.edu/abs/2026AJ....171...52K}
}

@ARTICLE{koppelman_2019,
       author = {{Koppelman}, Helmer H. and {Helmi}, Amina and {Massari}, Davide and {Price-Whelan}, Adrian M. and {Starkenburg}, Tjitske K.},
        title = "{Multiple retrograde substructures in the Galactic halo: A shattered view of Galactic history}",
      journal = {A&A},
         year = 2019,
        month = nov,
       volume = {631},
          eid = {L9},
        pages = {L9},
          doi = {10.1051/0004-6361/201936738},
archivePrefix = {arXiv},
       eprint = {1909.08924},
 primaryClass = {astro-ph.GA},
       adsurl = {https://ui.adsabs.harvard.edu/abs/2019A&A...631L...9K}
}

@ARTICLE{kruijssen_2020,
       author = {{Kruijssen}, J.~M. Diederik and {Pfeffer}, Joel L. and {Chevance}, M{\'e}lanie and {Bonaca}, Ana and {Trujillo-Gomez}, Sebastian and {Bastian}, Nate and {Reina-Campos}, Marta and {Crain}, Robert A. and {Hughes}, Meghan E.},
        title = "{Kraken reveals itself - the merger history of the Milky Way reconstructed with the E-MOSAICS simulations}",
      journal = {MNRAS},
         year = 2020,
        month = oct,
       volume = {498},
       number = {2},
        pages = {2472-2491},
          doi = {10.1093/mnras/staa2452},
archivePrefix = {arXiv},
       eprint = {2003.01119},
 primaryClass = {astro-ph.GA},
       adsurl = {https://ui.adsabs.harvard.edu/abs/2020MNRAS.498.2472K}
}

@ARTICLE{lucey_2019,
       author = {{Lucey}, Madeline and {Hawkins}, Keith and {Ness}, Melissa and {Asplund}, Martin and {Bensby}, Thomas and {Casagrande}, Luca and {Feltzing}, Sofia and {Freeman}, Kenneth C. and {Kobayashi}, Chiaki and {Marino}, Anna F.},
        title = "{The COMBS survey - I. Chemical origins of metal-poor stars in the Galactic bulge}",
      journal = {\mnras},
         year = 2019,
        month = sep,
       volume = {488},
       number = {2},
        pages = {2283-2300},
          doi = {10.1093/mnras/stz1847},
archivePrefix = {arXiv},
       eprint = {1903.11615},
 primaryClass = {astro-ph.SR},
       adsurl = {https://ui.adsabs.harvard.edu/abs/2019MNRAS.488.2283L}
}

@ARTICLE{lucey_2021,
       author = {{Lucey}, Madeline and {Hawkins}, Keith and {Ness}, Melissa and {Debattista}, Victor P. and {Luna}, Alice and {Asplund}, Martin and {Bensby}, Thomas and {Casagrande}, Luca and {Feltzing}, Sofia and {Freeman}, Kenneth C. and {Kobayashi}, Chiaki and {Marino}, Anna F.},
        title = "{The COMBS Survey - II. Distinguishing the metal-poor bulge from the halo interlopers}",
      journal = {\mnras},
         year = 2021,
        month = mar,
       volume = {501},
       number = {4},
        pages = {5981-5996},
          doi = {10.1093/mnras/stab003},
archivePrefix = {arXiv},
       eprint = {2009.03886},
 primaryClass = {astro-ph.GA},
       adsurl = {https://ui.adsabs.harvard.edu/abs/2021MNRAS.501.5981L}
}

@ARTICLE{lucey_2022,
       author = {{Lucey}, Madeline and {Hawkins}, Keith and {Ness}, Melissa and {Nelson}, Tyler and {Debattista}, Victor P. and {Luna}, Alice and {Bensby}, Thomas and {Freeman}, Kenneth C. and {Kobayashi}, Chiaki},
        title = "{The COMBS Survey - III. The chemodynamical origins of metal-poor bulge stars}",
      journal = {\mnras},
         year = 2022,
        month = jan,
       volume = {509},
       number = {1},
        pages = {122-144},
          doi = {10.1093/mnras/stab2878},
archivePrefix = {arXiv},
       eprint = {2107.02793},
 primaryClass = {astro-ph.GA},
       adsurl = {https://ui.adsabs.harvard.edu/abs/2022MNRAS.509..122L}
}

@ARTICLE{lucey_2023,
       author = {{Lucey}, Madeline and {Pearson}, Sarah and {Hunt}, Jason A.~S. and {Hawkins}, Keith and {Ness}, Melissa and {Petersen}, Michael S. and {Price-Whelan}, Adrian M. and {Weinberg}, Martin D.},
        title = "{Dynamically constraining the length of the Milky way bar}",
      journal = {MNRAS},
         year = 2023,
        month = apr,
       volume = {520},
       number = {3},
        pages = {4779-4792},
          doi = {10.1093/mnras/stad406},
archivePrefix = {arXiv},
       eprint = {2206.01798},
 primaryClass = {astro-ph.GA},
       adsurl = {https://ui.adsabs.harvard.edu/abs/2023MNRAS.520.4779L}
}

@ARTICLE{majewski_2017,
       author = {{Majewski}, Steven R. and {Schiavon}, Ricardo P. and {Frinchaboy}, Peter M. and {Allende Prieto}, Carlos and {Barkhouser}, Robert and {Bizyaev}, Dmitry and {Blank}, Basil and {Brunner}, Sophia and {Burton}, Adam and {Carrera}, Ricardo and {Chojnowski}, S. Drew and {Cunha}, K{\'a}tia and {Epstein}, Courtney and {Fitzgerald}, Greg and {Garc{\'\i}a P{\'e}rez}, Ana E. and {Hearty}, Fred R. and {Henderson}, Chuck and {Holtzman}, Jon A. and {Johnson}, Jennifer A. and {Lam}, Charles R. and {Lawler}, James E. and {Maseman}, Paul and {M{\'e}sz{\'a}ros}, Szabolcs and {Nelson}, Matthew and {Nguyen}, Duy Coung and {Nidever}, David L. and {Pinsonneault}, Marc and {Shetrone}, Matthew and {Smee}, Stephen and {Smith}, Verne V. and {Stolberg}, Todd and {Skrutskie}, Michael F. and {Walker}, Eric and {Wilson}, John C. and {Zasowski}, Gail and {Anders}, Friedrich and {Basu}, Sarbani and {Beland}, Stephane and {Blanton}, Michael R. and {Bovy}, Jo and {Brownstein}, Joel R. and {Carlberg}, Joleen and {Chaplin}, William and {Chiappini}, Cristina and {Eisenstein}, Daniel J. and {Elsworth}, Yvonne and {Feuillet}, Diane and {Fleming}, Scott W. and {Galbraith-Frew}, Jessica and {Garc{\'\i}a}, Rafael A. and {Garc{\'\i}a-Hern{\'a}ndez}, D. An{\'\i}bal and {Gillespie}, Bruce A. and {Girardi}, L{\'e}o and {Gunn}, James E. and {Hasselquist}, Sten and {Hayden}, Michael R. and {Hekker}, Saskia and {Ivans}, Inese and {Kinemuchi}, Karen and {Klaene}, Mark and {Mahadevan}, Suvrath and {Mathur}, Savita and {Mosser}, Beno{\^\i}t and {Muna}, Demitri and {Munn}, Jeffrey A. and {Nichol}, Robert C. and {O'Connell}, Robert W. and {Parejko}, John K. and {Robin}, A.~C. and {Rocha-Pinto}, Helio and {Schultheis}, Matthias and {Serenelli}, Aldo M. and {Shane}, Neville and {Silva Aguirre}, Victor and {Sobeck}, Jennifer S. and {Thompson}, Benjamin and {Troup}, Nicholas W. and {Weinberg}, David H. and {Zamora}, Olga},
        title = "{The Apache Point Observatory Galactic Evolution Experiment (APOGEE)}",
      journal = {AJ},
         year = 2017,
        month = sep,
       volume = {154},
       number = {3},
          eid = {94},
        pages = {94},
          doi = {10.3847/1538-3881/aa784d},
archivePrefix = {arXiv},
       eprint = {1509.05420},
 primaryClass = {astro-ph.IM},
       adsurl = {https://ui.adsabs.harvard.edu/abs/2017AJ....154...94M}
}

@ARTICLE{merrow_2024,
       author = {{Merrow}, Alex and {Grand}, Robert J.~J. and {Fragkoudi}, Francesca and {Martig}, Marie},
        title = "{Did the Gaia Enceladus/Sausage merger form the Milky Way's bar?}",
      journal = {MNRAS},
         year = 2024,
        month = jun,
       volume = {531},
       number = {1},
        pages = {1520-1533},
          doi = {10.1093/mnras/stae1250},
archivePrefix = {arXiv},
       eprint = {2312.02318},
 primaryClass = {astro-ph.GA},
       adsurl = {https://ui.adsabs.harvard.edu/abs/2024MNRAS.531.1520M}
}

@ARTICLE{monty_2024,
       author = {{Monty}, Stephanie and {Belokurov}, Vasily and {Sanders}, Jason L. and {Hansen}, Terese T. and {Sakari}, Charli M. and {McKenzie}, Madeleine and {Myeong}, GyuChul and {Davies}, Elliot Y. and {Ardern-Arentsen}, Anke and {Massari}, Davide},
        title = "{The ratio of [Eu/{\ensuremath{\alpha}}] differentiates accreted/in situ Milky Way stars across metallicities, as indicated by both field stars and globular clusters}",
      journal = {\mnras},
         year = 2024,
        month = sep,
       volume = {533},
       number = {2},
        pages = {2420-2440},
          doi = {10.1093/mnras/stae1895},
archivePrefix = {arXiv},
       eprint = {2405.08963},
 primaryClass = {astro-ph.GA},
       adsurl = {https://ui.adsabs.harvard.edu/abs/2024MNRAS.533.2420M}
}

@ARTICLE{nataf_2024,
       author = {{Nataf}, David M. and {Schlaufman}, Kevin C. and {Reggiani}, Henrique and {Hahn}, Isabel},
        title = "{Accurate, Precise, and Physically Self-consistent Ages and Metallicities for 400,000 Solar Neighborhood Subgiant Branch Stars}",
      journal = {\apj},
         year = 2024,
        month = nov,
       volume = {976},
       number = {1},
          eid = {87},
        pages = {87},
          doi = {10.3847/1538-4357/ad7c4e},
archivePrefix = {arXiv},
       eprint = {2407.18307},
 primaryClass = {astro-ph.SR},
       adsurl = {https://ui.adsabs.harvard.edu/abs/2024ApJ...976...87N}
}

@article{ness_2016,
	title = {The {Metallicity} {Distribution} of the {Milky} {Way} {Bulge}},
	volume = {33},
	doi = {10.1017/pasa.2015.51},
	journal = {PASA},
	author = {Ness, M. and Freeman, K.},
	month = jun,
	year = {2016},
	note = {\_eprint: 1511.07438},
	pages = {e022},
}

@article{nissen_2010,
	title = {Two distinct halo populations in the solar neighborhood. {Evidence} from stellar abundance ratios and kinematics},
	volume = {511},
	doi = {10.1051/0004-6361/200913877},
	journal = {A\&A},
	author = {Nissen, P. E. and Schuster, W. J.},
	month = feb,
	year = {2010},
	note = {\_eprint: 1002.4514},
	pages = {L10},
}

@article{nissen_2011,
	title = {Two distinct halo populations in the solar neighborhood. {II}. {Evidence} from stellar abundances of {Mn}, {Cu}, {Zn}, {Y}, and {Ba}},
	volume = {530},
	doi = {10.1051/0004-6361/201116619},
	journal = {A\&A},
	author = {Nissen, P. E. and Schuster, W. J.},
	month = jun,
	year = {2011},
	note = {\_eprint: 1103.4755},
	pages = {A15},
}

@ARTICLE{orkney_2022,
       author = {{Orkney}, Matthew D.~A. and {Laporte}, Chervin F.~P. and {Grand}, Robert J.~J. and {G{\'o}mez}, Facundo A. and {van de Voort}, Freeke and {Marinacci}, Federico and {Fragkoudi}, Francesca and {Pakmor}, Ruediger and {Springel}, Volker},
        title = "{The impact of two massive early accretion events in a Milky Way-like galaxy: repercussions for the buildup of the stellar disc and halo}",
      journal = {\mnras},
         year = 2022,
        month = nov,
       volume = {517},
       number = {1},
        pages = {L138-L142},
          doi = {10.1093/mnrasl/slac126},
archivePrefix = {arXiv},
       eprint = {2206.09246},
 primaryClass = {astro-ph.GA},
       adsurl = {https://ui.adsabs.harvard.edu/abs/2022MNRAS.517L.138O}
}

@ARTICLE{orkney_2023,
       author = {{Orkney}, Matthew D.~A. and {Laporte}, Chervin F.~P. and {Grand}, Robert J.~J. and {G{\'o}mez}, Facundo A. and {van de Voort}, Freeke and {Fattahi}, Azadeh and {Marinacci}, Federico and {Pakmor}, R{\"u}diger and {Fragkoudi}, Francesca and {Springel}, Volker},
        title = "{Exploring the diversity and similarity of radially anisotropic Milky Way-like stellar haloes: implications for disrupted dwarf galaxy searches}",
      journal = {\mnras},
         year = 2023,
        month = oct,
       volume = {525},
       number = {1},
        pages = {683-705},
          doi = {10.1093/mnras/stad2361},
archivePrefix = {arXiv},
       eprint = {2303.02147},
 primaryClass = {astro-ph.GA},
       adsurl = {https://ui.adsabs.harvard.edu/abs/2023MNRAS.525..683O}
}

@ARTICLE{orkney_2026,
       author = {{Orkney}, Matthew D.~A. and {Laporte}, Chervin F.~P.},
        title = "{Build-up and survival of the disc: from numerical models of galaxy formation to the Milky Way}",
      journal = {\mnras},
         year = 2026,
        month = jun,
       volume = {548},
       number = {4},
          eid = {staf2154},
        pages = {staf2154},
          doi = {10.1093/mnras/staf2154},
archivePrefix = {arXiv},
       eprint = {2509.09576},
 primaryClass = {astro-ph.GA},
       adsurl = {https://ui.adsabs.harvard.edu/abs/2026MNRAS.548f2154O}
}

@ARTICLE{pakmor_2016,
       author = {{Pakmor}, R{\"u}diger and {Springel}, Volker and {Bauer}, Andreas and {Mocz}, Philip and {Munoz}, Diego J. and {Ohlmann}, Sebastian T. and {Schaal}, Kevin and {Zhu}, Chenchong},
        title = "{Improving the convergence properties of the moving-mesh code AREPO}",
      journal = {MNRAS},
         year = 2016,
        month = jan,
       volume = {455},
       number = {1},
        pages = {1134-1143},
          doi = {10.1093/mnras/stv2380},
archivePrefix = {arXiv},
       eprint = {1503.00562},
 primaryClass = {astro-ph.GA},
       adsurl = {https://ui.adsabs.harvard.edu/abs/2016MNRAS.455.1134P}
}

@ARTICLE{pakmor_2025_a,
       author = {{Pakmor}, R{\"u}diger and {Bieri}, Rebekka and {Fragkoudi}, Francesca and {G{\'o}mez}, Facundo A. and {Grand}, Robert J.~J. and {Simpson}, Christine M. and {Talbot}, Rosie Y. and {van de Voort}, Freeke and {Werhahn}, Maria},
        title = "{Quantifying the intrinsic variability due to randomness of the Auriga galaxy formation model}",
      journal = {\mnras},
         year = 2025,
        month = oct,
       volume = {543},
       number = {2},
        pages = {1761-1774},
          doi = {10.1093/mnras/staf1542},
archivePrefix = {arXiv},
       eprint = {2507.13440},
 primaryClass = {astro-ph.GA},
       adsurl = {https://ui.adsabs.harvard.edu/abs/2025MNRAS.543.1761P}
}

@ARTICLE{pakmor_2025,
       author = {{Pakmor}, R{\"u}diger and {Fragkoudi}, Francesca and {Grand}, Robert J.~J. and {Simpson}, Christine M. and {G{\'o}mez}, Facundo A. and {van de Voort}, Freeke and {Bieri}, Rebekka and {Trick}, Wilma and {Werhahn}, Maria and {Talbot}, Rosie Y.},
        title = "{Auriga Superstars: Improving the resolution and fidelity of stellar dynamics in cosmological galaxy simulations}",
      journal = {\mnras},
         year = 2025,
        month = nov,
       volume = {543},
       number = {4},
        pages = {4355-4368},
          doi = {10.1093/mnras/staf1755},
archivePrefix = {arXiv},
       eprint = {2507.22104},
 primaryClass = {astro-ph.GA},
       adsurl = {https://ui.adsabs.harvard.edu/abs/2025MNRAS.543.4355P}
}

@ARTICLE{planck_2013,
       author = {{Planck Collaboration} and {Ade}, P.~A.~R. and {Aghanim}, N. and {Armitage-Caplan}, C. and {Arnaud}, M. and {Ashdown}, M. and {Atrio-Barandela}, F. and {Aumont}, J. and {Baccigalupi}, C. and {Banday}, A.~J. and {Barreiro}, R.~B. and {Bartlett}, J.~G. and {Battaner}, E. and {Benabed}, K. and {Beno{\^\i}t}, A. and {Benoit-L{\'e}vy}, A. and {Bernard}, J. -P. and {Bersanelli}, M. and {Bielewicz}, P. and {Bobin}, J. and {Bock}, J.~J. and {Bonaldi}, A. and {Bond}, J.~R. and {Borrill}, J. and {Bouchet}, F.~R. and {Bridges}, M. and {Bucher}, M. and {Burigana}, C. and {Butler}, R.~C. and {Calabrese}, E. and {Cappellini}, B. and {Cardoso}, J. -F. and {Catalano}, A. and {Challinor}, A. and {Chamballu}, A. and {Chary}, R. -R. and {Chen}, X. and {Chiang}, H.~C. and {Chiang}, L. -Y. and {Christensen}, P.~R. and {Church}, S. and {Clements}, D.~L. and {Colombi}, S. and {Colombo}, L.~P.~L. and {Couchot}, F. and {Coulais}, A. and {Crill}, B.~P. and {Curto}, A. and {Cuttaia}, F. and {Danese}, L. and {Davies}, R.~D. and {Davis}, R.~J. and {de Bernardis}, P. and {de Rosa}, A. and {de Zotti}, G. and {Delabrouille}, J. and {Delouis}, J. -M. and {D{\'e}sert}, F. -X. and {Dickinson}, C. and {Diego}, J.~M. and {Dolag}, K. and {Dole}, H. and {Donzelli}, S. and {Dor{\'e}}, O. and {Douspis}, M. and {Dunkley}, J. and {Dupac}, X. and {Efstathiou}, G. and {Elsner}, F. and {En{\ss}lin}, T.~A. and {Eriksen}, H.~K. and {Finelli}, F. and {Forni}, O. and {Frailis}, M. and {Fraisse}, A.~A. and {Franceschi}, E. and {Gaier}, T.~C. and {Galeotta}, S. and {Galli}, S. and {Ganga}, K. and {Giard}, M. and {Giardino}, G. and {Giraud-H{\'e}raud}, Y. and {Gjerl{\o}w}, E. and {Gonz{\'a}lez-Nuevo}, J. and {G{\'o}rski}, K.~M. and {Gratton}, S. and {Gregorio}, A. and {Gruppuso}, A. and {Gudmundsson}, J.~E. and {Haissinski}, J. and {Hamann}, J. and {Hansen}, F.~K. and {Hanson}, D. and {Harrison}, D. and {Henrot-Versill{\'e}}, S. and {Hern{\'a}ndez-Monteagudo}, C. and {Herranz}, D. and {Hildebrandt}, S.~R. and {Hivon}, E. and {Hobson}, M. and {Holmes}, W.~A. and {Hornstrup}, A. and {Hou}, Z. and {Hovest}, W. and {Huffenberger}, K.~M. and {Jaffe}, A.~H. and {Jaffe}, T.~R. and {Jewell}, J. and {Jones}, W.~C. and {Juvela}, M. and {Keih{\"a}nen}, E. and {Keskitalo}, R. and {Kisner}, T.~S. and {Kneissl}, R. and {Knoche}, J. and {Knox}, L. and {Kunz}, M. and {Kurki-Suonio}, H. and {Lagache}, G. and {L{\"a}hteenm{\"a}ki}, A. and {Lamarre}, J. -M. and {Lasenby}, A. and {Lattanzi}, M. and {Laureijs}, R.~J. and {Lawrence}, C.~R. and {Leach}, S. and {Leahy}, J.~P. and {Leonardi}, R. and {Le{\'o}n-Tavares}, J. and {Lesgourgues}, J. and {Lewis}, A. and {Liguori}, M. and {Lilje}, P.~B. and {Linden-V{\o}rnle}, M. and {L{\'o}pez-Caniego}, M. and {Lubin}, P.~M. and {Mac{\'\i}as-P{\'e}rez}, J.~F. and {Maffei}, B. and {Maino}, D. and {Mandolesi}, N. and {Maris}, M. and {Marshall}, D.~J. and {Martin}, P.~G. and {Mart{\'\i}nez-Gonz{\'a}lez}, E. and {Masi}, S. and {Massardi}, M. and {Matarrese}, S. and {Matthai}, F. and {Mazzotta}, P. and {Meinhold}, P.~R. and {Melchiorri}, A. and {Melin}, J. -B. and {Mendes}, L. and {Menegoni}, E. and {Mennella}, A. and {Migliaccio}, M. and {Millea}, M. and {Mitra}, S. and {Miville-Desch{\^e}nes}, M. -A. and {Moneti}, A. and {Montier}, L. and {Morgante}, G. and {Mortlock}, D. and {Moss}, A. and {Munshi}, D. and {Murphy}, J.~A. and {Naselsky}, P. and {Nati}, F. and {Natoli}, P. and {Netterfield}, C.~B. and {N{\o}rgaard-Nielsen}, H.~U. and {Noviello}, F. and {Novikov}, D. and {Novikov}, I. and {O'Dwyer}, I.~J. and {Osborne}, S. and {Oxborrow}, C.~A. and {Paci}, F. and {Pagano}, L. and {Pajot}, F. and {Paladini}, R. and {Paoletti}, D. and {Partridge}, B. and {Pasian}, F. and {Patanchon}, G. and {Pearson}, D. and {Pearson}, T.~J. and {Peiris}, H.~V. and {Perdereau}, O. and {Perotto}, L. and {Perrotta}, F. and {Pettorino}, V. and {Piacentini}, F. and {Piat}, M. and {Pierpaoli}, E. and {Pietrobon}, D. and {Plaszczynski}, S. and {Platania}, P. and {Pointecouteau}, E.},
        title = "{Planck 2013 results. XVI. Cosmological parameters}",
      journal = {A&A},
         year = 2014,
        month = nov,
       volume = {571},
          eid = {A16},
        pages = {A16},
          doi = {10.1051/0004-6361/201321591},
archivePrefix = {arXiv},
       eprint = {1303.5076},
 primaryClass = {astro-ph.CO},
       adsurl = {https://ui.adsabs.harvard.edu/abs/2014A&A...571A..16P}
}

@ARTICLE{portail_2017,
       author = {{Portail}, Matthieu and {Gerhard}, Ortwin and {Wegg}, Christopher and {Ness}, Melissa},
        title = "{Dynamical modelling of the galactic bulge and bar: the Milky Way's pattern speed, stellar and dark matter mass distribution}",
      journal = {\mnras},
         year = 2017,
        month = feb,
       volume = {465},
       number = {2},
        pages = {1621-1644},
          doi = {10.1093/mnras/stw2819},
archivePrefix = {arXiv},
       eprint = {1608.07954},
 primaryClass = {astro-ph.GA},
       adsurl = {https://ui.adsabs.harvard.edu/abs/2017MNRAS.465.1621P}
}

@ARTICLE{queiroz_2018,
       author = {{Queiroz}, A.~B.~A. and {Anders}, F. and {Santiago}, B.~X. and {Chiappini}, C. and {Steinmetz}, M. and {Dal Ponte}, M. and {Stassun}, K.~G. and {da Costa}, L.~N. and {Maia}, M.~A.~G. and {Crestani}, J. and {Beers}, T.~C. and {Fern{\'a}ndez-Trincado}, J.~G. and {Garc{\'\i}a-Hern{\'a}ndez}, D.~A. and {Roman-Lopes}, A. and {Zamora}, O.},
        title = "{StarHorse: a Bayesian tool for determining stellar masses, ages, distances, and extinctions for field stars}",
      journal = {\mnras},
         year = 2018,
        month = may,
       volume = {476},
       number = {2},
        pages = {2556-2583},
          doi = {10.1093/mnras/sty330},
archivePrefix = {arXiv},
       eprint = {1710.09970},
 primaryClass = {astro-ph.IM},
       adsurl = {https://ui.adsabs.harvard.edu/abs/2018MNRAS.476.2556Q}
}

@ARTICLE{rix_2022,
       author = {{Rix}, Hans-Walter and {Chandra}, Vedant and {Andrae}, Ren{\'e} and {Price-Whelan}, Adrian M. and {Weinberg}, David H. and {Conroy}, Charlie and {Fouesneau}, Morgan and {Hogg}, David W. and {De Angeli}, Francesca and {Naidu}, Rohan P. and {Xiang}, Maosheng and {Ruz-Mieres}, Daniela},
        title = "{The Poor Old Heart of the Milky Way}",
      journal = {\apj},
         year = 2022,
        month = dec,
       volume = {941},
       number = {1},
          eid = {45},
        pages = {45},
          doi = {10.3847/1538-4357/ac9e01},
archivePrefix = {arXiv},
       eprint = {2209.02722},
 primaryClass = {astro-ph.GA},
       adsurl = {https://ui.adsabs.harvard.edu/abs/2022ApJ...941...45R}
}

@ARTICLE{sanders_2019,
       author = {{Sanders}, Jason L. and {Smith}, Leigh and {Evans}, N. Wyn},
        title = "{The pattern speed of the Milky Way bar from transverse velocities}",
      journal = {MNRAS},
         year = 2019,
        month = oct,
       volume = {488},
       number = {4},
        pages = {4552-4564},
          doi = {10.1093/mnras/stz1827},
archivePrefix = {arXiv},
       eprint = {1903.02009},
 primaryClass = {astro-ph.GA},
       adsurl = {https://ui.adsabs.harvard.edu/abs/2019MNRAS.488.4552S}
}

@ARTICLE{sante_2025,
       author = {{Sante}, Andrea and {Kawata}, Daisuke and {Font}, Andreea S. and {Grand}, Robert J.~J.},
        title = "{GalactiKit: reconstructing mergers from z = 0 debris using simulation-based inference in Auriga}",
      journal = {\mnras},
         year = 2025,
        month = sep,
       volume = {542},
       number = {3},
        pages = {1776-1790},
          doi = {10.1093/mnras/staf1343},
archivePrefix = {arXiv},
       eprint = {2502.14972},
 primaryClass = {astro-ph.GA},
       adsurl = {https://ui.adsabs.harvard.edu/abs/2025MNRAS.542.1776S}
}

@ARTICLE{sante_2026,
       author = {{Sante}, Andrea and {Font}, Andreea S. and {Kawata}, Daisuke and {Makinen}, T. Lucas and {Grand}, Robert J.~J.},
        title = "{A simulation-based inference of the Milky Way merger history}",
      journal = {\mnras},
         year = 2026,
        month = aug,
       volume = {550},
       number = {2},
          eid = {stag1162},
        pages = {stag1162},
          doi = {10.1093/mnras/stag1162},
archivePrefix = {arXiv},
       eprint = {2603.12317},
 primaryClass = {astro-ph.GA},
       adsurl = {https://ui.adsabs.harvard.edu/abs/2026MNRAS.550g1162S}
}

@ARTICLE{santiago_2016,
       author = {{Santiago}, Bas{\'\i}lio X. and {Brauer}, Doroth{\'e}e E. and {Anders}, Friedrich and {Chiappini}, Cristina and {Queiroz}, Anna B. and {Girardi}, L{\'e}o and {Rocha-Pinto}, Helio J. and {Balbinot}, Eduardo and {da Costa}, Luiz N. and {Maia}, Marcio A.~G. and {Schultheis}, Mathias and {Steinmetz}, Matthias and {Miglio}, Andrea and {Montalb{\'a}n}, Josefina and {Schneider}, Donald P. and {Beers}, Timothy C. and {Frinchaboy}, Peter M. and {Lee}, Young Sun and {Zasowski}, Gail},
        title = "{Spectro-photometric distances to stars: A general purpose Bayesian approach}",
      journal = {\aap},
         year = 2016,
        month = jan,
       volume = {585},
          eid = {A42},
        pages = {A42},
          doi = {10.1051/0004-6361/201323177},
archivePrefix = {arXiv},
       eprint = {1501.05500},
 primaryClass = {astro-ph.IM},
       adsurl = {https://ui.adsabs.harvard.edu/abs/2016A&A...585A..42S}
}

@ARTICLE{schaye_2015,
       author = {{Schaye}, Joop and {Crain}, Robert A. and {Bower}, Richard G. and {Furlong}, Michelle and {Schaller}, Matthieu and {Theuns}, Tom and {Dalla Vecchia}, Claudio and {Frenk}, Carlos S. and {McCarthy}, I.~G. and {Helly}, John C. and {Jenkins}, Adrian and {Rosas-Guevara}, Y.~M. and {White}, Simon D.~M. and {Baes}, Maarten and {Booth}, C.~M. and {Camps}, Peter and {Navarro}, Julio F. and {Qu}, Yan and {Rahmati}, Alireza and {Sawala}, Till and {Thomas}, Peter A. and {Trayford}, James},
        title = "{The EAGLE project: simulating the evolution and assembly of galaxies and their environments}",
      journal = {MNRAS},
         year = 2015,
        month = jan,
       volume = {446},
       number = {1},
        pages = {521-554},
          doi = {10.1093/mnras/stu2058},
archivePrefix = {arXiv},
       eprint = {1407.7040},
 primaryClass = {astro-ph.GA},
       adsurl = {https://ui.adsabs.harvard.edu/abs/2015MNRAS.446..521S}
}

@ARTICLE{schonrich_2010,
       author = {{Sch{\"o}nrich}, Ralph and {Binney}, James and {Dehnen}, Walter},
        title = "{Local kinematics and the local standard of rest}",
      journal = {\mnras},
         year = 2010,
        month = apr,
       volume = {403},
       number = {4},
        pages = {1829-1833},
          doi = {10.1111/j.1365-2966.2010.16253.x},
archivePrefix = {arXiv},
       eprint = {0912.3693},
 primaryClass = {astro-ph.GA},
       adsurl = {https://ui.adsabs.harvard.edu/abs/2010MNRAS.403.1829S}
}

@ARTICLE{sharma_2019,
       author = {{Sharma}, Sanjib and {Stello}, Dennis and {Bland-Hawthorn}, Joss and {Hayden}, Michael R. and {Zinn}, Joel C. and {Kallinger}, Thomas and {Hon}, Marc and {Asplund}, Martin and {Buder}, Sven and {De Silva}, Gayandhi M. and {D'Orazi}, Valentina and {Freeman}, Ken and {Kos}, Janez and {Lewis}, Geraint F. and {Lin}, Jane and {Lind}, Karin and {Martell}, Sarah and {Simpson}, Jeffrey D. and {Wittenmyer}, Rob A. and {Zucker}, Daniel B. and {Zwitter}, Tomaz and {Bedding}, Timothy R. and {Chen}, Boquan and {Cotar}, Klemen and {Esdaile}, James and {Horner}, Jonathan and {Huber}, Daniel and {Kafle}, Prajwal R. and {Khanna}, Shourya and {Li}, Tanda and {Ting}, Yuan-Sen and {Nataf}, David M. and {Nordlander}, Thomas and {Saadon}, Mohd Hafiz Mohd and {Traven}, Gregor and {Wright}, Duncan and {Wyse}, Rosemary F.~G.},
        title = "{The K2-HERMES Survey: age and metallicity of the thick disc}",
      journal = {\mnras},
         year = 2019,
        month = dec,
       volume = {490},
       number = {4},
        pages = {5335-5352},
          doi = {10.1093/mnras/stz2861},
archivePrefix = {arXiv},
       eprint = {1904.12444},
 primaryClass = {astro-ph.GA},
       adsurl = {https://ui.adsabs.harvard.edu/abs/2019MNRAS.490.5335S}
}

@phdthesis{sipma_2022,
  author = {Sipma, Sten},
  title = {Kinematics of Very Metal-Poor Stars in the Inner Regions of Aurigaia Galaxies},
  school = {University of Groningen},
  year = {2022},
  type = {Bachelor's thesis},
  url = {https://fse.studenttheses.ub.rug.nl/id/eprint/28009}
}

@ARTICLE{springel_2001a,
       author = {{Springel}, Volker and {Yoshida}, Naoki and {White}, Simon D.~M.},
        title = "{GADGET: a code for collisionless and gasdynamical cosmological simulations}",
      journal = {New Astron.},
         year = 2001,
        month = apr,
       volume = {6},
       number = {2},
        pages = {79-117},
          doi = {10.1016/S1384-1076(01)00042-2},
archivePrefix = {arXiv},
       eprint = {astro-ph/0003162},
 primaryClass = {astro-ph},
       adsurl = {https://ui.adsabs.harvard.edu/abs/2001NewA....6...79S}
}

@ARTICLE{springel_2001b,
       author = {{Springel}, Volker and {White}, Simon D.~M. and {Tormen}, Giuseppe and {Kauffmann}, Guinevere},
        title = "{Populating a cluster of galaxies - I. Results at z=0}",
      journal = {MNRAS},
         year = 2001,
        month = dec,
       volume = {328},
       number = {3},
        pages = {726-750},
          doi = {10.1046/j.1365-8711.2001.04912.x},
archivePrefix = {arXiv},
       eprint = {astro-ph/0012055},
 primaryClass = {astro-ph},
       adsurl = {https://ui.adsabs.harvard.edu/abs/2001MNRAS.328..726S}
}

@ARTICLE{springel_2010,
       author = {{Springel}, Volker},
        title = "{E pur si muove: Galilean-invariant cosmological hydrodynamical simulations on a moving mesh}",
      journal = {MNRAS},
         year = 2010,
        month = jan,
       volume = {401},
       number = {2},
        pages = {791-851},
          doi = {10.1111/j.1365-2966.2009.15715.x},
archivePrefix = {arXiv},
       eprint = {0901.4107},
 primaryClass = {astro-ph.CO},
       adsurl = {https://ui.adsabs.harvard.edu/abs/2010MNRAS.401..791S}
}

@article{starkenburg_2017a,
	title = {The oldest and most metal-poor stars in the {APOSTLE} {Local} {Group} simulations},
	volume = {465},
	doi = {10.1093/mnras/stw2873},
	number = {2},
	journal = {MNRAS},
	author = {Starkenburg, Else and Oman, Kyle A. and Navarro, Julio F. and Crain, Robert A. and Fattahi, Azadeh and Frenk, Carlos S. and Sawala, Till and Schaye, Joop},
	month = feb,
	year = {2017},
	note = {\_eprint: 1609.05214},
	pages = {2212--2224},
}

@ARTICLE{starkenburg_2017b,
       author = {{Starkenburg}, Else and {Martin}, Nicolas and {Youakim}, Kris and {Aguado}, David S. and {Allende Prieto}, Carlos and {Arentsen}, Anke and {Bernard}, Edouard J. and {Bonifacio}, Piercarlo and {Caffau}, Elisabetta and {Carlberg}, Raymond G. and {C{\^o}t{\'e}}, Patrick and {Fouesneau}, Morgan and {Fran{\c{c}}ois}, Patrick and {Franke}, Oliver and {Gonz{\'a}lez Hern{\'a}ndez}, Jonay I. and {Gwyn}, Stephen D.~J. and {Hill}, Vanessa and {Ibata}, Rodrigo A. and {Jablonka}, Pascale and {Longeard}, Nicolas and {McConnachie}, Alan W. and {Navarro}, Julio F. and {S{\'a}nchez-Janssen}, Rub{\'e}n and {Tolstoy}, Eline and {Venn}, Kim A.},
        title = "{The Pristine survey - I. Mining the Galaxy for the most metal-poor stars}",
      journal = {\mnras},
         year = 2017,
        month = nov,
       volume = {471},
       number = {3},
        pages = {2587-2604},
          doi = {10.1093/mnras/stx1068},
archivePrefix = {arXiv},
       eprint = {1705.01113},
 primaryClass = {astro-ph.GA},
       adsurl = {https://ui.adsabs.harvard.edu/abs/2017MNRAS.471.2587S}
}

@ARTICLE{thomas_2025,
       author = {{Thomas}, Guillaume F. and {Battaglia}, Giuseppina and {Grand}, Robert J.~J. and {Aguiar {\'A}lvarez}, Amanda},
        title = "{How well can we unravel the accreted constituents of the Milky Way stellar halo?: A test on cosmological hydrodynamical simulations}",
      journal = {\aap},
         year = 2025,
        month = dec,
       volume = {704},
          eid = {A40},
        pages = {A40},
          doi = {10.1051/0004-6361/202452449},
archivePrefix = {arXiv},
       eprint = {2504.10398},
 primaryClass = {astro-ph.GA},
       adsurl = {https://ui.adsabs.harvard.edu/abs/2025A&A...704A..40T}
}

@ARTICLE{tomlinson_2026,
       author = {{Tomlinson}, Thomas and {Fragkoudi}, Francesca and {Carrillo}, Andreia and {Fattahi}, Azadeh and {Gherghinescu}, Paula and {Deason}, Alis and {Pakmor}, R{\"u}diger and {Grand}, Robert J.~J. and {G{\'o}mez}, Facundo A. and {van de Voort}, Freeke and {Bieri}, Rebekka},
        title = "{Stirring things up: Bar-induced substructures in the stellar halo of a cosmological Milky Way analogue}",
      journal = {\mnras},
         year = 2026,
        month = jun,
       volume = {549},
       number = {1},
          eid = {stag841},
        pages = {stag841},
          doi = {10.1093/mnras/stag841},
archivePrefix = {arXiv},
       eprint = {2601.14409},
 primaryClass = {astro-ph.GA},
       adsurl = {https://ui.adsabs.harvard.edu/abs/2026MNRAS.549ag841T}
}

@ARTICLE{venn_2004,
       author = {{Venn}, Kim A. and {Irwin}, Mike and {Shetrone}, Matthew D. and {Tout}, Christopher A. and {Hill}, Vanessa and {Tolstoy}, Eline},
        title = "{Stellar Chemical Signatures and Hierarchical Galaxy Formation}",
      journal = {\aj},
         year = 2004,
        month = sep,
       volume = {128},
       number = {3},
        pages = {1177-1195},
          doi = {10.1086/422734},
archivePrefix = {arXiv},
       eprint = {astro-ph/0406120},
 primaryClass = {astro-ph},
       adsurl = {https://ui.adsabs.harvard.edu/abs/2004AJ....128.1177V}
}

@misc{vizier_data,
  author= {{Ardern-Arentsen}, A. and {Monari}, G. and {Queiroz}, A.B.A. and {Starkenburg}, E. and {Martin}, N.F. and {Chiappini}, C.and {Aguado}, D.S. and {Belokurov}, V. and {Carlberg}, R. and {Monty}, S. and {Myeong}, G. and {Schultheis}, M. and {Sestito}, F. and {Venn}, K.A. and {Vitali}, S. and {Yuan}, Z. and {Zhang}, H. and {Buder}, S. and {Lewis}, G.F. and {Oliver}, W.H. and {Wan}, Z. and {Zucker}, D.B.},
  title = "{VizieR Catalogue: Pristine Inner Galaxy Survey (PIGS) VIII}",
  year = 2024,
  doi = {10.26093/cds/vizier.75303391}
}

@ARTICLE{white_1978,
       author = {{White}, S.~D.~M. and {Rees}, M.~J.},
        title = "{Core condensation in heavy halos: a two-stage theory for galaxy formation and clustering.}",
      journal = {\mnras},
         year = 1978,
        month = may,
       volume = {183},
        pages = {341-358},
          doi = {10.1093/mnras/183.3.341},
       adsurl = {https://ui.adsabs.harvard.edu/abs/1978MNRAS.183..341W}
}

@ARTICLE{white_1991,
       author = {{White}, Simon D.~M. and {Frenk}, Carlos S.},
        title = "{Galaxy Formation through Hierarchical Clustering}",
      journal = {\apj},
         year = 1991,
        month = sep,
       volume = {379},
        pages = {52},
          doi = {10.1086/170483},
       adsurl = {https://ui.adsabs.harvard.edu/abs/1991ApJ...379...52W}
}

%\begin{thebibliography}{99}
%\end{thebibliography}

%%%%%%%%%%%%%%%%%%%%%%%%%%%%%%%%%%%%%%%%%%%%%%%%%%

%%%%%%%%%%%%%%%%% APPENDICES %%%%%%%%%%%%%%%%%%%%%

\clearpage

\appendix

\section{MDF in bins of $E-L_z$ in halo 22 and halo 27 } \label{sec:appendix a}

To make robust predictions for the Milky Way, we want to show results for two other haloes in the Superstars suite: halo 22 and halo 27. These were selected to have different accreted fractions compared to halo 18. For the fiducial Auriga simulations, accreted stellar fractions can be found in the supplementary material of \citet{orkney_2023}. Halo 18 has an accreted fraction in the inner 4\,kpc of $0.6\%$, halo 22 has an accreted fraction of $0.2\%$, and halo 27 has an accreted fraction of $5\%$. Morphologically, haloes 22 and 27 are less similar to the Milky Way than halo 18. Halo 22 has a shorter, weaker bar, while halo 27 is an unbarred galaxy. The rotation of stars in the inner Milky Way is also a better match for halo 18.

In \figref{fig:mdf_grid_au22} and \figref{fig:mdf_grid_au27}, we show histograms of [Fe/H] in different bins of $E-L_z$ space, similar to \figref{fig:e_lz_bin_plot}. The histograms are  normalised with respect to the total number of stars in each bin, to show the contributions from the proto-galaxy, in-situ, and accreted stars. The fractions in the boxes in each histogram panel indicate the contribution from each component in this bin of $E-L_z$ space. In the top right are shown the total contributions, whereas in the top left we show the contribution for stars with $[\mathrm{Fe/H}] < -1$.

We can see that there is significant halo-to-halo variation in the individual histograms. We want to note in particular that halo 22 has a much higher proto-galaxy fraction at high energy compared to haloes 18 and 27. Halo 27, on the other hand, has a much higher accreted fraction at high energy compared to the other two haloes. In all three haloes, at low energies, the proto-galaxy is the dominant component for $[\mathrm{Fe/H}] < -1$. In the lowest two energy bins the proto-galaxy fraction for $[\mathrm{Fe/H}] < -1$ is always above $96\%$ for all three haloes considered in this paper.

\renewcommand\thefigure{A.\arabic{figure}} 

\begin{figure*}
	\includegraphics[width=\textwidth]{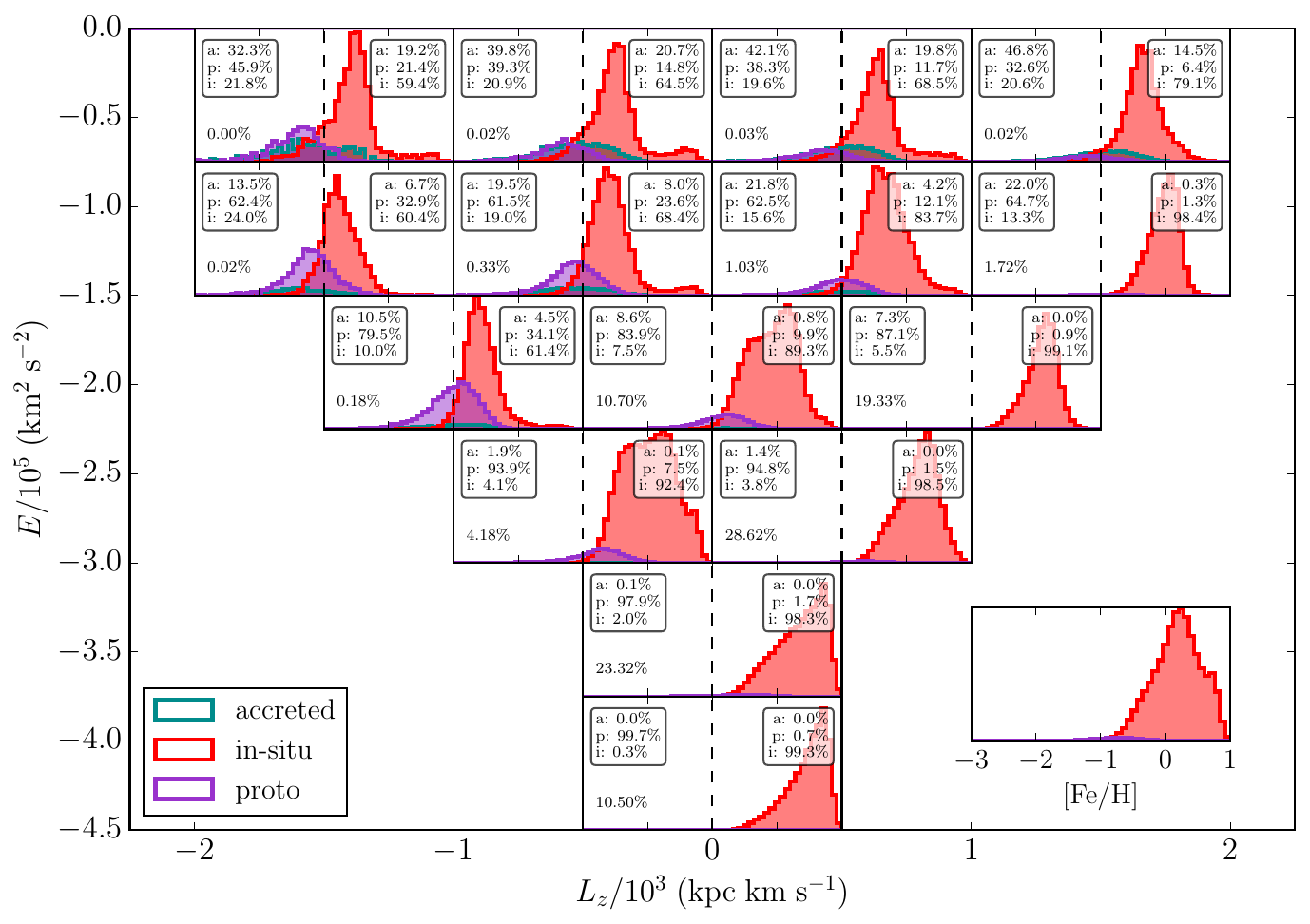}
    \caption{We show the metallicity distribution functions of stars, within 4\,kpc of the centre, in different bins of $E-L_z$ space of halo 22, normalised with respect to the number of stars in each bin. We separate stars into different bins in $E-L_z$ of equal size with an extent of $1 \times 10^3 \rm \;kpc \,km \, s^{-1}$ in $L_z$ and $0.75 \times 10^5 \; \rm km^2 \, s^{-2}$ in E. The borders of the subplots in $E-L_z$ correspond to the boundaries of the bin in $E-L_z$. In each subplot we separate stars into accreted (turquoise), in-situ (red), and proto-galaxy (purple). The range of the y-axes of the different subplots changes between subplots, and thus the size of the peaks in the subplots should not be used to compare the number of stars present in different regions of $E-L_z$ space. To do that we include the percentage of stars that are found in each bin in $E-L_z$ (compared to all stars within 4\,kpc of the centre) in the bottom left of each subplot. The percentages in the top right show the fractional contributions of accreted (a), proto-galaxy (p), and in-situ (i) to the total number of stars in this $E-L_z$ bin. The percentages in the top left show the same fractions for all stars in this bin in $E-L_z$ with $[\rm Fe/H] < -1$. The inset in the lower right of the plot shows the MDFs of stars in the inner 4\,kpc, with no cuts in $E$ and $L_z$. For halo 22, the proto-galaxy fraction for $[\rm Fe/H] < -1$ is generally much larger than in halo 18 and 27, with a proto-galaxy contribution of more than $30\%$ for all of the highest energy bins.} 
    \label{fig:mdf_grid_au22}
\end{figure*}

\begin{figure*}
	\includegraphics[width=\textwidth]{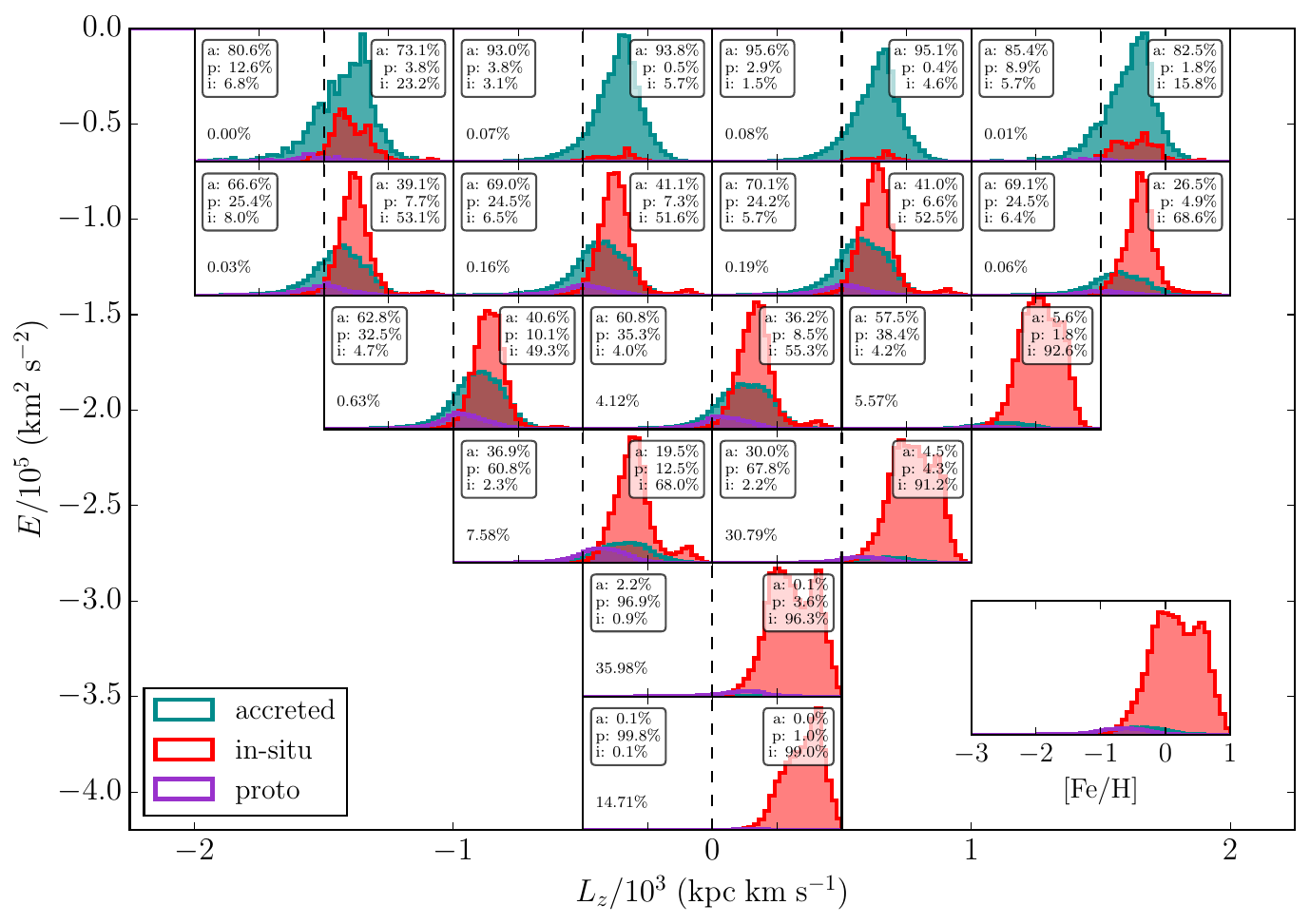}
    \caption{We show the metallicity distribution functions of stars, within 4\,kpc of the centre, in different bins of $E-L_z$ space of halo 27, normalised with respect to the number of stars in each bin. We separate stars into different bins in $E-L_z$ of equal size with an extent of $1 \times 10^3 \rm \;kpc \,km \, s^{-1}$ in $L_z$ and $0.7 \times 10^5 \; \rm km^2 \, s^{-2}$ in E. The borders of the subplots in $E-L_z$ correspond to the boundaries of the bin in $E-L_z$. In each subplot we separate stars into accreted (turquoise), in-situ (red), and proto-galaxy (purple). The range of the y-axes of the different subplots changes between subplots, and thus the size of the peaks in the subplots should not be used to compare the number of stars present in different regions of $E-L_z$ space. To do that we include the percentage of stars that are found in each bin in $E-L_z$ (compared to all stars within 4\,kpc of the centre) in the bottom left of each subplot. The percentages in the top right show the fractional contributions of accreted (a), proto-galaxy (p), and in-situ (i) to the total number of stars in this $E-L_z$ bin. The percentages in the top left show the same fractions for all stars in this bin in $E-L_z$ with $[\rm Fe/H] < -1$. The inset in the lower right of the plot shows the MDFs of stars in the inner 4\,kpc, with no cuts in $E$ and $L_z$. Halo 27 has a much higher accreted fraction compared to haloes 18 and 22, which is reflected in this figure. Due to an ongoing merger, halo 27 has a particularly elevated accreted fraction at energies above $-1 \times 10^5 \; \rm km^2 \, s^{-2}$.}
    \label{fig:mdf_grid_au27}
\end{figure*}

%%%%%%%%%%%%%%%%%%%%%%%%%%%%%%%%%%%%%%%%%%%%%%%%%%

% Don't change these lines
\bsp	% typesetting comment
\label{lastpage}
\end{document}